\documentclass[twocolumn]{aastex631}
\pdfoutput=1
\usepackage[utf8]{inputenc}
\usepackage{array}
\usepackage{xspace}
\usepackage{xstring}
\usepackage{acronym}
\usepackage{tabularx}
\usepackage{comment}
\usepackage{multirow}
\usepackage{rotating}
\usepackage{amsmath,amssymb,graphicx}
\usepackage{comment}
\usepackage{xcolor}
\usepackage{graphicx}
\usepackage[caption=false]{subfig}
\usepackage{makerobust}
\usepackage{ulem}
\usepackage{physics}
\usepackage{lipsum}
\usepackage{xfp}

\newcommand{\gwTenElevenSpinOneZMedian}{\ensuremath 0.66}
\newcommand{\gwTenElevenSpinOneZUpperError}{\ensuremath 0.08}
\newcommand{\gwTenElevenSpinOneZLowerError}{\ensuremath 0.09}

\newcommand{\gwTenElevenSpinOnePlaneMedian}{\ensuremath 0.39}
\newcommand{\gwTenElevenSpinOnePlaneUpperError}{\ensuremath 0.19}
\newcommand{\gwTenElevenSpinOnePlaneLowerError}{\ensuremath 0.14}

\newcommand{\gwTenElevenChiEffectiveMedian}{\ensuremath 0.49}
\newcommand{\gwTenElevenChiEffectiveUpperError}{\ensuremath 0.06}
\newcommand{\gwTenElevenChiEffectiveLowerError}{\ensuremath 0.07}

\newcommand{\gwTenElevenChiPMedian}{\ensuremath 0.39}
\newcommand{\gwTenElevenChiPUpperError}{\ensuremath 0.19}
\newcommand{\gwTenElevenChiPLowerError}{\ensuremath 0.14}

\newcommand{\gwTenElevenSpinOneMagMedian}{\ensuremath 0.78}
\newcommand{\gwTenElevenSpinOneMagUpperError}{\ensuremath 0.09}
\newcommand{\gwTenElevenSpinOneMagLowerError}{\ensuremath 0.09}
\newcommand{\gwTenElevenSpinOneMagLowerLimit}{\ensuremath 0.69}

\newcommand{\gwTenElevenSpinOneTiltMedian}{\ensuremath 31}

\newcommand{\gwTenElevenSpinOneTiltUpperErrorHPD}{\ensuremath 11}
\newcommand{\gwTenElevenSpinOneTiltLowerErrorHPD}{\ensuremath 14}

\newcommand{\gwTenElevenMassOne}{\ensuremath 19.6}
\newcommand{\gwTenElevenMassOneUpperError}{\ensuremath 3.6}
\newcommand{\gwTenElevenMassOneLowerError}{\ensuremath 2.5}

\newcommand{\gwTenElevenMassTwo}{\ensuremath 5.9}
\newcommand{\gwTenElevenMassTwoUpperError}{\ensuremath 0.8}
\newcommand{\gwTenElevenMassTwoLowerError}{\ensuremath 0.8}

\newcommand{\gwTenElevenMassRatio}{\ensuremath 0.30}
\newcommand{\gwTenElevenMassRatioUpperError}{\ensuremath 0.09}
\newcommand{\gwTenElevenMassRatioLowerError}{\ensuremath 0.08}

\newcommand{\gwTenElevenChirpMass}{\ensuremath 9.1}
\newcommand{\gwTenElevenChirpMassUpperError}{\ensuremath 0.1}
\newcommand{\gwTenElevenChirpMassLowerError}{\ensuremath 0.1}

\newcommand{\gwTenElevenDistance}{\ensuremath 214}
\newcommand{\gwTenElevenDistanceUpperError}{\ensuremath 44}
\newcommand{\gwTenElevenDistanceLowerError}{\ensuremath 48}
\newcommand{\gwTenElevenDistanceUpperLimit}{\ensuremath 248}

\newcommand{\gwTenElevenSEOBSpinOneZMedian}{\ensuremath 0.64}
\newcommand{\gwTenElevenSEOBSpinOneZUpperError}{\ensuremath 0.06}
\newcommand{\gwTenElevenSEOBSpinOneZLowerError}{\ensuremath 0.08}

\newcommand{\gwTenElevenSEOBSpinOnePlaneMedian}{\ensuremath 0.45}
\newcommand{\gwTenElevenSEOBSpinOnePlaneUpperError}{\ensuremath 0.19}
\newcommand{\gwTenElevenSEOBSpinOnePlaneLowerError}{\ensuremath 0.15}

\newcommand{\gwTenElevenSEOBChiEffectiveMedian}{\ensuremath 0.50}
\newcommand{\gwTenElevenSEOBChiEffectiveUpperError}{\ensuremath 0.05}
\newcommand{\gwTenElevenSEOBChiEffectiveLowerError}{\ensuremath 0.06}

\newcommand{\gwTenElevenSEOBChiPMedian}{\ensuremath 0.45}
\newcommand{\gwTenElevenSEOBChiPUpperError}{\ensuremath 0.19}
\newcommand{\gwTenElevenSEOBChiPLowerError}{\ensuremath 0.15}

\newcommand{\gwTenElevenSEOBSpinOneMagMedian}{\ensuremath 0.79}
\newcommand{\gwTenElevenSEOBSpinOneMagUpperError}{\ensuremath 0.09}
\newcommand{\gwTenElevenSEOBSpinOneMagLowerError}{\ensuremath 0.08}

\newcommand{\gwTenElevenSEOBSpinOneTiltMedian}{\ensuremath 35}

\newcommand{\gwTenElevenSEOBSpinOneTiltUpperErrorHPD}{\ensuremath 11}
\newcommand{\gwTenElevenSEOBSpinOneTiltLowerErrorHPD}{\ensuremath 14}

\newcommand{\gwTenElevenSEOBMassOne}{\ensuremath 21.1}
\newcommand{\gwTenElevenSEOBMassOneUpperError}{\ensuremath 2.8}
\newcommand{\gwTenElevenSEOBMassOneLowerError}{\ensuremath 3.6}

\newcommand{\gwTenElevenSEOBMassTwo}{\ensuremath 5.5}
\newcommand{\gwTenElevenSEOBMassTwoUpperError}{\ensuremath 0.9}
\newcommand{\gwTenElevenSEOBMassTwoLowerError}{\ensuremath 0.5}

\newcommand{\gwTenElevenSEOBMassRatio}{\ensuremath 0.26}
\newcommand{\gwTenElevenSEOBMassRatioUpperError}{\ensuremath 0.11}
\newcommand{\gwTenElevenSEOBMassRatioLowerError}{\ensuremath 0.05}

\newcommand{\gwTenElevenSEOBChirpMass}{\ensuremath 9.0}
\newcommand{\gwTenElevenSEOBChirpMassUpperError}{\ensuremath 0.1}
\newcommand{\gwTenElevenSEOBChirpMassLowerError}{\ensuremath 0.1}

\newcommand{\gwTenElevenSEOBDistance}{\ensuremath 225}
\newcommand{\gwTenElevenSEOBDistanceUpperError}{\ensuremath 39}
\newcommand{\gwTenElevenSEOBDistanceLowerError}{\ensuremath 40}

\newcommand{\gwTenElevenXOFourASpinOneZMedian}{\ensuremath 0.70}
\newcommand{\gwTenElevenXOFourASpinOneZUpperError}{\ensuremath 0.05}
\newcommand{\gwTenElevenXOFourASpinOneZLowerError}{\ensuremath 0.08}

\newcommand{\gwTenElevenXOFourASpinOnePlaneMedian}{\ensuremath 0.35}
\newcommand{\gwTenElevenXOFourASpinOnePlaneUpperError}{\ensuremath 0.17}
\newcommand{\gwTenElevenXOFourASpinOnePlaneLowerError}{\ensuremath 0.13}

\newcommand{\gwTenElevenXOFourAChiEffectiveMedian}{\ensuremath 0.44}
\newcommand{\gwTenElevenXOFourAChiEffectiveUpperError}{\ensuremath 0.05}
\newcommand{\gwTenElevenXOFourAChiEffectiveLowerError}{\ensuremath 0.04}

\newcommand{\gwTenElevenXOFourAChiPMedian}{\ensuremath 0.35}
\newcommand{\gwTenElevenXOFourAChiPUpperError}{\ensuremath 0.17}
\newcommand{\gwTenElevenXOFourAChiPLowerError}{\ensuremath 0.13}

\newcommand{\gwTenElevenXOFourASpinOneMagMedian}{\ensuremath 0.79}
\newcommand{\gwTenElevenXOFourASpinOneMagUpperError}{\ensuremath 0.08}
\newcommand{\gwTenElevenXOFourASpinOneMagLowerError}{\ensuremath 0.10}

\newcommand{\gwTenElevenXOFourASpinOneTiltMedian}{\ensuremath 27}

\newcommand{\gwTenElevenXOFourASpinOneTiltUpperErrorHPD}{\ensuremath 9}
\newcommand{\gwTenElevenXOFourASpinOneTiltLowerErrorHPD}{\ensuremath 12}

\newcommand{\gwTenElevenXOFourAMassOne}{\ensuremath 18.6}
\newcommand{\gwTenElevenXOFourAMassOneUpperError}{\ensuremath 2.1}
\newcommand{\gwTenElevenXOFourAMassOneLowerError}{\ensuremath 2.1}

\newcommand{\gwTenElevenXOFourAMassTwo}{\ensuremath 6.2}
\newcommand{\gwTenElevenXOFourAMassTwoUpperError}{\ensuremath 0.7}
\newcommand{\gwTenElevenXOFourAMassTwoLowerError}{\ensuremath 0.5}

\newcommand{\gwTenElevenXOFourAMassRatio}{\ensuremath 0.33}
\newcommand{\gwTenElevenXOFourAMassRatioUpperError}{\ensuremath 0.08}
\newcommand{\gwTenElevenXOFourAMassRatioLowerError}{\ensuremath 0.06}

\newcommand{\gwTenElevenXOFourAChirpMass}{\ensuremath 9.1}
\newcommand{\gwTenElevenXOFourAChirpMassUpperError}{\ensuremath 0.1}
\newcommand{\gwTenElevenXOFourAChirpMassLowerError}{\ensuremath 0.1}

\newcommand{\gwTenElevenXOFourADistance}{\ensuremath 201}
\newcommand{\gwTenElevenXOFourADistanceUpperError}{\ensuremath 44}
\newcommand{\gwTenElevenXOFourADistanceLowerError}{\ensuremath 46}

\newcommand{\gwTenElevenXPHMSpinOneZMedian}{\ensuremath 0.64}
\newcommand{\gwTenElevenXPHMSpinOneZUpperError}{\ensuremath 0.06}
\newcommand{\gwTenElevenXPHMSpinOneZLowerError}{\ensuremath 0.09}

\newcommand{\gwTenElevenXPHMSpinOnePlaneMedian}{\ensuremath 0.38}
\newcommand{\gwTenElevenXPHMSpinOnePlaneUpperError}{\ensuremath 0.15}
\newcommand{\gwTenElevenXPHMSpinOnePlaneLowerError}{\ensuremath 0.13}

\newcommand{\gwTenElevenXPHMChiEffectiveMedian}{\ensuremath 0.51}
\newcommand{\gwTenElevenXPHMChiEffectiveUpperError}{\ensuremath 0.05}
\newcommand{\gwTenElevenXPHMChiEffectiveLowerError}{\ensuremath 0.04}

\newcommand{\gwTenElevenXPHMChiPMedian}{\ensuremath 0.38}
\newcommand{\gwTenElevenXPHMChiPUpperError}{\ensuremath 0.15}
\newcommand{\gwTenElevenXPHMChiPLowerError}{\ensuremath 0.13}

\newcommand{\gwTenElevenXPHMSpinOneMagMedian}{\ensuremath 0.75}
\newcommand{\gwTenElevenXPHMSpinOneMagUpperError}{\ensuremath 0.07}
\newcommand{\gwTenElevenXPHMSpinOneMagLowerError}{\ensuremath 0.07}

\newcommand{\gwTenElevenXPHMSpinOneTiltMedian}{\ensuremath 31}

\newcommand{\gwTenElevenXPHMSpinOneTiltUpperErrorHPD}{\ensuremath 10}
\newcommand{\gwTenElevenXPHMSpinOneTiltLowerErrorHPD}{\ensuremath 12}

\newcommand{\gwTenElevenXPHMMassOne}{\ensuremath 19.8}
\newcommand{\gwTenElevenXPHMMassOneUpperError}{\ensuremath 2.4}
\newcommand{\gwTenElevenXPHMMassOneLowerError}{\ensuremath 2.4}

\newcommand{\gwTenElevenXPHMMassTwo}{\ensuremath 5.9}
\newcommand{\gwTenElevenXPHMMassTwoUpperError}{\ensuremath 0.7}
\newcommand{\gwTenElevenXPHMMassTwoLowerError}{\ensuremath 0.5}

\newcommand{\gwTenElevenXPHMMassRatio}{\ensuremath 0.30}
\newcommand{\gwTenElevenXPHMMassRatioUpperError}{\ensuremath 0.08}
\newcommand{\gwTenElevenXPHMMassRatioLowerError}{\ensuremath 0.06}

\newcommand{\gwTenElevenXPHMChirpMass}{\ensuremath 9.1}
\newcommand{\gwTenElevenXPHMChirpMassUpperError}{\ensuremath 0.1}
\newcommand{\gwTenElevenXPHMChirpMassLowerError}{\ensuremath 0.1}

\newcommand{\gwTenElevenXPHMDistance}{\ensuremath 214}
\newcommand{\gwTenElevenXPHMDistanceUpperError}{\ensuremath 41}
\newcommand{\gwTenElevenXPHMDistanceLowerError}{\ensuremath 43}

\newcommand{\gwElevenTenSpinOneZMedian}{\ensuremath -0.39}
\newcommand{\gwElevenTenSpinOneZUpperError}{\ensuremath 0.34}
\newcommand{\gwElevenTenSpinOneZLowerError}{\ensuremath 0.37}

\newcommand{\gwElevenTenSpinOneZProbabilityNegative}{\ensuremath 97.7}

\newcommand{\gwElevenTenSpinOnePlaneMedian}{\ensuremath 0.40}
\newcommand{\gwElevenTenSpinOnePlaneUpperError}{\ensuremath 0.34}
\newcommand{\gwElevenTenSpinOnePlaneLowerError}{\ensuremath 0.30}

\newcommand{\gwElevenTenChiEffectiveMedian}{\ensuremath -0.28}
\newcommand{\gwElevenTenChiEffectiveUpperError}{\ensuremath 0.23}
\newcommand{\gwElevenTenChiEffectiveLowerError}{\ensuremath 0.20}

\newcommand{\gwElevenTenChiEffectiveProbabilityPositive}{\ensuremath 1.9}
\newcommand{\gwElevenTenChiEffectiveProbabilityLessThanZero}{\ensuremath 98.1}
\newcommand{\gwElevenTenRestrictedChiEffectiveProbabilityLessThanZero}{\ensuremath 99.9}

\newcommand{\gwElevenTenChiPMedian}{\ensuremath 0.42}
\newcommand{\gwElevenTenChiPUpperError}{\ensuremath 0.33}
\newcommand{\gwElevenTenChiPLowerError}{\ensuremath 0.27}

\newcommand{\gwElevenTenSpinOneMagMedian}{\ensuremath 0.61}
\newcommand{\gwElevenTenSpinOneMagUpperError}{\ensuremath 0.33}
\newcommand{\gwElevenTenSpinOneMagLowerError}{\ensuremath 0.40}

\newcommand{\gwElevenTenSpinOneTiltMedian}{\ensuremath 133}

\newcommand{\gwElevenTenSpinOneTiltUpperErrorHPD}{\ensuremath 47}
\newcommand{\gwElevenTenSpinOneTiltLowerErrorHPD}{\ensuremath 25}
\newcommand{\gwElevenTenSpinOneTiltLowerLimit}{\ensuremath 108}

\newcommand{\gwElevenTenSpinOneTiltProbabilityMoreThanNinteyDegrees}{\ensuremath 97.7}

\newcommand{\gwElevenTenMassOne}{\ensuremath 17.2}
\newcommand{\gwElevenTenMassOneUpperError}{\ensuremath 5.0}
\newcommand{\gwElevenTenMassOneLowerError}{\ensuremath 4.4}

\newcommand{\gwElevenTenMassTwo}{\ensuremath 7.7}
\newcommand{\gwElevenTenMassTwoUpperError}{\ensuremath 2.2}
\newcommand{\gwElevenTenMassTwoLowerError}{\ensuremath 1.5}

\newcommand{\gwElevenTenMassRatio}{\ensuremath 0.45}
\newcommand{\gwElevenTenMassRatioUpperError}{\ensuremath 0.32}
\newcommand{\gwElevenTenMassRatioLowerError}{\ensuremath 0.17}

\newcommand{\gwElevenTenChirpMass}{\ensuremath 9.8}
\newcommand{\gwElevenTenChirpMassUpperError}{\ensuremath 0.5}
\newcommand{\gwElevenTenChirpMassLowerError}{\ensuremath 0.4}

\newcommand{\gwElevenTenDistance}{\ensuremath 736}
\newcommand{\gwElevenTenDistanceUpperError}{\ensuremath 270}
\newcommand{\gwElevenTenDistanceLowerError}{\ensuremath 267}

\newcommand{\gwElevenTenSEOBSpinOneZMedian}{\ensuremath -0.37}
\newcommand{\gwElevenTenSEOBSpinOneZUpperError}{\ensuremath 0.34}
\newcommand{\gwElevenTenSEOBSpinOneZLowerError}{\ensuremath 0.38}

\newcommand{\gwElevenTenSEOBSpinOnePlaneMedian}{\ensuremath 0.39}
\newcommand{\gwElevenTenSEOBSpinOnePlaneUpperError}{\ensuremath 0.34}
\newcommand{\gwElevenTenSEOBSpinOnePlaneLowerError}{\ensuremath 0.29}

\newcommand{\gwElevenTenSEOBChiEffectiveMedian}{\ensuremath -0.27}
\newcommand{\gwElevenTenSEOBChiEffectiveUpperError}{\ensuremath 0.23}
\newcommand{\gwElevenTenSEOBChiEffectiveLowerError}{\ensuremath 0.20}

\newcommand{\gwElevenTenSEOBChiPMedian}{\ensuremath 0.40}
\newcommand{\gwElevenTenSEOBChiPUpperError}{\ensuremath 0.33}
\newcommand{\gwElevenTenSEOBChiPLowerError}{\ensuremath 0.26}

\newcommand{\gwElevenTenSEOBSpinOneMagMedian}{\ensuremath 0.59}
\newcommand{\gwElevenTenSEOBSpinOneMagUpperError}{\ensuremath 0.34}
\newcommand{\gwElevenTenSEOBSpinOneMagLowerError}{\ensuremath 0.39}

\newcommand{\gwElevenTenSEOBSpinOneTiltMedian}{\ensuremath 133}

\newcommand{\gwElevenTenSEOBSpinOneTiltUpperErrorHPD}{\ensuremath 47}
\newcommand{\gwElevenTenSEOBSpinOneTiltLowerErrorHPD}{\ensuremath 26}

\newcommand{\gwElevenTenSEOBMassOne}{\ensuremath 17.4}
\newcommand{\gwElevenTenSEOBMassOneUpperError}{\ensuremath 5.1}
\newcommand{\gwElevenTenSEOBMassOneLowerError}{\ensuremath 4.7}

\newcommand{\gwElevenTenSEOBMassTwo}{\ensuremath 7.7}
\newcommand{\gwElevenTenSEOBMassTwoUpperError}{\ensuremath 2.3}
\newcommand{\gwElevenTenSEOBMassTwoLowerError}{\ensuremath 1.5}

\newcommand{\gwElevenTenSEOBMassRatio}{\ensuremath 0.44}
\newcommand{\gwElevenTenSEOBMassRatioUpperError}{\ensuremath 0.33}
\newcommand{\gwElevenTenSEOBMassRatioLowerError}{\ensuremath 0.17}

\newcommand{\gwElevenTenSEOBChirpMass}{\ensuremath 9.8}
\newcommand{\gwElevenTenSEOBChirpMassUpperError}{\ensuremath 0.5}
\newcommand{\gwElevenTenSEOBChirpMassLowerError}{\ensuremath 0.4}

\newcommand{\gwElevenTenSEOBDistance}{\ensuremath 736}
\newcommand{\gwElevenTenSEOBDistanceUpperError}{\ensuremath 277}
\newcommand{\gwElevenTenSEOBDistanceLowerError}{\ensuremath 280}

\newcommand{\gwElevenTenXOFourASpinOneZMedian}{\ensuremath -0.43}
\newcommand{\gwElevenTenXOFourASpinOneZUpperError}{\ensuremath 0.35}
\newcommand{\gwElevenTenXOFourASpinOneZLowerError}{\ensuremath 0.35}

\newcommand{\gwElevenTenXOFourASpinOnePlaneMedian}{\ensuremath 0.43}
\newcommand{\gwElevenTenXOFourASpinOnePlaneUpperError}{\ensuremath 0.32}
\newcommand{\gwElevenTenXOFourASpinOnePlaneLowerError}{\ensuremath 0.32}

\newcommand{\gwElevenTenXOFourAChiEffectiveMedian}{\ensuremath -0.30}
\newcommand{\gwElevenTenXOFourAChiEffectiveUpperError}{\ensuremath 0.22}
\newcommand{\gwElevenTenXOFourAChiEffectiveLowerError}{\ensuremath 0.20}

\newcommand{\gwElevenTenXOFourAChiPMedian}{\ensuremath 0.45}
\newcommand{\gwElevenTenXOFourAChiPUpperError}{\ensuremath 0.31}
\newcommand{\gwElevenTenXOFourAChiPLowerError}{\ensuremath 0.29}

\newcommand{\gwElevenTenXOFourASpinOneMagMedian}{\ensuremath 0.65}
\newcommand{\gwElevenTenXOFourASpinOneMagUpperError}{\ensuremath 0.30}
\newcommand{\gwElevenTenXOFourASpinOneMagLowerError}{\ensuremath 0.41}

\newcommand{\gwElevenTenXOFourASpinOneTiltMedian}{\ensuremath 133}

\newcommand{\gwElevenTenXOFourASpinOneTiltUpperErrorHPD}{\ensuremath 47}
\newcommand{\gwElevenTenXOFourASpinOneTiltLowerErrorHPD}{\ensuremath 22}

\newcommand{\gwElevenTenXOFourAMassOne}{\ensuremath 16.9}
\newcommand{\gwElevenTenXOFourAMassOneUpperError}{\ensuremath 4.9}
\newcommand{\gwElevenTenXOFourAMassOneLowerError}{\ensuremath 4.2}

\newcommand{\gwElevenTenXOFourAMassTwo}{\ensuremath 7.8}
\newcommand{\gwElevenTenXOFourAMassTwoUpperError}{\ensuremath 2.1}
\newcommand{\gwElevenTenXOFourAMassTwoLowerError}{\ensuremath 1.6}

\newcommand{\gwElevenTenXOFourAMassRatio}{\ensuremath 0.46}
\newcommand{\gwElevenTenXOFourAMassRatioUpperError}{\ensuremath 0.31}
\newcommand{\gwElevenTenXOFourAMassRatioLowerError}{\ensuremath 0.17}

\newcommand{\gwElevenTenXOFourAChirpMass}{\ensuremath 9.8}
\newcommand{\gwElevenTenXOFourAChirpMassUpperError}{\ensuremath 0.5}
\newcommand{\gwElevenTenXOFourAChirpMassLowerError}{\ensuremath 0.4}

\newcommand{\gwElevenTenXOFourADistance}{\ensuremath 731}
\newcommand{\gwElevenTenXOFourADistanceUpperError}{\ensuremath 268}
\newcommand{\gwElevenTenXOFourADistanceLowerError}{\ensuremath 260}

\newcommand{\gwElevenTenXPHMSpinOneZMedian}{\ensuremath -0.37}
\newcommand{\gwElevenTenXPHMSpinOneZUpperError}{\ensuremath 0.33}
\newcommand{\gwElevenTenXPHMSpinOneZLowerError}{\ensuremath 0.38}

\newcommand{\gwElevenTenXPHMSpinOnePlaneMedian}{\ensuremath 0.39}
\newcommand{\gwElevenTenXPHMSpinOnePlaneUpperError}{\ensuremath 0.34}
\newcommand{\gwElevenTenXPHMSpinOnePlaneLowerError}{\ensuremath 0.29}

\newcommand{\gwElevenTenXPHMChiEffectiveMedian}{\ensuremath -0.27}
\newcommand{\gwElevenTenXPHMChiEffectiveUpperError}{\ensuremath 0.22}
\newcommand{\gwElevenTenXPHMChiEffectiveLowerError}{\ensuremath 0.20}

\newcommand{\gwElevenTenXPHMChiPMedian}{\ensuremath 0.40}
\newcommand{\gwElevenTenXPHMChiPUpperError}{\ensuremath 0.33}
\newcommand{\gwElevenTenXPHMChiPLowerError}{\ensuremath 0.26}

\newcommand{\gwElevenTenXPHMSpinOneMagMedian}{\ensuremath 0.58}
\newcommand{\gwElevenTenXPHMSpinOneMagUpperError}{\ensuremath 0.35}
\newcommand{\gwElevenTenXPHMSpinOneMagLowerError}{\ensuremath 0.39}

\newcommand{\gwElevenTenXPHMSpinOneTiltMedian}{\ensuremath 132}

\newcommand{\gwElevenTenXPHMSpinOneTiltUpperErrorHPD}{\ensuremath 48}
\newcommand{\gwElevenTenXPHMSpinOneTiltLowerErrorHPD}{\ensuremath 26}

\newcommand{\gwElevenTenXPHMMassOne}{\ensuremath 17.2}
\newcommand{\gwElevenTenXPHMMassOneUpperError}{\ensuremath 5.0}
\newcommand{\gwElevenTenXPHMMassOneLowerError}{\ensuremath 4.5}

\newcommand{\gwElevenTenXPHMMassTwo}{\ensuremath 7.7}
\newcommand{\gwElevenTenXPHMMassTwoUpperError}{\ensuremath 2.2}
\newcommand{\gwElevenTenXPHMMassTwoLowerError}{\ensuremath 1.5}

\newcommand{\gwElevenTenXPHMMassRatio}{\ensuremath 0.45}
\newcommand{\gwElevenTenXPHMMassRatioUpperError}{\ensuremath 0.33}
\newcommand{\gwElevenTenXPHMMassRatioLowerError}{\ensuremath 0.17}

\newcommand{\gwElevenTenXPHMChirpMass}{\ensuremath 9.9}
\newcommand{\gwElevenTenXPHMChirpMassUpperError}{\ensuremath 0.5}
\newcommand{\gwElevenTenXPHMChirpMassLowerError}{\ensuremath 0.4}

\newcommand{\gwElevenTenXPHMDistance}{\ensuremath 738}
\newcommand{\gwElevenTenXPHMDistanceUpperError}{\ensuremath 263}
\newcommand{\gwElevenTenXPHMDistanceLowerError}{\ensuremath 261}

\newcommand{\gwNineteenOhFiveSeventeenLowerLimit}{\ensuremath 0.52}

\newcommand{\gwTwentyThreeElevenTwentyThreeSpinOneMagLowerLimit}{\ensuremath 0.63}

\newcommand{\gwTenElevenSNRlho}{\ensuremath 35.4}
\newcommand{\gwTenElevenSNRvirgo}{\ensuremath 9.1}

\newcommand{\gwElevenTenOfflineFARmbtaRounded}{\ensuremath 0.85\,\mathrm{yr}^{-1}}

\newcommand{\gwElevenTenOfflineFARpycbcRounded}{\ensuremath 6.3\times10^{-4}\, \mathrm{yr}^{-1}}
\newcommand{\gwElevenTenOnlineFARgstlal}{\ensuremath 0.15\,\mathrm{yr}^{-1}}
\newcommand{\gwElevenTenOfflineFARgstlal}{\ensuremath 0.099\,\mathrm{yr}^{-1}}

\newcommand{\GWTCFourCosMaxTiltUpperLimit}{\ensuremath -0.57}
\newcommand{\GWTCFourMaxTiltLowerLimit}{\ensuremath 125}
\newcommand{\GWTCFourPlusTenElevenCosMaxTiltUpperLimit}{\ensuremath -0.55}
\newcommand{\GWTCFourPlusTenElevenMaxTiltLowerLimit}{\ensuremath 124}
\newcommand{\GWTCFourPlusElevenTenCosMaxTiltUpperLimit}{\ensuremath -0.59}
\newcommand{\GWTCFourPlusElevenTenMaxTiltLowerLimit}{\ensuremath 126}

\newcommand{\GWTCFourFracHighSpinUpperLimit}{\ensuremath 0.32}
\newcommand{\GWTCFourPlusSpecialEventsFracHighSpinUpperLimit}{\ensuremath 0.36}

\newcommand{\ElevenTenPrimaryCosTiltUpperLimitDEFAULT}{\ensuremath -0.43}
\newcommand{\ElevenTenPrimaryCosTiltUpperLimitMAXTILT}{\ensuremath -0.26}
\newcommand{\ElevenTenPrimaryCosTiltUpperLimitHIGHSPIN}{\ensuremath -0.37}
\newcommand{\TenElevenPrimarySpinMagLowerLimitDEFAULT}{\ensuremath 0.68}
\newcommand{\TenElevenPrimarySpinMagLowerLimitMAXTILT}{\ensuremath 0.68}
\newcommand{\TenElevenPrimarySpinMagLowerLimitHIGHSPIN}{\ensuremath 0.69}

\newcommand{\gwTenElevenProgMassOneMedianMahapatra}{\ensuremath 11.0}
\newcommand{\gwTenElevenProgMassOneUpperErrorMahapatra}{\ensuremath 1.9}
\newcommand{\gwTenElevenProgMassOneLowerErrorMahapatra}{\ensuremath 1.6}

\newcommand{\gwTenElevenProgMassOneMedianWong}{\ensuremath 13.3}
\newcommand{\gwTenElevenProgMassOneUpperErrorWong}{\ensuremath 4.8}
\newcommand{\gwTenElevenProgMassOneLowerErrorWong}{\ensuremath 3.2}

\newcommand{\gwTenElevenProgMassTwoMedianMahapatra}{\ensuremath 9.6}
\newcommand{\gwTenElevenProgMassTwoUpperErrorMahapatra}{\ensuremath 1.7}
\newcommand{\gwTenElevenProgMassTwoLowerErrorMahapatra}{\ensuremath 1.8}

\newcommand{\gwTenElevenProgMassTwoMedianWong}{\ensuremath 7.5}
\newcommand{\gwTenElevenProgMassTwoUpperErrorWong}{\ensuremath 3.2}
\newcommand{\gwTenElevenProgMassTwoLowerErrorWong}{\ensuremath 3.9}

\newcommand{\gwTenElevenProgChiEffMedianMahapatra}{\ensuremath 0.02}
\newcommand{\gwTenElevenProgChiEffUpperErrorMahapatra}{\ensuremath 0.08}
\newcommand{\gwTenElevenProgChiEffLowerErrorMahapatra}{\ensuremath 0.06}

\newcommand{\gwTenElevenProgChiEffMedianWong}{\ensuremath 0.23}
\newcommand{\gwTenElevenProgChiEffUpperErrorWong}{\ensuremath 0.29}
\newcommand{\gwTenElevenProgChiEffLowerErrorWong}{\ensuremath 0.28}

\newcommand{\gwTenElevenProgRecoilMedianMahapatra}{\ensuremath 100}
\newcommand{\gwTenElevenProgRecoilUpperErrorMahapatra}{\ensuremath 190}
\newcommand{\gwTenElevenProgRecoilLowerErrorMahapatra}{\ensuremath 70}

\newcommand{\gwTenElevenProgRecoilMedianWong}{\ensuremath 750}
\newcommand{\gwTenElevenProgRecoilUpperErrorWong}{\ensuremath 1400}
\newcommand{\gwTenElevenProgRecoilLowerErrorWong}{\ensuremath 630}

\newcommand{\gwElevenTenProgMassOneMedianMahapatra}{\ensuremath 9.6}
\newcommand{\gwElevenTenProgMassOneUpperErrorMahapatra}{\ensuremath 3.1}
\newcommand{\gwElevenTenProgMassOneLowerErrorMahapatra}{\ensuremath 2.0}

\newcommand{\gwElevenTenProgMassOneMedianWong}{\ensuremath 12.3}
\newcommand{\gwElevenTenProgMassOneUpperErrorWong}{\ensuremath 5.6}
\newcommand{\gwElevenTenProgMassOneLowerErrorWong}{\ensuremath 4.2}

\newcommand{\gwElevenTenProgMassTwoMedianMahapatra}{\ensuremath 8.0}
\newcommand{\gwElevenTenProgMassTwoUpperErrorMahapatra}{\ensuremath 2.1}
\newcommand{\gwElevenTenProgMassTwoLowerErrorMahapatra}{\ensuremath 2.3}

\newcommand{\gwElevenTenProgMassTwoMedianWong}{\ensuremath 5.1}
\newcommand{\gwElevenTenProgMassTwoUpperErrorWong}{\ensuremath 3.6}
\newcommand{\gwElevenTenProgMassTwoLowerErrorWong}{\ensuremath 1.9}

\newcommand{\gwElevenTenProgChiEffMedianMahapatra}{\ensuremath -0.00}
\newcommand{\gwElevenTenProgChiEffUpperErrorMahapatra}{\ensuremath 0.06}
\newcommand{\gwElevenTenProgChiEffLowerErrorMahapatra}{\ensuremath 0.06}

\newcommand{\gwElevenTenProgChiEffMedianWong}{\ensuremath -0.04}
\newcommand{\gwElevenTenProgChiEffUpperErrorWong}{\ensuremath 0.65}
\newcommand{\gwElevenTenProgChiEffLowerErrorWong}{\ensuremath 0.57}

\newcommand{\gwElevenTenProgRecoilMedianMahapatra}{\ensuremath 100}
\newcommand{\gwElevenTenProgRecoilUpperErrorMahapatra}{\ensuremath 160}
\newcommand{\gwElevenTenProgRecoilLowerErrorMahapatra}{\ensuremath 80}

\newcommand{\gwElevenTenProgRecoilMedianWong}{\ensuremath 480}
\newcommand{\gwElevenTenProgRecoilUpperErrorWong}{\ensuremath 1270}
\newcommand{\gwElevenTenProgRecoilLowerErrorWong}{\ensuremath 330}

\newcommand{\gwTenElevenSEOBEccentricityUpperLimit}{\ensuremath 0.05}
\newcommand{\gwElevenTenSEOBEccentricityUpperLimit}{\ensuremath 0.17}

\newcommand{\gwElevenTenTEOBEccentricityUpperLimit}{\ensuremath 0.14}

\newcommand{\gwTenElevenSymmetricSiqmPhenomMedian}{\ensuremath 0.10}
\newcommand{\gwTenElevenSymmetricSiqmPhenomUpperError}{\ensuremath 0.09}
\newcommand{\gwTenElevenSymmetricSiqmPhenomLowerError}{\ensuremath 0.11}
\newcommand{\gwTenElevenSymmetricSiqmPhenomUpperLimit}{\ensuremath 0.17}

\newcommand{\gwTenElevenSiqmPhenomMedian}{\ensuremath 0.10}
\newcommand{\gwTenElevenSiqmPhenomUpperError}{\ensuremath 0.82}
\newcommand{\gwTenElevenSiqmPhenomLowerError}{\ensuremath 0.82}
\newcommand{\gwOhFourTwelveSymmetricSiqmPhenomMedian}{\ensuremath 0}
\newcommand{\gwOhFourTwelveSymmetricSiqmPhenomUpperError}{\ensuremath 2}
\newcommand{\gwOhFourTwelveSymmetricSiqmPhenomLowerError}{\ensuremath 91}

\newcommand{\gwTenElevenSMAUpperBound}{\ensuremath 0.5}
\newcommand{\gwTenElevenSMALowerBound}{\ensuremath -1.9}

\newcommand{\gwTenElevenSMAUpperModeMedian}{\ensuremath 0.0}
\newcommand{\gwTenElevenSMAUpperModeUpperError}{\ensuremath 0.5}
\newcommand{\gwTenElevenSMAUpperModeLowerError}{\ensuremath 0.3}

\newcommand{\gwTenElevenSMALowerModeMedian}{\ensuremath -2.1}
\newcommand{\gwTenElevenSMALowerModeUpperError}{\ensuremath 0.4}
\newcommand{\gwTenElevenSMALowerModeLowerError}{\ensuremath 0.5}

\newcommand{\gwOhEightFourteenSMAUpperBound}{\ensuremath 1.6}
\newcommand{\gwOhEightFourteenSMALowerBound}{\ensuremath -3.6}

\newcommand{\gwOhFourTwelveSMAUpperBound}{\ensuremath 4.0}
\newcommand{\gwOhFourTwelveSMALowerBound}{\ensuremath -5.3}

\newcommand{\gwTenElevenPrecessingSNRMedian}{\ensuremath 5.3}
\newcommand{\gwTenElevenPrecessingSNRUpperError}{\ensuremath 2.1}
\newcommand{\gwTenElevenPrecessingSNRLowerError}{\ensuremath 1.9}
\newcommand{\gwTenElevenPrecessingSNRPercentageAboveNull}{\ensuremath 99.5}

\newcommand{\gwTenElevenHOMSNRMedian}{\ensuremath 5.9}
\newcommand{\gwTenElevenHOMSNRUpperError}{\ensuremath 1.0}
\newcommand{\gwTenElevenHOMSNRLowerError}{\ensuremath 1.1}

\newcommand{\gwTenElevenMatchedFilterSNRMedian}{\ensuremath 36.0}

\newcommand{\logBayesHOM}{\ensuremath 5.2}
\newcommand{\logBayesPrecession}{\ensuremath 5.4}

\newcommand{\SuperradPower}{\ensuremath -12}
\newcommand{\TenElevenScalarLowerConstraint}{\ensuremath 0.3}
\newcommand{\TenElevenScalarUpperConstraint}{\ensuremath 2.6}

\newcommand{\TenElevenVectorLowerConstraint}{\ensuremath 0.1}
\newcommand{\TenElevenVectorUpperConstraint}{\ensuremath 5.3}

\newcommand{\CMCNumberOfMergersFirstGenTotal}{\ensuremath 11005}
\newcommand{\CMCNumberOfMergersNGenTotal}{\ensuremath 2263}

\newcommand{\CMCNumberOfMergersFirstGenZOOOTwo}{\ensuremath 3681}
\newcommand{\CMCNumberOfMergersNGenZOOOTwo}{\ensuremath 660}

\newcommand{\CMCNumberOfMergersFirstGenZOOTwo}{\ensuremath 3566}
\newcommand{\CMCNumberOfMergersNGenZOOTwo}{\ensuremath 625}

\newcommand{\CMCNumberOfMergersFirstGenZOTwo}{\ensuremath 3758}
\newcommand{\CMCNumberOfMergersNGenZOTwo}{\ensuremath 978}

\newcommand{\CBHBDNumberOfMergersFirstGenTotal}{\ensuremath 5276}
\newcommand{\CBHBDNumberOfMergersNGenTotal}{\ensuremath 1305}

\newcommand{\CBHBDNumberOfMergersFirstGenZOOOTwo}{\ensuremath 1690}
\newcommand{\CBHBDNumberOfMergersNGenZOOOTwo}{\ensuremath 316}

\newcommand{\CBHBDNumberOfMergersFirstGenZOOTwo}{\ensuremath 1761}
\newcommand{\CBHBDNumberOfMergersNGenZOOTwo}{\ensuremath 351}

\newcommand{\CBHBDNumberOfMergersFirstGenZOTwo}{\ensuremath 1825}
\newcommand{\CBHBDNumberOfMergersNGenZOTwo}{\ensuremath 638}

\acrodef{LSC}[LSC]{LIGO Scientific Collaboration}
\acrodef{LVC}[LVC]{LIGO Scientific and Virgo Collaboration}
\acrodef{LVK}[LVK]{LIGO Scientific, Virgo and KAGRA Collaboration}
\acrodef{aLIGO}{Advanced Laser Interferometer Gravitational-Wave Observatory}
\acrodef{aVirgo}{Advanced Virgo}
\acrodef{LIGO}[LIGO]{Laser Interferometer Gravitational-Wave Observatory}
\acrodef{IFO}[IFO]{interferometer}
\acrodef{LHO}[LHO]{LIGO-Hanford}
\acrodef{LLO}[LLO]{LIGO-Livingston}
\acrodef{O2}[O2]{second observing run}
\acrodef{O1}[O1]{first observing run}
\acrodef{O3}[O3]{third observing run}
\acrodef{O3a}[O3a]{first half of the third observing run}
\acrodef{O3b}[O3b]{second half of the third observing run}
\acrodef{o4a}[O4a]{first part of the fourth observing run}

\acrodef{BH}[BH]{black hole}
\acrodef{bbh}[BBH]{binary black hole}
\acrodef{BNS}[BNS]{binary neutron star}
\acrodef{IMBH}[IMBH]{intermediate-mass black hole}
\acrodef{NS}[NS]{neutron star}
\acrodef{BHNS}[BHNS]{black hole--neutron star binaries}
\acrodef{NSBH}[NSBH]{neutron star--black hole binary}
\acrodef{PBH}[PBH]{primordial black hole binaries}
\acrodef{CBC}[CBC]{compact binary coalescence}
\acrodef{gw}[GW]{gravitational wave}
\acrodef{GWH}[GW]{gravitational-wave}

\acrodef{cwb}[cWB]{coherent WaveBurst}

\acrodef{snr}[SNR]{signal-to-noise ratio}
\acrodef{FAR}[FAR]{false alarm rate}
\acrodef{IFAR}[IFAR]{inverse false alarm rate}
\acrodef{FAP}[FAP]{false alarm probability}
\acrodef{PSD}[PSD]{power spectral density}
\acrodef{far}[FAR]{false alarm rate}

\acrodef{GR}[GR]{general relativity}
\acrodef{NR}[NR]{numerical relativity}
\acrodef{PN}[PN]{post-Newtonian}
\acrodef{EOB}[EOB]{effective-one-body}
\acrodef{ROM}[ROM]{reduced-order model}
\acrodef{IMR}[IMR]{inspiral--merger--ringdown}

\acrodef{PDF}[PDF]{probability density function}
\acrodef{PE}[PE]{parameter estimation}
\acrodef{CL}[CL]{credible level}

\acrodef{EOS}[EoS]{equation of state}

\acrodef{LAL}[LAL]{LIGO Algorithm Library}

\acrodef{KLD}[KLD]{Kullback--Leibler divergence}
\acrodef{JSD}[JSD]{Jensen--Shannon divergence}

\newcommand{\gwTenElevenFull}{GW241011\_233834\xspace}
\newcommand{\gwElevenTenFull}{GW241110\_124123\xspace}
\newcommand{\gwTenEleven}{GW241011\xspace}
\newcommand{\gwElevenTen}{GW241110\xspace}

\newcommand{\gstlal}{\textsc{GstLAL}\xspace}
\newcommand{\BAYESTAR}{BAYESTAR\xspace}

\newcommand{\pycbc}{\textsc{PyCBC}\xspace}
\newcommand{\mbta}{\textsc{MBTA}\xspace}

\newcommand{\BAYESWAVE}{\textsc{BayesWave}\xspace}
\newcommand{\BILBY}{\textsc{Bilby}\xspace}
\newcommand{\RIFT}{\textsc{RIFT}\xspace}

\newcommand{\ASIMOV}{{Asimov}\xspace}
\newcommand{\PESUMMARY}{PESummary\xspace}

\newcommand{\NUMPY}{\textsc{NumPy}\xspace}
\newcommand{\SCIPY}{\textsc{SciPy}\xspace}
\newcommand{\PLT}{\textsc{Matplotlib}\xspace}

\newcommand{\DYNESTY}{\textsc{Dynesty}\xspace}
\newcommand{\DMT}{{DMT}\xspace}
\newcommand{\DQR}{{DQR}\xspace}
\newcommand{\DQSEGDB}{{DQSEGDB}\xspace}
\newcommand{\GWDETCHAR}{{gwdetchar}\xspace}
\newcommand{\HVETO}{{hveto}\xspace}
\newcommand{\PYTHONVIRGOTOOLS}{{PythonVirgoTools}\xspace}
\newcommand{\OMICRONSCAN}{{Omicron}\xspace}
\newcommand{\IDQ}{iDQ\xspace}

\renewcommand{\today}{\number\day\space\ifcase\month\or
  January\or February\or March\or April\or May\or June\or
  July\or August\or September\or October\or November\or December\fi
  \space\number\year}

\begin{document}

\title{GW241011 and GW241110: Exploring Binary Formation and Fundamental Physics with Asymmetric, High-Spin Black Hole Coalescences}

\author[0000-0003-4786-2698]{A.~G.~Abac}
\affiliation{Max Planck Institute for Gravitational Physics (Albert Einstein Institute), D-14476 Potsdam, Germany}
\author{I.~Abouelfettouh}
\affiliation{LIGO Hanford Observatory, Richland, WA 99352, USA}
\author{F.~Acernese}
\affiliation{Dipartimento di Farmacia, Universit\`a di Salerno, I-84084 Fisciano, Salerno, Italy}
\affiliation{INFN, Sezione di Napoli, I-80126 Napoli, Italy}
\author[0000-0002-8648-0767]{K.~Ackley}
\affiliation{University of Warwick, Coventry CV4 7AL, United Kingdom}
\author[0000-0001-5525-6255]{C.~Adamcewicz}
\affiliation{OzGrav, School of Physics \& Astronomy, Monash University, Clayton 3800, Victoria, Australia}
\author[0009-0004-2101-5428]{S.~Adhicary}
\affiliation{The Pennsylvania State University, University Park, PA 16802, USA}
\author{D.~Adhikari}
\affiliation{Max Planck Institute for Gravitational Physics (Albert Einstein Institute), D-30167 Hannover, Germany}
\affiliation{Leibniz Universit\"{a}t Hannover, D-30167 Hannover, Germany}
\author[0000-0002-4559-8427]{N.~Adhikari}
\affiliation{University of Wisconsin-Milwaukee, Milwaukee, WI 53201, USA}
\author[0000-0002-5731-5076]{R.~X.~Adhikari}
\affiliation{LIGO Laboratory, California Institute of Technology, Pasadena, CA 91125, USA}
\author{V.~K.~Adkins}
\affiliation{Louisiana State University, Baton Rouge, LA 70803, USA}
% ...

\collaboration{3000}{The LIGO Scientific Collaboration, the Virgo Collaboration, and the KAGRA Collaboration}

\date[\relax]{Compiled: \today}

\begin{abstract}

We report the observation of gravitational waves from two binary black hole coalescences during the fourth observing run of the LIGO--Virgo--KAGRA detector network, \gwTenEleven and \gwElevenTen.
The sources of these two signals are characterized by rapid and precisely measured primary spins, non-negligible spin--orbit misalignment, and unequal mass ratios between their constituent black holes.
These properties are characteristic of binaries in which the more massive object was itself formed from a previous binary black hole merger, and suggest that the sources of \gwTenEleven and \gwElevenTen may have formed in dense stellar environments in which repeated mergers can take place.
As the third loudest gravitational-wave event published to date, with a median network signal-to-noise ratio of $\gwTenElevenMatchedFilterSNRMedian$, \gwTenEleven furthermore yields stringent constraints on the Kerr nature of black holes, the multipolar structure of gravitational-wave generation, and the existence of ultralight bosons within the mass range $10^{-13}$--$10^{-12}$ eV.

\end{abstract}


\section{Introduction}

It has been a decade since the inception of practical gravitational-wave astronomy
In the years following the first direct observation of gravitational waves from a binary black hole coalescence in 2015~\citep{LIGOScientific:2016aoc}, the Advanced LIGO, Advanced Virgo, and KAGRA experiments~\citep{LIGOScientific:2014pky, VIRGO:2014yos, KAGRA:2020tym} have operated in tandem to regularly identify an ever-increasing number of gravitational-wave signals~\citep{LIGOScientific:2021usb, KAGRA:2021vkt}.
The recently-released fourth Gravitational-Wave Transient Catalog (GWTC-4.0), including signals discovered through January 2024, contains hundreds of gravitational-wave signals~\citep{gwtc4}, and the rate of discoveries continues to accelerate as further upgrades improvement broadband instrumental sensitivity to gravitational waves~\citep{LIGOO4Detector:2023wmz, membersoftheLIGOScientific:2024elc, Capote:2024rmo, LIGO:2024kkz, VIRGO:2023elp, gwtc4-introduction}.
The growing collection of observed gravitational-wave sources includes binary neutron stars~\citep{LIGOScientific:2017zic, LIGOScientific:2020aai}, one of which was accompanied by transient multimessenger emission seen across the electromagnetic spectrum~\citep[e.g., ][]{LIGOScientific:2017ync, LIGOScientific:2017zic, Goldstein:2017mmi, Coulter:2017wya, Hallinan:2017woc, Margutti:2017cjl}, as well as likely neutron star--black hole binaries~\citep{LIGOScientific:2021qlt, KAGRA:2021vkt, LIGOScientific:2024elc, gwtc4}.
And it includes a growing number of black holes whose masses, whether unexpectedly large, unexpectedly small, or unexpectedly unequal, challenge present understanding of compact binary formation and evolution~\citep{LIGOScientific:2016vpg, LIGOScientific:2020iuh, LIGOScientific:2020zkf, LIGOScientific:2024elc, LIGOScientific:2025rsn}.

\begin{figure*}[t!]
    \centering
    \includegraphics[width=0.92\textwidth]{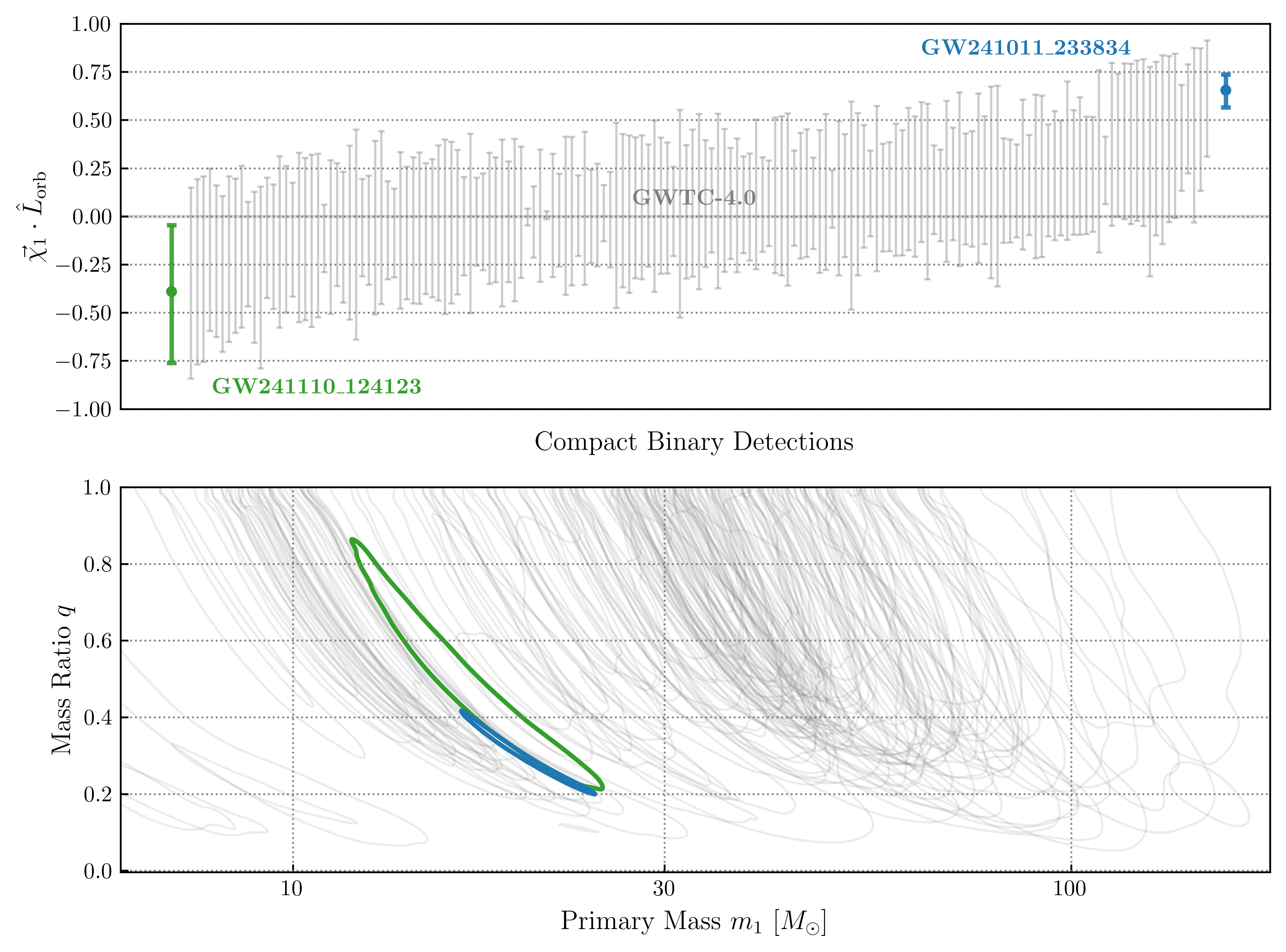}
    \caption{
    \textit{Top:} Central $90\%$ credible bounds on the dimensionless primary spins $\vec\chi_1$ of \gwTenEleven (blue) and \gwElevenTen (green), projected parallel to the direction $\hat L_\mathrm{N}$ of each binary's Newtonian orbital angular momentum.
    Shown in grey for comparison are $90\%$ credible bounds on the projected primary spins of previously-published compact binary coalescences in the fourth Gravitational-Wave Transient Catalog~\citep[GWTC-4.0;][]{gwtc4}, sorted by their median posterior values of $\chi_{1,z} \equiv \vec\chi_1 \cdot \hat L_\mathrm{N}$.
    We specifically show events with false alarm rates (FARs) below $1\,\mathrm{yr}^{-1}$, consistent with the significance threshold adopted for compact binary population studies in GWTC-3.0 and GWTC-4.0~\citep{KAGRA:2021duu, gwtc4-astrodist}.
    The source of \gwTenEleven contains one of the most rapidly spinning black holes observed by LIGO--Virgo--KAGRA to date, possessing the largest lower limit on both its primary spin magnitude and projected spin $\vec \chi_{1,z}$.
    \gwElevenTen, conversely, yields the most confident measurement to date of a black hole spinning retrograde with respect to its orbit, with $\chi_{1,z}<0$ at $\gwElevenTenSpinOneZProbabilityNegative\%$ credibility.
    \textit{Bottom:} $90\%$ credible posterior bounds on the primary masses and mass ratios of \gwTenEleven and \gwElevenTen, together with all binary mergers in GWTC-4.0 with $\mathrm{FAR}<1\,\mathrm{yr}^{-1}$.
    Despite the opposite nature of their spins, \gwTenEleven and \gwElevenTen are likely to have very similar masses, each favoring a primary of $15$--$20\,M_\odot$ and unequal mass ratios.
    }
    \label{fig:spin-violin}
\end{figure*}

Here, we report a pair of gravitational-wave events discovered in late 2024, \gwTenElevenFull and \gwElevenTenFull, arising from binary black hole coalescences that each contain at least one rapidly rotating black hole.
As illustrated in Fig.~\ref{fig:spin-violin}, the primary (more massive) source black hole of \gwTenElevenFull (hereafter abbreviated \gwTenEleven) exhibits one of the most rapid and precisely measured spins observed to date, with $\chi_1 = \gwTenElevenSpinOneMagMedian^{+\gwTenElevenSpinOneMagUpperError}_{-\gwTenElevenSpinOneMagLowerError}$.
This event provides a larger lower bound on both spin magnitude and $\vec \chi_1 \cdot \hat L_\mathrm{N}$, the spin projected parallel to a binary's Newtonian orbital angular momentum $\vec L_\mathrm{N}$, than any other binary previously published in GWTC-4.0~\citep{gwtc4}.
The primary spin of \gwTenEleven's source is tilted by $\sim 30\,\mathrm{deg}$ with respect to $\hat L_\mathrm{N}$, and the gravitational-wave signal confidently exhibits relativistic spin-orbit precession.
In contrast, the primary component of \gwElevenTenFull (hereafter \gwElevenTen) is measured to be spinning in a direction antiparallel to its orbital angular momentum vector, the most confidently antiparallel spin observed. 
Despite the opposite character of their spins, the sources of \gwTenEleven and \gwElevenTen possess similar masses.
Each is inferred to contain a primary mass between approximately $15$--$20\,M_\odot$ and both favor unequal component black hole masses, with \gwTenEleven in particular requiring an approximately $3$:$1$ mass ratio.

The rapid primary spins, significant spin--orbit misalignment, and unequal mass ratios of \gwTenEleven and \gwElevenTen are in tension with expectations from isolated evolution of massive stellar binaries~\citep{Kalogera:1999tq,  Belczynski:2016obo, deMink:2016vkw, Marchant:2016wow, Rodriguez:2016vmx, Callister:2020vyz, Broekgaarden:2022nst, Zevin:2022wrw}.
The source properties of \gwTenEleven and \gwElevenTen are consistent, however, with those expected from hierarchical mergers in dense stellar clusters.
Binary black hole mergers yield remnant black holes that are rapidly rotating, with spin magnitudes of $\chi \approx 0.7$~\citep{Pretorius:2005gq, Buonanno:2007sv, Berti:2008af}.
Asymmetric gravitational-wave emission can cause these remnant black holes to receive large kicks, with velocities that can reach thousands of kilometers per second~\citep{Fitchett:1983qzq, Favata:2004wz, Schnittman:2007sn, Campanelli:2007cga, Gonzalez:2007hi}.
In environments with sufficiently high escape velocities, however, remnant black holes may remain gravitationally bound, capture new partners, and participate in subsequent binary mergers~\citep{Lee:1994nq, OLeary:2005vqo, Giersz:2015mlk, Antonini:2016gqe, Gerosa:2017kvu, Fishbach:2017dwv, Rodriguez:2017pec, Antonini:2018auk, Rodriguez:2019huv, Baibhav:2020xdf, Fragione:2020nib, Kimball:2020qyd, Baibhav:2021qzw, Doctor:2021qfn, Fragione:2021nhb, Gerosa:2021mno, Mahapatra:2021hme, Mapelli:2021syv, Antonini:2022vib, ArcaSedda:2023mlv, Mahapatra:2022ngs, Chattopadhyay:2023pil}.
Under this hypothesis, the primary black holes of \gwTenEleven and \gwElevenTen may themselves each be a product of a previous binary black hole merger.

The rapid spins and unequal mass ratios of \gwTenEleven and \gwElevenTen furthermore make them prime laboratories with which to test fundamental physics.
\gwTenEleven, in particular, exhibits both significant relativistic spin precession~\citep{Apostolatos:1994mx, Kidder:1995zr} and gravitational radiation from higher-order multipole moments~\citep{Thorne:1980ru}.
By virtue of these features, \gwTenEleven offers one of the most precise confirmations to date of the Kerr nature of spinning black holes~\citep{Kerr:1963ud, Carter:1971zc, Hansen:1974zz} and the multipolar emission pattern of gravitational waves.

The rest of this paper is organized as follows.
In Section~\ref{sec:detection}, we describe the low-latency identification and subsequent validation of \gwTenEleven and \gwElevenTen.
In Section~\ref{sec:pe}, we discuss the measured properties of these two events.
In Section~\ref{sec:astro}, we describe the possible astrophysical interpretation and implications of \gwTenEleven and \gwElevenTen, and in Section~\ref{sec:physics} present tests of general relativity (GR) using \gwTenEleven.
We conclude in Section~\ref{sec:conc}.
Additional details and results are provided in the Appendices.

\section{Detection and Significance}
\label{sec:detection}

\subsection{\gwTenEleven}

\gwTenEleven passed through the Earth's geocenter on October 11, 2024 at 23:38:34.9 UTC.
It was detected in low latency in LIGO Hanford and Virgo data~\citep{2024GCN.37776....1L} by the \gstlal matched-filter search pipeline~\citep{Messick:2016aqy, Sachdev:2019vvd, Hanna:2019ezx, Cannon:2020qnf, Sakon:2022ibh, Ewing:2023qqe, Tsukada:2023edh, Ray:2023nhx, Joshi:2025nty, Joshi:2025zdu}.
The signal was measured with optimized signal-to-noise ratios (SNRs) of $\gwTenElevenSNRlho$ and $\gwTenElevenSNRvirgo$ in LIGO Hanford and Virgo, respectively, and a false alarm rate (FAR) of $<10^{-5}\,\mathrm{yr}$.
LIGO Livingston was not operating at the time of the event.
Significant candidates were also identified by the \mbta~\citep{Adams:2015ulm, Aubin:2020goo, Allene:2025saz} and \pycbc~\citep{Allen:2004gu, Allen:2005fk, Usman:2015kfa, Nitz:2017lco, Nitz:2017svb, DalCanton:2020vpm, Davies:2020tsx} pipelines at the time of \gwTenEleven, but fell outside the mass range for which these searches report candidates in low latency seen in only one LIGO instrument~\citep{gwtc4-methods}.
LIGO Hanford and Virgo were each operating normally at the time of detection, with stable angle-averaged binary neutron star inspiral ranges of approximately $160$ and $50\,\mathrm{Mpc}$, respectively~\citep{Finn:1992xs, Chen:2017wpg}.
In subsequent high-latency searches, for which more comprehensive data quality assessment and precise background estimates were available, \gstlal, \mbta, and \pycbc each detected \gwTenEleven with false alarm rates below $3\times10^{-5}\,\mathrm{yr}^{-1}$.
\gwTenEleven's high network SNR makes it the third-loudest gravitational-wave event published to date, behind GW230814\_230901~\citep{gwtc4} and GW250114\_082203~\citep{gw250114}.

\subsection{\gwElevenTen}

\gwElevenTen arrived at Earth on November 10, 2024 at 12:41:23.6 UTC.
It was identified in low latency in LIGO Hanford, LIGO Livingston, and Virgo data~\citep{2024GCN.38155....1L}, assigned $\mathrm{FAR}=\gwElevenTenOnlineFARgstlal$ by the \gstlal search pipeline and lower significances by \mbta and \pycbc.
At the time of \gwElevenTen, the LIGO instruments had angle-averaged binary neutron star inspiral ranges of approximately $160\,\mathrm{Mpc}$, while Virgo's inspiral range was near $50\,\mathrm{Mpc}$.
High-latency analyses with the \pycbc, \gstlal, and \mbta pipelines later re-identified \gwElevenTen with false alarm rates of $\gwElevenTenOfflineFARpycbcRounded$, $\gwElevenTenOfflineFARgstlal$, and $\gwElevenTenOfflineFARmbtaRounded$ respectively.

For several hours around the arrival time of \gwElevenTen, microseismic ground motion near LIGO Livingston was elevated.
Elevated microseism is known to increase the rate of scattered light glitches in detector strain data, particularly below $30\,\mathrm{Hz}$~\citep{Soni:2020zwl,Soni:2023kqq}.
Between 0.3 and 2 s after the coalescence time, three noise transients are observed in the data, all below 30 Hz.
Two are low-SNR transients, while the third is a high-SNR scattered-light glitch occurring below $15\,\mathrm{Hz}$.
Although these transients are not expected to affect the observation and analysis of \gwElevenTen~\citep{Macas:2022afm, Hourihane:2025vxc}, all further studies of \gwElevenTen's properties (see Section~\ref{sec:pe}) use LIGO Livingston data only above $30\,\mathrm{Hz}$.

\section{Source Properties}
\label{sec:pe}

\begin{table*}[]
    \footnotesize
    \setlength{\tabcolsep}{4pt}
    \renewcommand{\arraystretch}{1.2}
    \centering
    \caption{
    Inferred source properties of \gwTenEleven and \gwElevenTen.
    Shown are each source's primary mass ($m_1$), secondary mass ($m_2$), mass ratio ($q$), chirp mass ($\mathcal{M}$), and luminosity distance $D_\mathrm{L}$, with mass parameters defined in the rest-frame of the source binary.
    We additionally quote a variety of spin measurements:
    primary dimensionless spin magnitude ($\chi_1$) and spin--orbit misalignment angle ($\theta_1$), primary spin components projected parallel ($\chi_{1,z} = \chi_1 \cos\theta_1$) and perpendicular ($\chi_{1,\perp} = \chi_1\sin\theta_1$) to the binaries' Newtonian orbital angular momenta, the effective inspiral spin $\chi_\mathrm{eff}$~\citep{Racine:2008qv, Ajith:2009bn, Santamaria:2010yb}, and the effective precessing spin $\chi_\mathrm{p}$~\citep{Hannam:2013oca, Schmidt:2014iyl}.
    We do not obtain informative constraints on the secondary spins of each source; see Appendix~\ref{app:pe} for secondary spin measurements.
    We follow parameter conventions as defined in Table 3 of~\citep{gwtc4-introduction}.
    For all but one parameter, we quote posterior medians and central $90\%$ credible uncertainties.
    For $\theta_1$ alone, uncertainties instead correspond to values containing the $90\%$ credible highest posterior density interval for $\cos\theta_1$.
    This is done to minimize the influence of the uniform-in-$\cos\theta$ priors, which exclude $\theta_1 = 0$.
    Spin parameters are quoted at infinite binary separation.
    }
    \begin{tabular}{r  c c c c c c c c c c c}
    \hline \hline
    Event & $m_1\,[M_\odot]$ & $m_2\,[M_\odot]$ & $q$ & $\mathcal{M}\,[M_\odot]$ & $D_\mathrm{L}\,[\mathrm{Mpc}]$ & $\chi_1$ & $\theta_1\,[\mathrm{deg}]$ & $\chi_{1,z}$ & $\chi_{1,\perp}$ & $\chi_\mathrm{eff}$ & $\chi_\mathrm{p}$ \\
    \hline
    \gwTenEleven 
    	& $\gwTenElevenMassOne^{+\gwTenElevenMassOneUpperError}_{-\gwTenElevenMassOneLowerError}$
	& $\gwTenElevenMassTwo^{+\gwTenElevenMassTwoUpperError}_{-\gwTenElevenMassTwoLowerError}$
	& $\gwTenElevenMassRatio^{+\gwTenElevenMassRatioUpperError}_{-\gwTenElevenMassRatioLowerError}$ 
	& $\gwTenElevenChirpMass^{+\gwTenElevenChirpMassUpperError}_{-\gwTenElevenChirpMassLowerError}$ 
	& $\gwTenElevenDistance^{+\gwTenElevenDistanceUpperError}_{-\gwTenElevenDistanceLowerError}$ 
	& $\gwTenElevenSpinOneMagMedian^{+\gwTenElevenSpinOneMagUpperError}_{-\gwTenElevenSpinOneMagLowerError}$
	& $\gwTenElevenSpinOneTiltMedian^{+\gwTenElevenSpinOneTiltUpperErrorHPD}_{-\gwTenElevenSpinOneTiltLowerErrorHPD}$
	& $\gwTenElevenSpinOneZMedian^{+\gwTenElevenSpinOneZUpperError}_{-\gwTenElevenSpinOneZLowerError}$
	& $\gwTenElevenSpinOnePlaneMedian^{+\gwTenElevenSpinOnePlaneUpperError}_{-\gwTenElevenSpinOnePlaneLowerError}$
	& $\gwTenElevenChiEffectiveMedian^{+\gwTenElevenChiEffectiveUpperError}_{-\gwTenElevenChiEffectiveLowerError}$
	& $\gwTenElevenChiPMedian^{+\gwTenElevenChiPUpperError}_{-\gwTenElevenChiPLowerError}$ \\
    \gwElevenTen
    	& $\gwElevenTenMassOne^{+\gwElevenTenMassOneUpperError}_{-\gwElevenTenMassOneLowerError}$
	& $\gwElevenTenMassTwo^{+\gwElevenTenMassTwoUpperError}_{-\gwElevenTenMassTwoLowerError}$
	& $\gwElevenTenMassRatio^{+\gwElevenTenMassRatioUpperError}_{-\gwElevenTenMassRatioLowerError}$
	& $\gwElevenTenChirpMass^{+\gwElevenTenChirpMassUpperError}_{-\gwElevenTenChirpMassLowerError}$ 
	& $\gwElevenTenDistance^{+\gwElevenTenDistanceUpperError}_{-\gwElevenTenDistanceLowerError}$
	& $\gwElevenTenSpinOneMagMedian^{+\gwElevenTenSpinOneMagUpperError}_{-\gwElevenTenSpinOneMagLowerError}$
	& $\gwElevenTenSpinOneTiltMedian^{+\gwElevenTenSpinOneTiltUpperErrorHPD}_{-\gwElevenTenSpinOneTiltLowerErrorHPD}$
	& $\gwElevenTenSpinOneZMedian^{+\gwElevenTenSpinOneZUpperError}_{-\gwElevenTenSpinOneZLowerError}$
	& $\gwElevenTenSpinOnePlaneMedian^{+\gwElevenTenSpinOnePlaneUpperError}_{-\gwElevenTenSpinOnePlaneLowerError}$
	& $\gwElevenTenChiEffectiveMedian^{+\gwElevenTenChiEffectiveUpperError}_{-\gwElevenTenChiEffectiveLowerError}$
	& $\gwElevenTenChiPMedian^{+\gwElevenTenChiPUpperError}_{-\gwElevenTenChiPLowerError}$ \\[1pt]
    \hline
    \hline
    \end{tabular}
    \label{tab:pe-params}
\end{table*}

\begin{figure}[t!]
    \centering
    \includegraphics[width=0.49\textwidth]{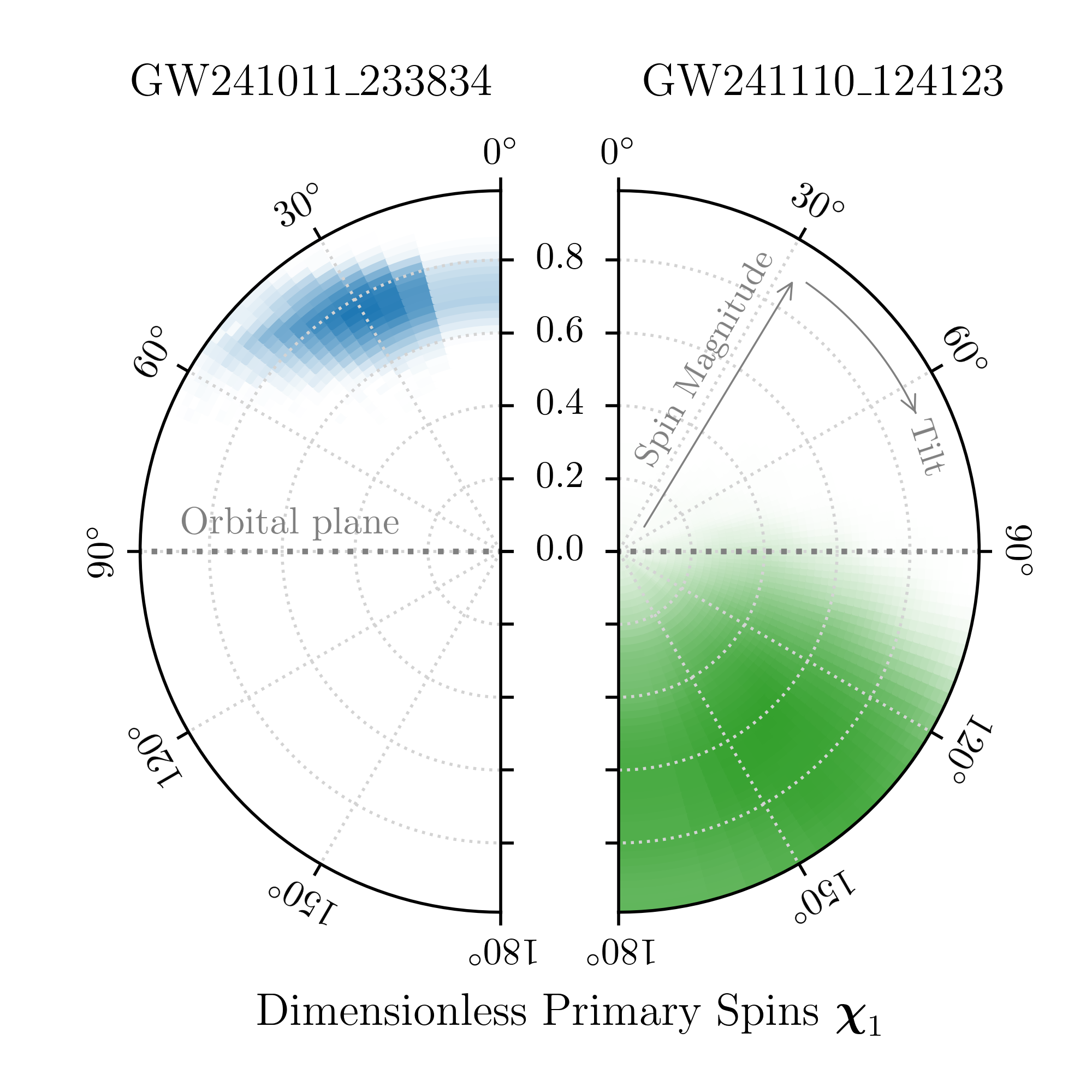}
    \caption{
    Posterior on the primary spin vector of \gwTenEleven (left) and \gwElevenTen (right).
    Within each subplot, radial coordinates span the range $0$ to $1$ and correspond to dimensionless spin magnitudes.
    The polar angles, spanning $0$ to $180\,\mathrm{deg}$, correspond to spin--orbit misalignment angles.
    Color saturation indicates posterior probability as a function of spin magnitude and orientation.
    Pixels are spaced linearly in spin magnitude and cosine tilt angle such that they each contain equal prior probability;
    a completely uninformative spin measurement would therefore yield a uniformly colored disk.
    \gwTenEleven has a precisely measured spin magnitude of $\chi_1  = \gwTenElevenSpinOneMagMedian^{+\gwTenElevenSpinOneMagUpperError}_{-\gwTenElevenSpinOneMagLowerError}$ that is misaligned by $\gwTenElevenSpinOneTiltMedian^{+\gwTenElevenSpinOneTiltUpperErrorHPD}_{-\gwTenElevenSpinOneTiltLowerErrorHPD}\,\mathrm{degrees}$ with respect to its orbital angular momentum.
    The primary spin of \gwElevenTen, in contrast, is constrained to be misaligned by more than $110\,\mathrm{deg}$ from its orbital angular momentum.
    }
    \label{fig:spin-disk}
\end{figure}

\begin{figure*}
    \centering
    \includegraphics[width=\textwidth]{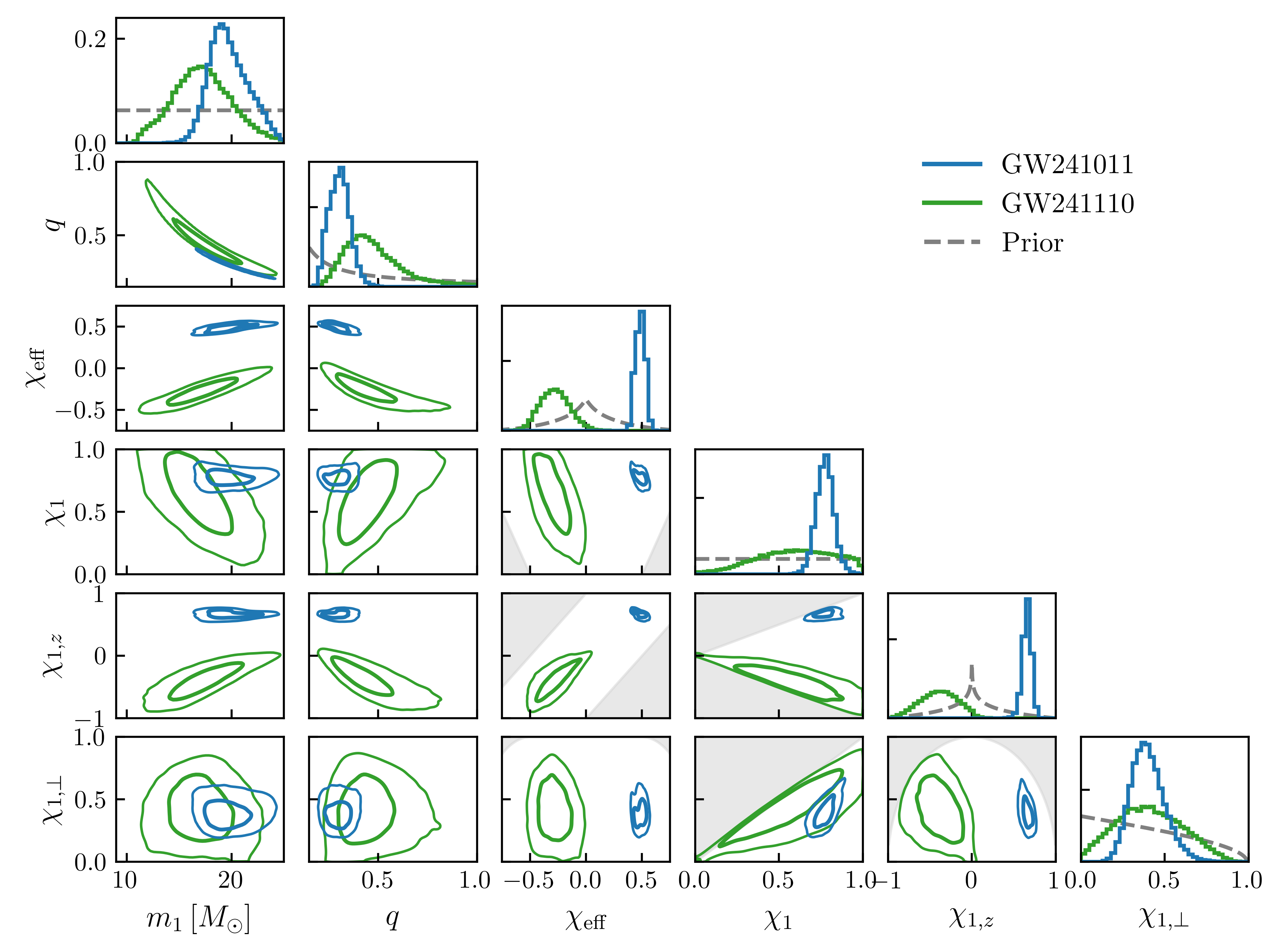}
    \caption{
    Posterior probabilities on selected properties of \gwTenEleven (blue) and \gwElevenTen (green):
    their primary masses $m_1$, mass ratios $q$, effective inspiral spins $\chi_\mathrm{eff}$, primary spin magnitudes $\chi_1$, and components of their primary spins projected parallel ($\chi_{1,z}$) and perpendicular ($\chi_{1,\perp}$) to each binary's Newtonian orbital angular momentum.
    Panels along the diagonal show marginalized posteriors on each parameter; the dashed histograms correspond to the prior probability distributions adopted during parameter estimation.
    Off-diagonal panels illustrate joint two-dimensional posteriors on each pair of parameters; thick and thin contours denote central $50\%$ and $90\%$ credible bounds.
    Shaded grey regions indicate to parts of parameter space that the constraint that spin magnitudes be less than or equal to one.
    Despite their extreme and opposite spins, \gwTenEleven and \gwElevenTen have consistent component mass measurements, each favoring a $15$--$20\,M_\odot$ primary and an unequal mass ratio.
    }
    \label{fig:corner}
\end{figure*}

\begin{figure}[ht!]
    \centering
    \includegraphics[width=0.48\textwidth]{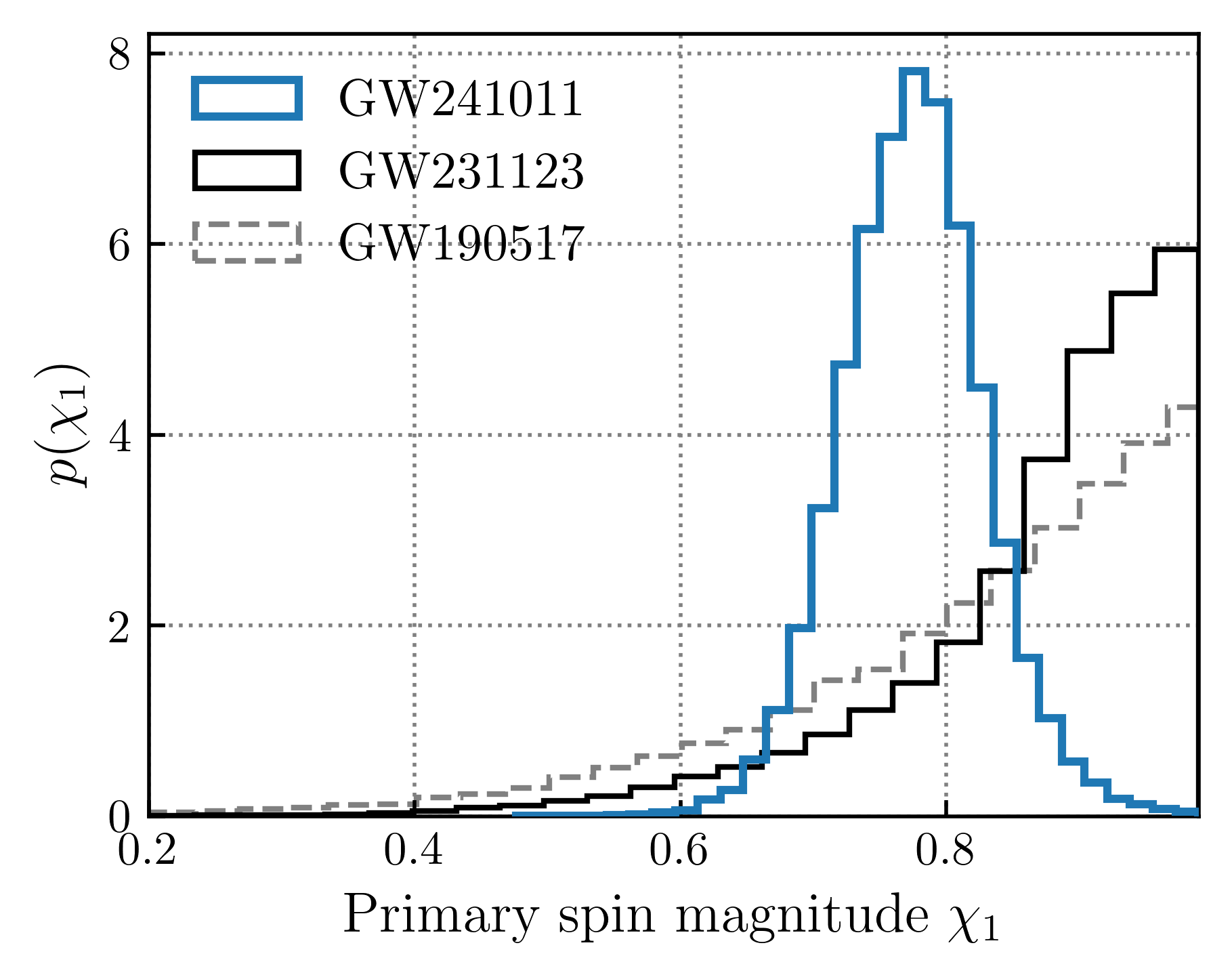}
    \caption{
    Posterior on the primary spin magnitude of \gwTenEleven (blue).
    The spin of \gwTenEleven's primary black hole has the largest lower bound than that of all other merging compact objects observed to date.
    We obtain  $\chi_1 > \gwTenElevenSpinOneMagLowerLimit$ at $95\%$ credibility.
    For comparison, also shown are the primary spin measurements from GW231123~\citep[black;][]{LIGOScientific:2025rsn} and GW190517~\citep[dashed grey]{KAGRA:2021vkt}, which provide the two next-largest lower limits of $\chi_1 > \gwTwentyThreeElevenTwentyThreeSpinOneMagLowerLimit$ and $\chi_1 > \gwNineteenOhFiveSeventeenLowerLimit$, respectively.
    }
    \label{fig:mag-comparison}
\end{figure}

\begin{figure}[ht!]
    \centering
    \includegraphics[width=0.48\textwidth]{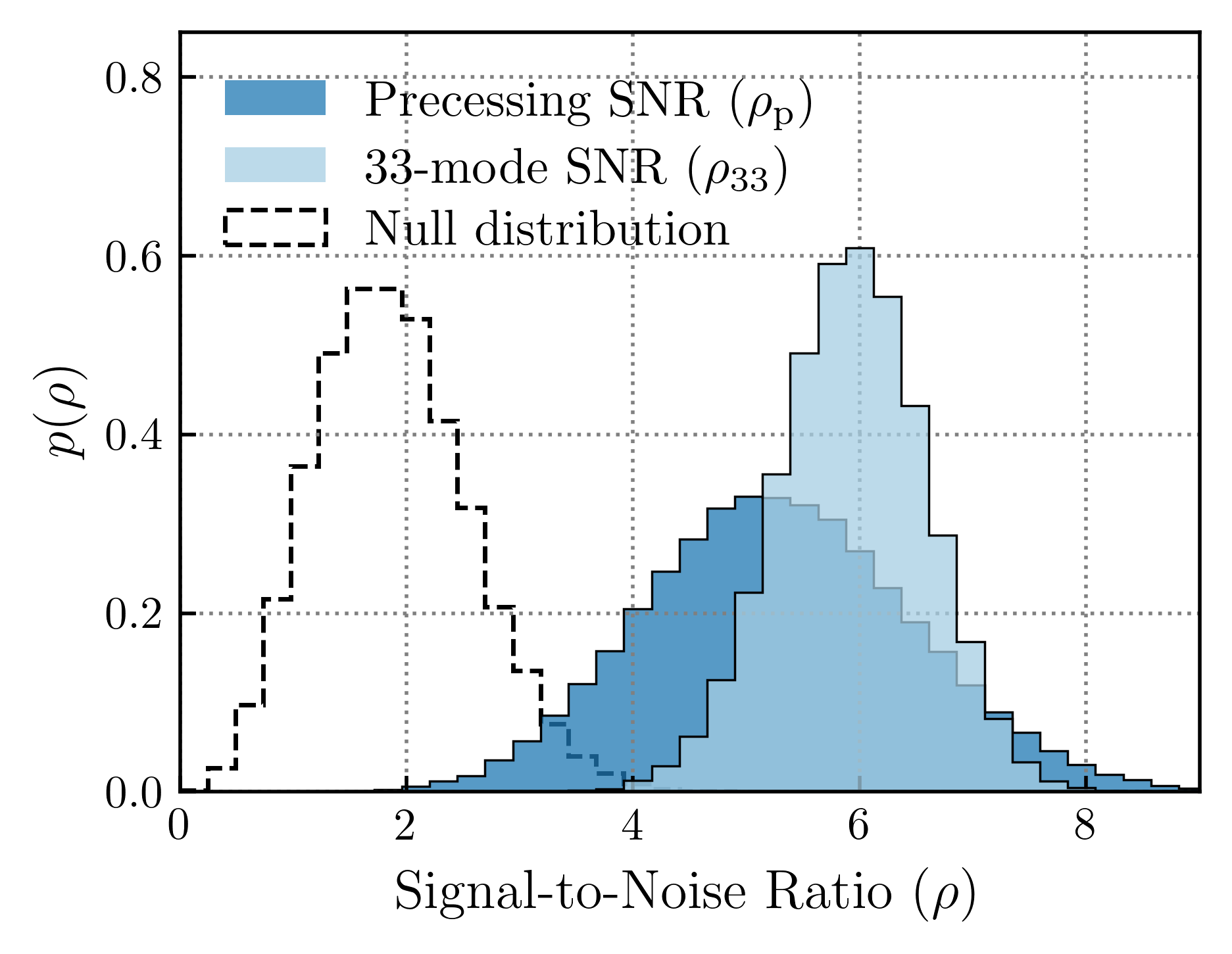}
    \caption{
    Posterior distribution on \gwTenEleven's precession SNR ratio and its SNR in $(\ell, m) = (3, \pm3)$ spherical harmonic modes.
    The precession SNR quantifies the observational strength of spin--orbit precession, while $(\ell, m) = (3, \pm3)$ mode radiation arises from source multipoles beyond the leading-order mass quadrupole.
    The unequal mass ratio and large, misaligned primary spin of \gwTenEleven together yield $\rho_\mathrm{p} = \gwTenElevenPrecessingSNRMedian^{+\gwTenElevenPrecessingSNRUpperError}_{-\gwTenElevenPrecessingSNRLowerError}$ and $\rho_{33} = \gwTenElevenHOMSNRMedian^{+\gwTenElevenHOMSNRUpperError}_{-\gwTenElevenHOMSNRLowerError}$.
    For comparison, the dashed histogram illustrates the expected null distribution (a $\chi$-distribution with four degrees of freedom, two per detector active during \gwTenEleven) of $\rho_\mathrm{p}$ and $\rho_{33}$ in the absence of spin--orbit precession or higher-order radiation modes~\citep{Prix:2007ks, Harry:2010fr}.
    This null distribution is conservative; in some cases the null distribution may take alternative forms (such as a $\chi$-distribution with fewer degrees of freedom) concentrated towards smaller SNRs~\citep{Fairhurst:2019vut, Hoy:2021dqg, Hoy:2024wkc}.
    }
    \label{fig:snrs}
\end{figure}

Bayesian parameter estimation is performed on \gwTenEleven and \gwElevenTen following the methodology described in~\cite{gwtc4-methods}.
Noise power spectral densities are obtained with the \BAYESWAVE algorithm~\citep{Cornish:2014kda, Littenberg:2014oda, Littenberg:2015kpb, Cornish:2020dwh, Gupta:2023jrn}
and astrophysical parameter inference is performed using the \RIFT~\citep{Pankow:2015cra, Lange:2017wki, Wysocki:2019grj} and \BILBY~\citep{Ashton:2018jfp, Romero-Shaw:2020owr} code packages, the latter of which invokes the \DYNESTY~\citep{Speagle:2019ivv} nested sampler.
Analysis of \gwTenEleven includes $32\,\mathrm{s}$ of data between $20$--$1792\,\mathrm{Hz}$ from LIGO Hanford and Virgo.
Analysis of \gwElevenTen incorporates $16\,\mathrm{s}$ of data from all three LIGO and Virgo instruments; data between $20$--$1792\,\mathrm{Hz}$ are used from LIGO Hanford and Virgo, whereas LIGO Livingston data are used only at frequencies above $30\,\mathrm{Hz}$ due to the presence of nonstationary low-frequency noise.
We adopt priors that are uniform in detector-frame component masses, uniform and isotropic in component spin magnitudes and orientations, and uniform in comoving volume and source-frame time~\citep{gwtc4-methods}.
Black hole spin orientations evolve over the course of an inspiral due to relativistic spin-orbit precession; we present spin measurements corresponding to the asymptotic values in the limit of infinite binary separation~\citep{Mould:2021xst, Johnson-McDaniel:2021rvv, Gerosa:2023xsx}.

The inferred source properties of \gwTenEleven and \gwElevenTen are summarized in Table~\ref{tab:pe-params}.
Posteriors on the primary black hole's spin vectors of each binary are shown in Figure~\ref{fig:spin-disk}, while posteriors on a subset of other binary parameters appear in Figure~\ref{fig:corner}.
These results comprise the union of posterior samples obtained using three different waveform models, \textsc{SEOBNRv5PHM}~\citep{Ramos-Buades:2023ehm}, \textsc{IMRPhenomXPHM-SpinTaylor}~\citep{Colleoni:2024knd} and \textsc{IMRPhenomXO4a}~\citep{Thompson:2023ase}, that each include the effects of spin--orbit precession and higher-order spherical harmonic modes.
Further details about parameter inference, including source properties inferred with each individual waveform model, are presented in Appendix~\ref{app:pe}.

\subsection{Properties of \gwTenEleven}

The source of \gwTenEleven is inferred to possess a $\gwTenElevenMassOne^{+\gwTenElevenMassOneUpperError}_{-\gwTenElevenMassOneLowerError}\,M_\odot$ primary mass and a confidently unequal mass ratio, $q=\gwTenElevenMassRatio^{+\gwTenElevenMassRatioUpperError}_{-\gwTenElevenMassRatioLowerError}$.
It has a large and precisely measured primary spin magnitude, $\chi_1 = \gwTenElevenSpinOneMagMedian^{+\gwTenElevenSpinOneMagUpperError}_{-\gwTenElevenSpinOneMagLowerError}$, and its primary spin is inferred to be misaligned with respect to its Newtonian orbital angular momentum by $\theta_1 = \gwTenElevenSpinOneTiltMedian^{+\gwTenElevenSpinOneTiltUpperErrorHPD}_{-\gwTenElevenSpinOneTiltLowerErrorHPD}\,\mathrm{deg}$.
The spin of \gwTenEleven's secondary black hole is unconstrained, as expected for an unequal-mass systems in which the primary's spin angular momentum dominates.
The inferred geometry of the spin vector of \gwTenEleven's primary black hole is depicted in the left-hand side of Figure~\ref{fig:spin-disk}.
Within this plot, the radial distance depicts the magnitude of \gwTenEleven's primary spin vector, while the polar coordinate denotes its spin--orbit misalignment angle; color saturation indicates posterior probability.
\gwTenEleven is, furthermore, probably the closest binary black hole merger observed to date, with a luminosity distance of $D_\mathrm{L} < \gwTenElevenDistanceUpperLimit\,\mathrm{Mpc}$ at $90\%$ credibility.

The primary of \gwTenEleven is among the most rapidly rotating black holes observed to date.
At $95\%$ credibility, \gwTenEleven has the largest lower limit obtained thus far on the spin of any merging black hole, with $\chi_1 > \gwTenElevenSpinOneMagLowerLimit$.
Figure~\ref{fig:mag-comparison} compares this primary spin measurement to two signals with similar spin magnitude limits, GW190517~\citep{KAGRA:2021vkt} and the recently-announced GW231123~\citep{LIGOScientific:2025rsn}, whose sources have $\chi_1 > \gwNineteenOhFiveSeventeenLowerLimit$ and $\chi_1 > \gwTwentyThreeElevenTwentyThreeSpinOneMagLowerLimit$ at $95\%$ credibility, respectively.
GW190403\_051519 favors similarly rapid spins, but this candidate does not meet the significance threshold adopted for population analyses.
Additionally, both the primary aligned spin $\chi_{1,z}$ and the effective inspiral spin of \gwTenEleven~\citep{Racine:2008qv, Ajith:2009bn, Santamaria:2010yb}, defined as
	\begin{equation}
	\chi_\mathrm{eff} = \frac{(m_1 \vec \chi_1 + m_2 \vec \chi_2) \cdot \hat L_\mathrm{N}}{m_1+m_2},
	\end{equation}
are bounded higher than any other gravitational-wave source.
Among black holes that are confidently rotating, the primary spin magnitude of \gwTenEleven is also the most precisely measured, with a $90\%$ credible region that is half as wide as the next-most-precise measurement, made using GW190412~\citep{LIGOScientific:2020stg}.
The binary neutron star source of GW170817~\citep{LIGOScientific:2017vwq}, the lower mass gap binary source of GW190814~\citep{LIGOScientific:2020zkf}, and the low-significance neutron star-black hole candidate GW191219\_163120 each have more precisely measured spins, but these spins are consistent with zero.
\gwTenEleven is therefore the gravitational-wave event that, taken individually, provides the strongest evidence to date that at least some component black holes in merging binaries spin rapidly, with dimensionless spins with magnitudes $\sim 0.7$ or greater.

\gwTenEleven additionally exhibits strong signatures of orbital plane precession and radiation in higher spherical harmonic modes.
Orbital plane precession occurs due to relativistic spin--orbit coupling, an effect that is maximized by the large and misaligned primary spin of \gwTenEleven.
The degree of spin-orbit precession may be parametrized using the effective precessing spin~\citep{Hannam:2013oca, Schmidt:2014iyl},
	\begin{equation}
	\chi_\mathrm{p} = \mathrm{Max}\left[ \chi_1 \sin \theta_1, \frac{3+4q}{4+3q} q \chi_2 \sin\theta_2\right],
	\end{equation}
related to the in-plane spin components $\chi_{i,\perp} = \chi_i \sin\theta_i$.
The source of \gwTenEleven has confidently non-zero effective precessing spin, with $\chi_\mathrm{p} = \gwTenElevenChiPMedian^{+\gwTenElevenChiPUpperError}_{-\gwTenElevenChiPLowerError}$,
and we obtain a log-Bayes factor of $\log_{10} \mathcal{B} = \logBayesPrecession$ in favor of a precessing source over a model in which spins are restricted to be co-aligned with their orbit.
As Bayes factors may, in general, be sensitive to one's choice of prior, we additionally quantify evidence for precession via the precession SNR $\rho_\mathrm{p}$~\citep{Fairhurst:2019srr, Fairhurst:2019vut}, the posterior distribution of which is shown in Figure~\ref{fig:snrs}.
We find $\rho_\mathrm{p} = \gwTenElevenPrecessingSNRMedian^{+\gwTenElevenPrecessingSNRUpperError}_{-\gwTenElevenPrecessingSNRLowerError}$.
In the absence of precession, $\rho_\mathrm{p}$ is expected to follow the null distribution indicated in Fig.~\ref{fig:snrs}.
Random draws from \gwTenEleven's $\rho_\mathrm{p}$ posterior exceed random draws from this null distribution $\gwTenElevenPrecessingSNRPercentageAboveNull\%$ of the time.

The significant mass asymmetry of \gwTenEleven also yields significant radiation in higher-order spherical harmonic modes.
Whereas gravitational-wave radiation is typically dominated by $(\ell, m) = (2, \pm 2)$ spherical harmonics, subdominant modes, such as $(\ell, m) = (2, \pm 1)$ or $(3, \pm 3)$, become increasingly important for systems with considerable mass asymmetry~\citep{Thorne:1980ru, Berti:2007fi, Blanchet:2013haa, Mills:2020thr}.
\gwTenEleven exhibits significant radiation in the $(\ell, m) = (3, \pm 3)$ spherical harmonics, here assuming symmetric contributions from $m=3$ and $-3$ modes thus neglecting small asymmetries arising from source precession.
We find a log-Bayes factor of $\log_{10} \mathcal{B}=\logBayesHOM$ in favor of a signal model including higher-order spherical harmonics, compared to a model including only contributions from $(\ell, m) = (2, \pm 2)$ modes.
A posterior distribution on the SNR $\rho_{33}$ measured in higher-order modes is shown in Figure~\ref{fig:snrs}, with $\rho_{33} = \gwTenElevenHOMSNRMedian^{+\gwTenElevenHOMSNRUpperError}_{-\gwTenElevenHOMSNRLowerError}$.
Random draws from the posterior on $\rho_{33}$ exceed draws from the null distribution over $99.9\%$ of the time.
No detection is made of other spherical harmonic modes.

\subsection{Properties of \gwElevenTen}

\gwElevenTen is inferred to have component masses consistent with those of \gwTenEleven, albeit with larger uncertainties due to its lower SNR:
the primary mass and mass ratio of \gwElevenTen are measured to be $m_1 = \gwElevenTenMassOne^{+\gwElevenTenMassOneUpperError}_{-\gwElevenTenMassOneLowerError}\,M_\odot$ and $q=\gwElevenTenMassRatio^{+\gwElevenTenMassRatioUpperError}_{-\gwElevenTenMassRatioLowerError}$ (see Figure~\ref{fig:corner}).
Although \gwElevenTen favors unequal component masses, mass ratios of $q\approx1$ cannot be fully excluded.

In contrast to \gwTenEleven, though, \gwElevenTen is measured to have a primary spin that is likely significantly misaligned with respect to its orbital angular momentum.
The angle between the binary's orbital angular momentum and the spin vector of its more massive black hole is $\theta_1 = \gwElevenTenSpinOneTiltMedian^{+\gwElevenTenSpinOneTiltUpperErrorHPD}_{-\gwElevenTenSpinOneTiltLowerErrorHPD}$ degrees, with $\theta_1>\gwElevenTenSpinOneTiltLowerLimit\,\mathrm{degrees}$ at $90\%$ credibility and $\theta_1>90\,\mathrm{degrees}$ at $\gwElevenTenSpinOneTiltProbabilityMoreThanNinteyDegrees\%$ credibility.
The spin of \gwElevenTen's secondary black hole is unconstrained.
\gwElevenTen favors negative effective spin, with $\chi_\mathrm{eff} < 0$ at $\gwElevenTenChiEffectiveProbabilityLessThanZero\%$ credibility.
As illustrated in Figure~\ref{fig:corner}, \gwElevenTen inferred mass ratio is strongly anticorrelated with its inferred $\chi_\mathrm{eff}$ and $\chi_{1,z}$.
This effect arises from a degeneracy in the post-Newtonian expansion of a compact binary's phase evolution~\citep{Cutler:1994ys, Poisson:1995ef, Baird:2012cu, Purrer:2013ojf, Ng:2018neg}.
Thus, if one requires \gwElevenTen to have positive $\chi_\mathrm{eff}$ ($\gwElevenTenChiEffectiveProbabilityPositive\%$ probability), it must also be the case that its mass ratio is $q\lesssim 0.3$.
If \gwElevenTen's mass ratio is above $q=0.3$, then the event must have negative $\chi_\mathrm{eff}$ at $\gwElevenTenRestrictedChiEffectiveProbabilityLessThanZero\%$ credibility.

\gwElevenTen offers the most significant, although not necessarily conclusive, direct evidence to date that at least some merging black holes have spins antialigned with their orbital angular momentum.
A growing number of binary black holes have been observed with large spin--orbit misalignment angles (with, e.g., $\theta \gtrsim 50\,\mathrm{degrees}$ at high credibility).
Several detections, including, e.g., GW231230\_170116, GW230723\_101834, and GW230609\_064958, furthermore favor misalignment angles greater than $90\,\mathrm{degrees}$, but only at $\sim 80\%$ credibilities.~\citep{gwtc4}.
Analysis of the binary black hole GW191109\_010717~\citep[GW191109;][]{KAGRA:2021vkt} with a numerical-relativity surrogate waveform model, meanwhile, found a negative effective inspiral spin at $>99\%$ credibility~\citep{Islam:2023zzj}.
The robustness of GW191109's spin measurement is uncertain, however, due to significant contamination by noise transients~\citep{KAGRA:2021vkt}; reanalyses that simultaneously seek to model and subtract these glitches are inconclusive~\citep{Udall:2024ovp}.
The spin measurement with \gwElevenTen, in contrast, is not believed to be subject to data-quality concerns.
Under uniform and isotropic spin priors and standard waveform models, there does remain a $\gwElevenTenChiEffectiveProbabilityPositive\%$ probability that the source possesses zero or positive effective inspiral spin.
The confidence in \gwElevenTen's spin anti-alignment is further bolstered when adopting an astrophysically-informed prior (see Appendix~\ref{subsec:pop_reweighting}), but we cannot completely exclude the possibility that the source has zero or positive primary spin.

\subsection{Relationship with the binary black hole population}
\label{sec:astro:pop}

Although \gwTenEleven and \gwElevenTen are remarkable for their large and well-measured black hole spins, we do not conclude that they are outliers with respect to the known binary black hole population; see Appendix~\ref{app:pop} for details.
These events do, however, individually reinforce conclusions that have been previously drawn only on statistical grounds.
Past studies of the binary black hole spin distribution have concluded (\textit{i}) that merging black holes are usually, but not exclusively, slowly rotating, (\textit{ii}) that black hole spins are unlikely to be isotropic, statistically favoring spin--orbit alignment and positive effective spins, but that (\textit{iii}) black hole spins nevertheless exhibit a wide range of spin--orbit misalignment angles, with some component black holes misaligned by nearly or greater than $90\,\mathrm{deg}$ with respect to their orbit~\citep{Farr:2017uvj, Farr:2017gtv, LIGOScientific:2018jsj, Roulet:2018jbe, Miller:2020zox, LIGOScientific:2020stg, KAGRA:2021duu, Callister:2022qwb, Tong:2022iws, Vitale:2022dpa, Callister:2023tgi, Banagiri:2025dxo, gwtc4-astrodist}.
\gwElevenTen offers the strongest confirmation to date of this latter conclusion.
The tension between population-level conclusions and the paucity of individual binary black holes favoring anti-aligned spins can be understood as a confluence of three factors: the large uncertainties typically inherent in spin measurements~\citep{vanderSluys:2007st, Vitale:2014mka, Ghosh:2015jra, Vitale:2016avz, Chatziioannou:2018wqx, Pratten:2020igi, Biscoveanu:2021nvg}, a degeneracy between mass ratio and spins that asymmetrically biases $\chi_\mathrm{eff}$ measurements towards larger positive values~\citep{Cutler:1994ys, Poisson:1995ef, Baird:2012cu, Purrer:2013ojf, Ng:2018neg}, and selection effects that cause events with larger, positive $\chi_\mathrm{eff}$ to be more readily detected and more precisely characterized~\citep{Flanagan:1997sx, Campanelli:2006uy, Ng:2018neg}.
These effects together have been shown to resolve the apparent inconsistency between individual binary black hole properties and statistical population-level conclusions~\citep{Hoy:2024wkc, Payne:2024ywe}.

While we do not conclude that \gwTenEleven and \gwElevenTen are outliers, it is possible that they are members of an emerging subpopulation of binary black holes, characterized by low primary masses, unequal mass ratios, and a wide range of spin--orbit misalignment angles.
It remains unknown, however, whether such a subpopulation exists and can be formally characterized.
This question will be further explored in future work involving additional observations from LIGO--Virgo--KAGRA's fourth observing run.

\section{Astrophysical Interpretation and Implications}
\label{sec:astro}

\subsection{\gwTenEleven and \gwElevenTen through isolated binary evolution}

The primary spins of \gwTenEleven and \gwElevenTen are difficult to explain via isolated binary evolution.
Efficient angular momentum transport from stellar interiors is predicted to yield black holes that are born slowly rotating~\citep{Spruit:1999cc, Spruit:2001tz, Qin:2018vaa, Fuller:2019sxi}, and torques exerted via mass transfer or tides are expected to coalign residual spin with a binary's orbital angular momentum~\citep{Zaldarriaga:2017qkw, Gerosa:2018wbw, Qin:2018vaa, Bavera:2020inc, Bavera:2020uch, Ma:2023nrf}.
The natal spins of black holes do remain highly uncertain, however, with the prediction of small spins due to the Spruit--Tayler dynamo being only one of many possibilities~\citep{Miller:2014aaa}.
Models predicting larger natal spins may better accommodate the observed source properties of \gwTenEleven and \gwElevenTen.

Alternative scenarios can also potentially explain the spin properties of \gwTenEleven and \gwElevenTen in the context of isolated binary evolution.
Stochastic spin-up of stellar cores immediately preceding core collapse could impart large and misaligned birth spins to black holes~\citep{2014ApJ...796...17F, 2015ApJ...810..101F, 2016ApJ...827...40G, Ma:2019cpr, McNeill:2020hbp, Antoni:2021xzs, Antoni:2023yxs, Baibhav:2024rkn}.
Although black hole progenitors are generally predicted to experience small natal kicks due to near-complete fallback accretion~\citep[e.g.,][]{Fryer:2011cx, Zevin:2017evb, Mandel:2020qwb, Vigna-Gomez:2023euq}, stronger-than-expected natal kicks and/or asymmetric fallback might misalign spins either by tilting a binary's orbital plane or by torquing of a black hole's spin vector~\citep{Wongwathanarat:2012zp, Chan:2020lnd, Janka:2021deg, Tauris:2022ggv, Burrows:2024wqv}; spin--orbit misalignment among some black hole X-ray binaries may provide evidence for such effects~\citep{Zdziarski:2018dms, Salvesen:2020nkm, Poutanen:2021sag, Zdziarski:2023ygh}.
Finally, the von Zeipel--Lidov--Kozai mechanism~\citep{Zeipel:1910, Lidov:1962wjn, Kozai:1962zz} due to a tertiary companion can, when coupled to relativistic spin--orbit precession and gravitational-wave emission, tilt the orbital plane as well as component spins to yield a large spin--orbit misalignment~\citep{Liu:2017yhr, Antonini:2017tgo, Liu:2018nrf, Fragione:2019poq, Liu:2019gdc, Yu:2020iqj, Stegmann:2025clo}; however, this mechanism does not explain large spin \textit{magnitudes}.

\begin{figure*}[t!]
    \centering
    \includegraphics[width=\textwidth]{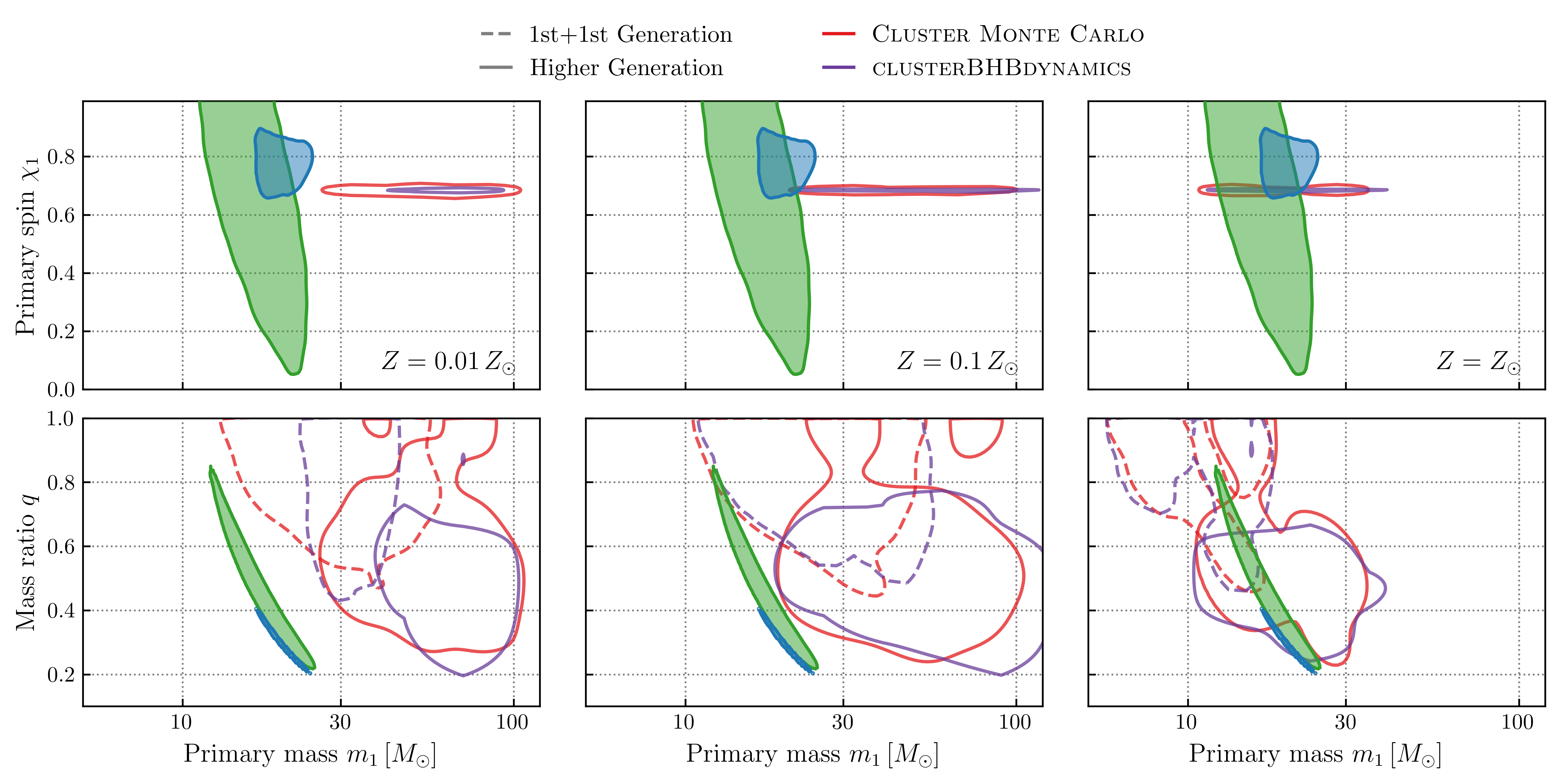}
    \caption{
    90\% credible bounds on the primary masses, mass ratios, spins of \gwTenEleven (blue) and \gwElevenTen (green), compared to predicted properties of merging black holes in dense star clusters from the \textsc{Cluster Monte Carlo} catalog~\citep{Kremer:2019iul, 2022ApJS..258...22R}, and \textsc{clusterBHBddynamics} models~\citep{Antonini:2022vib}.
    Dashed contours correspond to merging first-generation black holes, while solid contours correspond to higher-generation binaries containing remnants from previous mergers.
    Both models assume black holes to be born non-rotating.
    The three columns correspond to star cluster simulations with stellar populations at three different metallicities: $Z=0.01\,Z_\odot$, $0.1\,Z_\odot$, and $Z_\odot$.
    Under the modeling assumptions, the masses of \gwTenEleven and \gwElevenTen appear inconsistent with those predicted in low-metallicity stellar clusters with $Z=0.01\,Z_\odot$.
    Their masses and spins may be consistent, though, with those  predicted among  hierarchical binary black hole mergers that are formed in clusters of moderate or near-solar metallicities.
    }
    \label{fig:cluster-comparison}
\end{figure*}

\subsection{\gwTenEleven and \gwElevenTen as hierarchical mergers}

A more natural interpretation for \gwTenEleven and \gwElevenTen is that they involve the mergers of second-generation black holes in dense stellar environments, such as globular, nuclear, and young massive star clusters~\citep{PortegiesZwart:2010cly, Neumayer:2020gno}.
Compact binaries merging within clusters yield remnants that may be retained by the cluster, continue to interact dynamically, and themselves participate in subsequent mergers driven by gravitational wave emission~\citep[e.g.,][]{2001CQGra..18.3977L, OLeary:2005vqo, Miller:2008yw, Giersz:2015mlk, Antonini:2016gqe, Rodriguez:2019huv, Doctor:2021qfn, Fragione:2021nhb, Mahapatra:2021hme, Rizzuto:2021atw, Atallah:2022toy, Mahapatra:2022ngs, ArcaSedda:2023mlv}.
Such second-generation black holes are systematically more massive than their first-generation ancestors and are expected to be rapidly rotating; the spin distribution of remnant black holes is generically and robustly concentrated about $\chi \approx 0.7$~\citep{Pretorius:2005gq, Buonanno:2007sv, Berti:2008af}.
A remnant black hole's spin arises from the total remaining angular momentum of its ancestral binary at merger, which, for approximately equal-mass mergers, is dominated by the orbital angular momentum.
The spin angular momenta of the ancestral black holes do affect the remnant's spin, but their contribution is, in part, countered by the relationship between spin and inspiral duration.
Binaries with larger aligned spins undergo longer inspirals and radiate away more orbital angular momentum, while binaries with small or anti-aligned spins merge more promptly and thus retain more orbital angular momentum~\citep{Campanelli:2006uy}.

Observationally, mergers involving second-generation objects (often called hierarchical mergers) would distinguish themselves via their large spin magnitudes, statistically isotropic spin orientations, and mass ratios that are typically less than unity.
The spin magnitudes, spin--orbit misalignments, and unequal mass ratios of \gwTenEleven and \gwElevenTen therefore make these signals prime candidates for arising from a hierarchical origin.
Figure~\ref{fig:cluster-comparison}, for example, illustrates predictions for the possible primary spin magnitudes (upper row) and mass ratios (lower row) among binary black hole mergers in dense star clusters.
Dashed contours indicate predicted properties of mergers in which both components are first generation black holes, while solid contours correspond to systems in which at least one component is the product of a previous merger.
We show data from two models, the \textsc{Cluster Monte Carlo} catalog~\citep[\textsc{CMC};][]{Kremer:2019iul, 2022ApJS..258...22R} and the \textsc{clusterBHBdynamics} model~\citep[\textsc{cBHBd};][]{Antonini:2019ulv, Antonini:2022vib};
further details regarding both models are provided in Appendix~\ref{app:astromodels}.
The spin magnitude and mass ratio of \gwTenEleven, in particular, are inconsistent with the properties predicted of first-generation mergers, but lie within ranges predicted by both models for higher-generation hierarchical mergers.
Figure~\ref{fig:cluster-comparison} does not, however, convey relative numbers of first-generation and higher-generation mergers; both models predict approximately one higher-generation merger per five first-generation mergers.

Hierarchical black hole mergers are often associated with massive black holes.
At sub-solar metallicities, black holes masses are thought to be limited by (pulsational)-pair-instability processes~\citep{Barkat:1967zz, Woosley:2007qp}, 
preventing the formation of first-generation black holes with masses between $\sim50$ and $\sim 120\,M_\odot$ \cite[e.g.,][]{Spera:2017fyx, Woosley:2021xba, Hendriks:2023yrw}.
Hierarchical mergers may yield second-generation remnants with masses situated in this range, and are therefore a possible evolutionary origin for massive systems like the sources of GW190521~\citep{LIGOScientific:2020iuh} and GW231123\_135430~\citep{LIGOScientific:2025rsn}.
The sources of \gwTenEleven and \gwElevenTen are, in contrast, relatively light.
These masses are not inconsistent with a hierarchical origin, however.
The three columns of Figure~\ref{fig:cluster-comparison} correspond to predictions for clusters of differing stellar metallicities.
At high metallicities, stellar winds increasingly strip massive stars of their envelopes, preventing the formation of massive carbon-oxygen cores and reducing the final masses of black holes~\citep[e.g.,][]{Belczynski:2001uc, Fryer:2011cx, Belczynski:2016jno, Spera:2016slz}.
Therefore, although hierarchical mergers in low-metallicity environments predominantly involve high primary masses, clusters with stellar metallicities $Z\approx 0.1\,Z_\odot$ and above are predicted to readily yield hierarchical mergers with $\sim 20\,M_\odot$ primaries, consistent with the sources of \gwTenEleven and \gwElevenTen~\citep{Ye:2025ano}.

\subsection{The ancestors of \gwTenEleven and \gwElevenTen}

\begin{figure*}[t!]
    \centering
    \includegraphics[width=\textwidth]{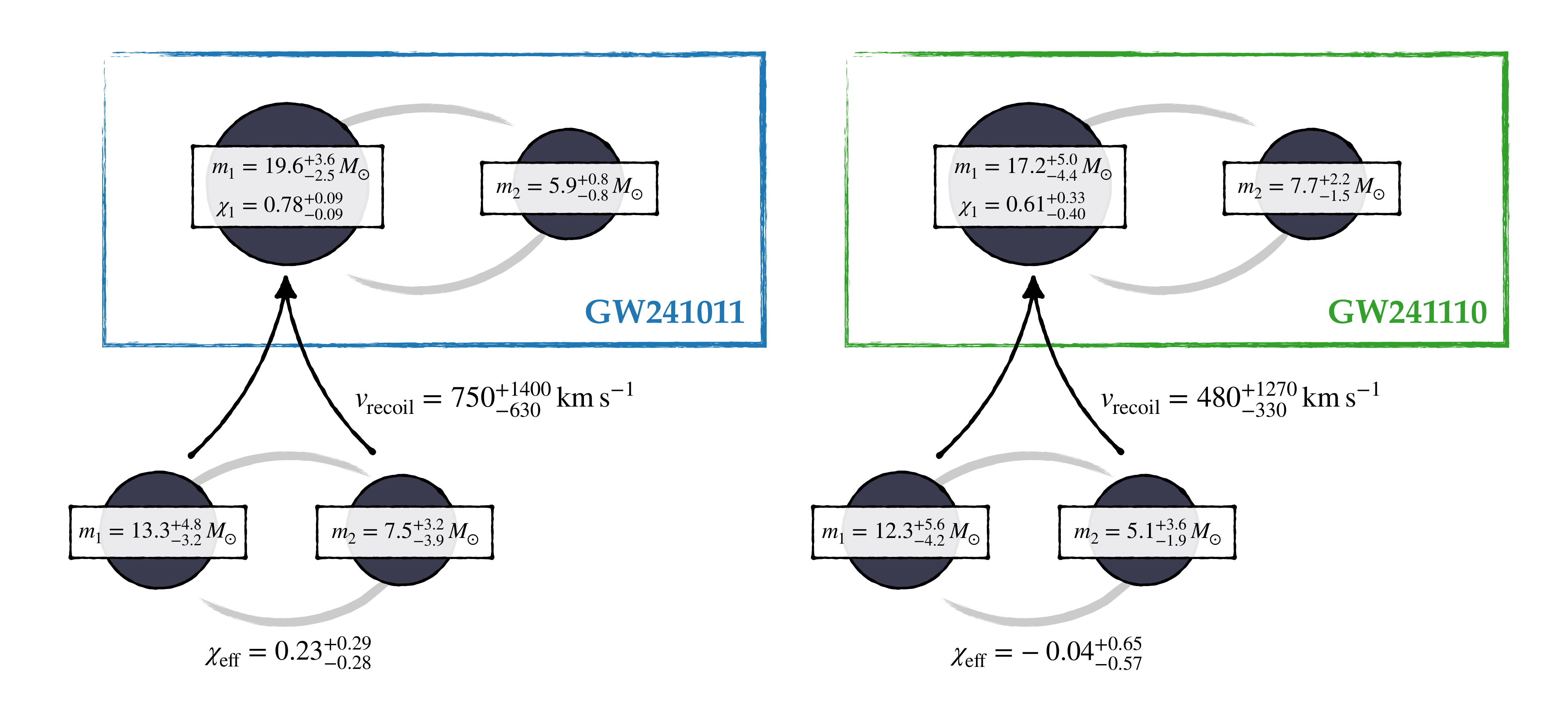}
    \caption{
   Inferred properties of the first-generation ancestors to the more massive black holes in \gwTenEleven and \gwElevenTen, under the hypothesis that these black holes were formed hierarchically from a previous merger.
   Shown are median and $90\%$ credible bounds inferred on ancestral component masses and effective inspiral spins, as well as the recoil kicks imparted to each remnant black hole due to asymmetric gravitational-wave-emission.
   We compute ancestral properties in two manners: one (the \textit{Backward} approach) that agnostically retains the same priors and posteriors on \gwTenEleven and \gwElevenTen's source properties as presented in Section~\ref{sec:pe}, and one (the \textit{Forward} approach) that adopts an astrophysically-informed prior on possible ancestral properties.
   This figure includes results from the agnostic \textit{Backward} approach; constraints obtained under the astrophysically-informed \textit{Forward} approach are described in the text.
    }
    \label{fig:progenitors}
\end{figure*}

If the primaries of \gwTenEleven and \gwElevenTen are second-generation remnants of previous mergers, we can indirectly constrain the masses, spins, and recoil kicks associated with their first-generation ancestors~\citep[e.g., ][]{Baibhav:2021qzw, Barrera:2022yfj, Paynter:2022mqw, Mahapatra:2024qsy, Alvarez:2024dpd, Mahapatra:2025agb}.
We explore this question using two complementary methods.
In the first method (the \textit{Forward} approach), we adopt astrophysically-informed priors on the ancestral masses and spins of \gwTenEleven and \gwElevenTen's hypothesized first-generation ancestors.
Priors are chosen to follow closely the results from stellar cluster simulations presented in Figure~\ref{fig:cluster-comparison}, strongly preferring equal masses and first-generation black hole spins near zero; these priors are described in more detail in Appendix~\ref{app:progenitor_appendix}.
We then proceed via a hierarchical Bayesian approach, using observed strain data to obtain posteriors on ancestral properties, marginalized over the source properties of \gwTenEleven and \gwElevenTen~\citep{Mahapatra:2024qsy}.
In the second method (the \textit{Backward} approach), we proceed more agnostically.
Beginning with the source masses and spins of \gwTenEleven and \gwElevenTen's primaries from Section~\ref{sec:pe}, for each posterior sample we identify an ancestral binary whose remnant mass and spin are consistent, within a small tolerance, with this posterior sample.
This approach allows us to construct posteriors on required ancestral properties while preserving the astrophysically-agnostic priors and posteriors on the source properties of \gwTenEleven and \gwElevenTen themselves~\citep{Alvarez:2024dpd}.
In both cases, numerical-relativity simulations are used to map between ancestral binaries and remnant properties~\citep{Varma:2019csw}.

Figure~\ref{fig:progenitors} illustrates the required properties of \gwTenEleven and \gwElevenTen's ancestors estimated in the more agnostic \textit{Backward} approach.
We infer a first-generation ancestor of \gwTenEleven to have had masses $\gwTenElevenProgMassOneMedianWong^{+\gwTenElevenProgMassOneUpperErrorWong}_{-\gwTenElevenProgMassOneLowerErrorWong}\,M_\odot$ and $\gwTenElevenProgMassTwoMedianWong^{+\gwTenElevenProgMassTwoUpperErrorWong}_{-\gwTenElevenProgMassTwoLowerErrorWong}\,M_\odot$.
A first generation ancestor to \gwElevenTen likely possessed similar masses: $\gwElevenTenProgMassOneMedianWong^{+\gwElevenTenProgMassOneUpperErrorWong}_{-\gwElevenTenProgMassOneLowerErrorWong}\,M_\odot$ and $\gwElevenTenProgMassTwoMedianWong^{+\gwElevenTenProgMassTwoUpperErrorWong}_{-\gwElevenTenProgMassTwoLowerErrorWong}\,M_\odot$.
The effective inspiral spin of \gwTenEleven's ancestor is constrained to $\chi_\mathrm{eff} = \gwTenElevenProgChiEffMedianWong^{+\gwTenElevenProgChiEffUpperErrorWong}_{-\gwTenElevenProgChiEffLowerErrorWong}$; larger or smaller values would over- or underpredict, respectively, the observed primary spin $\chi_1$.
The primary spin of \gwElevenTen is measured less precisely and so allows for a broader range of ancestral effective spins: $\chi_\mathrm{eff} = \gwElevenTenProgChiEffMedianWong^{+\gwElevenTenProgChiEffUpperErrorWong}_{-\gwElevenTenProgChiEffLowerErrorWong}$.
For both binaries, the remnant recoil kicks are inferred to lie between approximately $100$--$2000\,\mathrm{km}\,\mathrm{s}^{-1}$.
Constraints on recoil kicks are, however, almost entirely prior dominated, and it is therefore unclear if they yield meaningful constraints on environmental escape velocities required for successful remnant retention.

In the \textit{Forward} approach, the astrophysically-informed yields posteriors favoring more equal-mass ancestors.
The ancestor of \gwTenEleven is inferred to have component masses $\gwTenElevenProgMassOneMedianMahapatra^{+\gwTenElevenProgMassOneUpperErrorMahapatra}_{-\gwTenElevenProgMassOneLowerErrorMahapatra}\,M_\odot$ and $\gwTenElevenProgMassTwoMedianMahapatra^{+\gwTenElevenProgMassTwoUpperErrorMahapatra}_{-\gwTenElevenProgMassTwoLowerErrorMahapatra}\,M_\odot$,
while \gwElevenTen's ancestor is inferred to have masses $\gwElevenTenProgMassOneMedianMahapatra^{+\gwElevenTenProgMassOneUpperErrorMahapatra}_{-\gwElevenTenProgMassOneLowerErrorMahapatra}\,M_\odot$ and $\gwElevenTenProgMassTwoMedianMahapatra^{+\gwElevenTenProgMassTwoUpperErrorMahapatra}_{-\gwElevenTenProgMassTwoLowerErrorMahapatra}\,M_\odot$.
The global maximum of the binary black hole mass function is situated at approximately $10\,M_\odot$~\citep{Tiwari:2020otp, KAGRA:2021duu, Farah:2023vsc, gwtc4-astrodist}; the ancestral black holes of both \gwTenEleven and \gwElevenTen are inferred to lie near this peak.
Because the astrophysical prior adopted in the \textit{Forward} approach requires ancestral spins to be near zero, inferred recoil kicks are systematically lower, ranging between approximately $10$--$300\,\mathrm{km}\,\mathrm{s}^{-1}$.
As in the \textit{Backward} approach above, this range is a consequence of our prior.

A more detailed presentation of both sets of results and additional methodological detail is provided in Appendix~\ref{app:progenitor_appendix}.

\subsection{No evidence for eccentricity}
\label{sec:astro:ecc}

\begin{figure}[t!]
    \centering
    \includegraphics[width=0.48\textwidth]{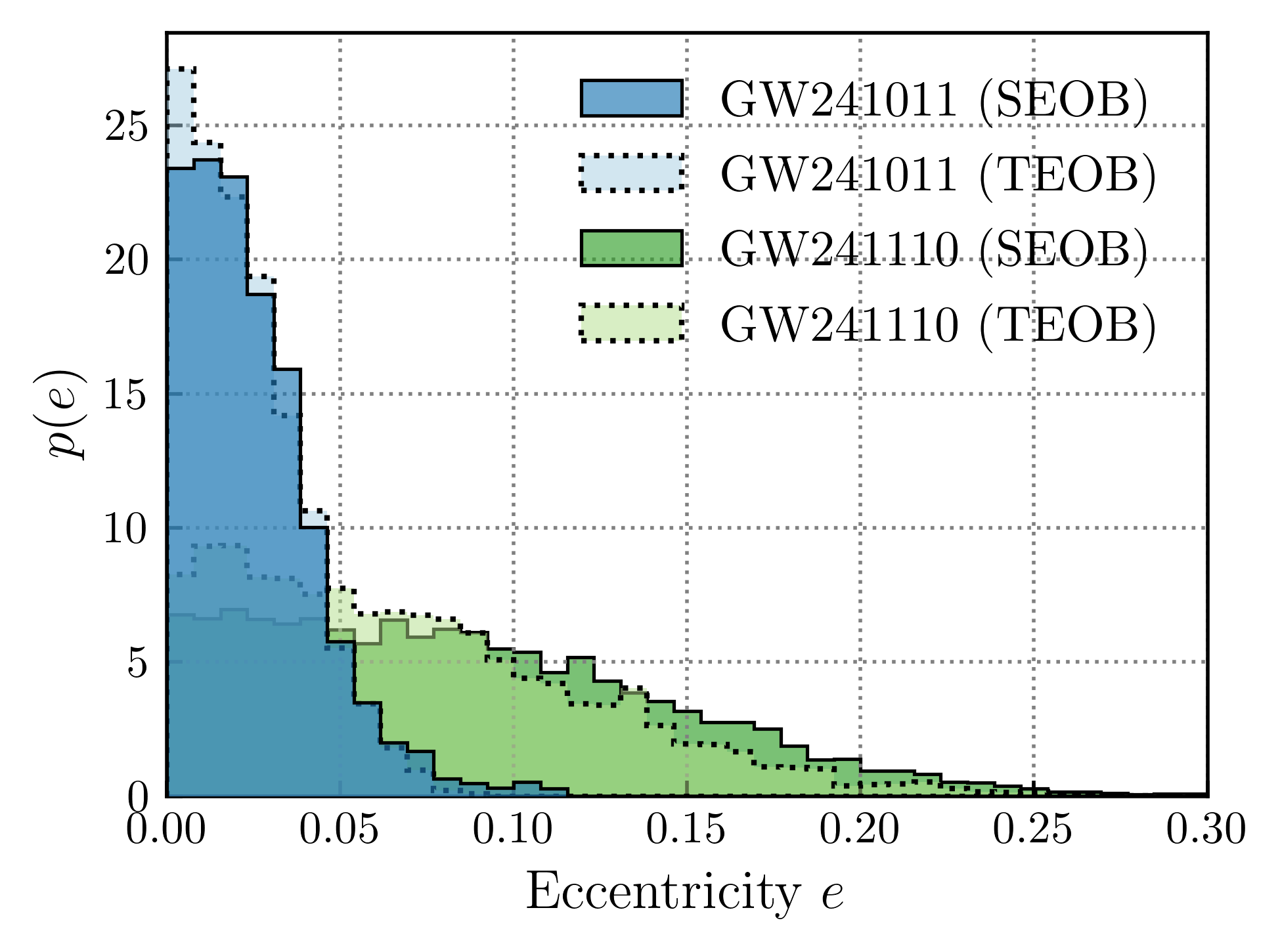}
    \caption{
    Posteriors on the orbital eccentricity of \gwTenEleven and \gwElevenTen.
    Results are obtained using the \textsc{SEOBNRv5EHM}~\citep{Gamboa:2024hli} and \textsc{TEOBRESUMS-DAL\'I}~\citep{Nagar:2024dzj} waveform models, and eccentricities are quoted at a reference frequency of $13.33\,\mathrm{Hz}$, when higher-order spherical harmonic modes first enter the observable frequency band.
    Neither event exhibits evidence for residual eccentricity.
    We bound $e<\gwTenElevenSEOBEccentricityUpperLimit$ at $90\%$ credibility for \gwTenEleven under both waveform models.
    \gwElevenTen, meanwhile, yields $90\%$ credible upper limits of $e<\gwElevenTenSEOBEccentricityUpperLimit$ and $e<\gwElevenTenTEOBEccentricityUpperLimit$ under the \textsc{SEOBNRv5EHM} and \textsc{TEOBRESUMS-DAL\'I} models, respectively.
    }
    \label{fig:eccentricity}
\end{figure}

Binary black hole coalescences arising dynamically in dense stellar environments may bear unique signatures of orbital eccentricity in their gravitational-wave emission.
Gravitational-wave emission rapidly circularizes initially eccentric orbits~\citep{Peters:1964zz}, and binaries evolving in isolation are expected to be nearly perfectly quasi-circular by the time their gravitational-wave emission enters the sensitivity band of ground-based detectors.
Following many-body encounters in dense clusters, however, binaries can be placed on nearly-hyperbolic trajectories and merge promptly.
Several percent of these binaries may retain observable eccentricity in the frequency band of ground-based
detectors.
In old, metal-poor globular clusters, $5$--$10\%$ of binary black hole mergers in the local universe are predicted to have eccentricities measurable by the LIGO, Virgo, and KAGRA experiments~\citep{Wen:2002km,Gultekin:2005fd, OLeary:2005vqo, Antonini:2012ad,Antonini:2013tea,Samsing:2017oij, Samsing:2017rat, Samsing:2017xmd, Rodriguez:2018pss, Zevin:2018kzq}, although this may decrease by a factor of a couple when assuming higher initial cluster densities~\citep{Antonini:2020xnd}.
The non-secular evolution of isolated triple systems may yield $\gtrsim 10\%$ of systems with measurable eccentricities, and potentially have a higher merger rate of eccentric sources in the local universe~\citep{Dorozsmai:2025jlu}.
The total merger rate in active galactic nuclei is uncertain with predictions that span multiple orders of magnitude~\citep{Grobner:2020drr}, with predictions for measurably eccentric fraction ranging from $\gtrsim 10\%$~\citep{Tagawa:2020jnc} up to $\sim 70\%$~\citep{Samsing:2020tda}.
There exists growing evidence that at least some observed compact binary mergers may possess residual eccentricity~\citep{Gayathri:2020coq, Romero-Shaw:2020thy, Gamba:2021gap, Romero-Shaw:2022xko, Iglesias:2022xfc, Gupte:2024jfe, Romero-Shaw:2025vbc, Planas:2025jny, Morras:2025xfu}, possibly indicating binary formation in one or more of these environments.

If \gwTenEleven and \gwElevenTen evolved dynamically in dense clusters, they may be prime candidates to exhibit measurable eccentricity.
We reanalyze \gwTenEleven and \gwElevenTen using a pair of alternative waveform models, \textsc{SEOBNRv5EHM}~\citep{Gamboa:2024hli} and \textsc{TEOBRESUMS-DAL\'I}~\citep{Nagar:2024dzj}, that describe gravitational-wave emission from eccentric compact binaries through binary inspiral, merger, and ringdown.
These waveform models are valid under restricted spin geometries, requiring component spins to be purely parallel or antiparallel to a binary's orbital angular momentum.
The gravitational-wave signatures of orbital eccentricity and spin--orbit misalignment are known to be degenerate~\citep[e.g., ][]{CalderonBustillo:2020xms, Romero-Shaw:2020thy, Romero-Shaw:2022fbf, Divyajyoti:2023rht, Planas:2025jny}.
Degeneracies between spin misalignment and eccentricity are weakest for binaries like \gwTenEleven and \gwElevenTen~\citep{Romero-Shaw:2022fbf, Divyajyoti:2023rht} with low chirp masses, which complete many observable cycles.
Nevertheless, eccentricity measurements that neglect effects of spin--orbit precession (or, conversely, spin measurements that neglect eccentricity, as in Section~\ref{sec:pe}) may be biased.

Constraints on the orbital eccentricity of \gwTenEleven and \gwElevenTen are presented in Figure~\ref{fig:eccentricity}.
Orbital eccentricity is an evolving function of time; results are quoted at the instant when the binaries' orbit-averaged quadrupole emission is observed at $13.33\,\mathrm{Hz}$, corresponding to the time at which $\ell=3$ spherical harmonic modes enter the observable band at $20\,\mathrm{Hz}$.
Neither event possesses measurable eccentricity.
Under both waveform models, \gwTenEleven is bounded to have $e<\gwTenElevenSEOBEccentricityUpperLimit$ at $90\%$ credibility, while \gwElevenTen has $e<\gwElevenTenSEOBEccentricityUpperLimit$ and $e<\gwElevenTenTEOBEccentricityUpperLimit$ under the \textsc{SEOBNRv5EHM} and \textsc{TEOBRESUMS-DAL\'I} models, respectively.
This is not inconsistent with a dynamical origin; the vast majority of mergers in clusters are expected to have eccentricities $e\lesssim 0.1$ at frequencies accessible to Advanced LIGO, Advanced Virgo, and KAGRA~\citep{Gultekin:2005fd, OLeary:2005vqo, Samsing:2013kua, Samsing:2017xmd, Rodriguez:2017pec, Gondan:2017wzd, Zevin:2018kzq, DallAmico:2023neb}.
On the basis of eccentricity alone, though, we cannot rule out any individual formation scenarios for \gwTenEleven and \gwElevenTen.
Further details are presented in Appendix~\ref{app:eccentricity}.

\section{Tests of Fundamental Physics}
\label{sec:physics}

\begin{figure}[t!]
    \centering
    \includegraphics[width=0.48\textwidth]{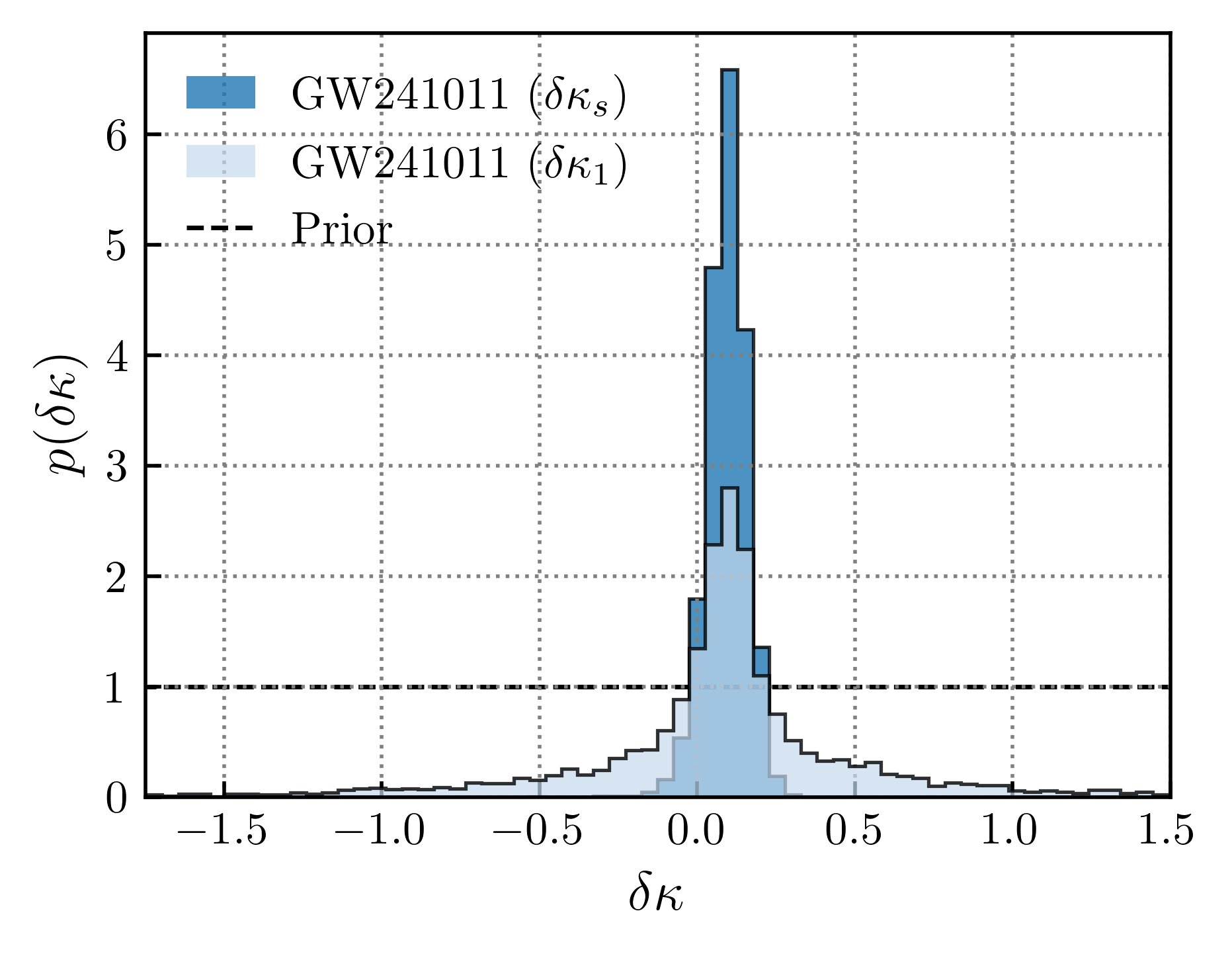}
    \caption{
    Deviations from the Kerr prediction for the spin-induced quadrupole moment of \gwTenEleven's primary black hole, $\delta\kappa_1$, as well as the deviation $\delta\kappa_s$ in the symmetric combination $\kappa_s = (\kappa_1 + \kappa_2)/2$~\citep{Krishnendu:2017shb, Krishnendu:2019tjp, LIGOScientific:2021sio}.
    We bound $\delta\kappa_1 = \gwTenElevenSiqmPhenomMedian^{+\gwTenElevenSiqmPhenomUpperError}_{-\gwTenElevenSiqmPhenomLowerError}$ and $\delta\kappa_s = \gwTenElevenSymmetricSiqmPhenomMedian^{+\gwTenElevenSymmetricSiqmPhenomUpperError}_{-\gwTenElevenSymmetricSiqmPhenomLowerError}$, consistent with expectations from GR.
       The next-most informative event, GW190412~\citep{LIGOScientific:2020stg}, yielded a constraint $\delta\kappa_s = \gwOhFourTwelveSymmetricSiqmPhenomMedian^{+\gwOhFourTwelveSymmetricSiqmPhenomUpperError}_{-\gwOhFourTwelveSymmetricSiqmPhenomLowerError}$~\citep{Divyajyoti:2023izl}.
    }
    \label{fig:siqm}
\end{figure}

Gravitational waveforms from compact binary coalescences encode detailed information about the nature and internal structure of the merging objects, enabling rigorous tests of general relativity (GR) and fundamental physics.
The large primary spin and significant mass asymmetry of \gwTenEleven, in particular, makes this event a uniquely powerful probe of the Kerr nature of black holes and the multipolar structure of gravitational-wave emission.

\subsection{Black hole spin-induced quadrupole moment}
\label{sec:physics:siqm}

Within GR, rotating and charge-neutral black holes are uniquely described by the Kerr solution~\citep{Kerr:1963ud, Carter:1971zc}.
In Kerr spacetime, the black hole spin-induced quadrupole moment, the leading contribution of spin to a black hole spacetime's multipolar expansion, is given by $Q = -\kappa S^2 / (m c^2)$.
Here, $m$ is the black hole's mass, $S$ is its spin angular momentum, $c$ is the speed of light, and $\kappa = 1$ exactly~\citep{Hansen:1974zz}.
Non-black hole spacetimes, including neutron stars~\citep{Laarakkers:1997hb,Pappas:2012qg,Pappas:2012ns,Harry:2018hke}, boson stars~\citep{Ryan:1996nk,Herdeiro:2014goa,Baumann:2018vus,Chia:2020psj}, and other exotic compact objects, may in contrast exhibit significantly different values of $\kappa$ owing to differences in internal structure and composition.
Gravitational-wave sources containing rapidly-spinning black holes enable direct measurements of the spin-induced quadrupole moment and tests of the Kerr hypothesis~\citep{Arun:2008kb, Mishra:2016whh}; any measured deviation from $\kappa=1$ would strongly suggest the presence of non-black hole constituents or indicate new physics beyond the predictions of GR.

We define $\kappa_1 = 1 + \delta \kappa_1$ and $\kappa_2 = 1 + \delta \kappa_2$ as the spin-induced quadrupole coefficients of each compact object in \gwTenEleven's source binary.
We repeat parameter estimation with a modified \textsc{IMRPhenomXPHM} waveform model~\citep{Pratten:2020ceb}, allowing for non-zero $\delta \kappa_1$ and $\delta \kappa_2$~\citep{Divyajyoti:2023izl}.
The resulting posterior is shown in Fig.~\ref{fig:siqm}.
The spin-induced quadrupole coefficient of \gwTenEleven's primary deviates from the Kerr prediction by $\delta\kappa_1 = \gwTenElevenSiqmPhenomMedian^{+\gwTenElevenSiqmPhenomUpperError}_{-\gwTenElevenSiqmPhenomLowerError}$, consistent with GR (constraints on $\delta \kappa_2$ are uninformative).
This is the most stringent constraint to date on the spin-induced quadrupole of a compact object.
The constraints offered by \gwTenEleven on $\delta \kappa_1$ are unusual, enabled by the event's high SNR, large primary spin, and unequal mass ratio.
Typically, gravitational-wave signals primarily constrain only the symmetric combination $\kappa_s = (\kappa_1 + \kappa_2)/2$~\citep{Krishnendu:2017shb, Krishnendu:2019tjp, LIGOScientific:2021sio, Divyajyoti:2023izl}.
\gwTenEleven constrains this symmetric combination to be within $\delta\kappa_s = \gwTenElevenSymmetricSiqmPhenomMedian^{+\gwTenElevenSymmetricSiqmPhenomUpperError}_{-\gwTenElevenSymmetricSiqmPhenomLowerError}$ of the Kerr hypothesis, under the assumption that $\kappa_1 = \kappa_2$.
The previous best constraint on $\kappa_s$ was obtained from the binary black hole GW190412~\citep{Divyajyoti:2023izl}, which gave $\delta\kappa_s = \gwOhFourTwelveSymmetricSiqmPhenomMedian^{+\gwOhFourTwelveSymmetricSiqmPhenomUpperError}_{-\gwOhFourTwelveSymmetricSiqmPhenomLowerError}$.
\gwTenEleven improves on this constraint by approximately three orders of magnitude.
A different implementation for the estimation of the spin-induced quadrupole moment, utilizing the \textsc{SEOBNRv5HM\_ROM}~\citep{Pompili:2023tna} waveform model, yields consistent results and is discussed in Appendix~\ref{app:tgr:siqm}. 

Measurement of \gwTenEleven{}’s spin-induced quadrupole moment may rule out a wide range of exotic compact objects or black hole mimickers. 
Massive boson star models~\citep{Ryan:1996nk, Pacilio:2020jza} predict spin-induced quadrupole moment parameters of order $\sim 10$--$150$ for self-interacting spinning boson stars with quadratic coupling.
The measurement of $\delta\kappa_s \leq \gwTenElevenSymmetricSiqmPhenomUpperLimit$ at $90\%$ credibility from \gwTenEleven likely rules out all the massive boson star models described in e.g.~\cite{Pacilio:2020jza}.
Other models, such as minimal boson stars~\citep{Kaup:1968zz, Ruffini:1969qy, Vaglio:2022flq} and solitonic boson stars~\citep{Friedberg:1986tq}, remain poorly understood in terms of their spin-induced multipole moments~\citep{Cardoso:2017cfl, Cardoso:2019rvt}.
Another class of exotic compact objects are the gravastars~\citep{Mottola:2023jxl}, where the spin-induced multipole moments can take negative values due to the prolate deformation induced by their spinning motion. 
Although spin-induced quadrupole moment values have been predicted for thin-shell gravastar models~\citep{Uchikata:2015yma}, the \gwTenEleven data is insufficient to make definitive conclusions.

\subsection{Radiation beyond the quadrupole approximation}

\begin{figure}[t!]
    \centering
    \includegraphics[width=0.48\textwidth]{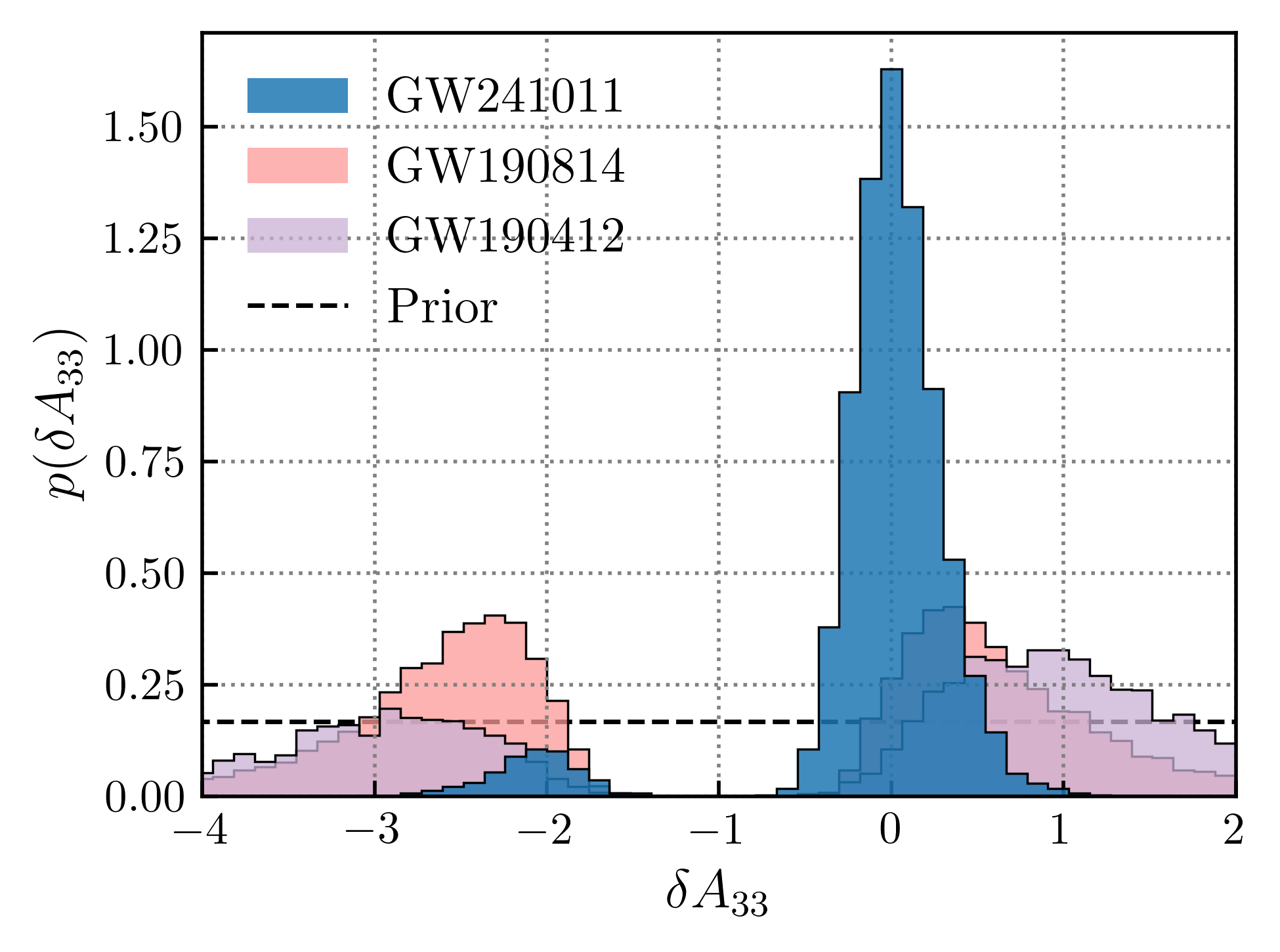}
    \caption{
    Posterior constraints on the amplitude of \gwTenEleven's gravitational radiation in $(\ell, m) = (3, \pm 3)$ spherical harmonic modes, relative to the prediction from GR.
    \gwTenEleven is consistent with expectation, with deviations from the GR limited to the interval $\gwTenElevenSMALowerBound \leq \delta A_{33}  \leq \gwTenElevenSMAUpperBound$ at $90\%$ credibility.
    Degeneracy with orbital inclination yields a strongly bimodal structure in the posterior for $\delta A_{33}$.
    The dominant mode has $\delta A_{33} = \gwTenElevenSMAUpperModeMedian^{+\gwTenElevenSMAUpperModeUpperError}_{-\gwTenElevenSMAUpperModeLowerError}$, consistent with GR, while the subdominant mode has $\delta A_{33} = \gwTenElevenSMALowerModeMedian^{+\gwTenElevenSMALowerModeUpperError}_{-\gwTenElevenSMALowerModeLowerError}$.
    For comparison, also shown are the posteriors obtained from the next-most-informative gravitational-wave events, GW190814 and GW190412~\citep{LIGOScientific:2020stg, LIGOScientific:2020zkf, Puecher:2022sfm, gwtc4-tgr}.
    }
    \label{fig:tgr_sma}
\end{figure}

Gravitational-wave radiation may be generically decomposed into an expansion over spin-weight $-2$ spherical harmonics, ${}_{-2}Y_{lm}$.
The gravitational waves from merging compact binaries are dominated by the $(\ell, m) = (2, \pm 2)$ spherical harmonic, sourced by a binary's mass quadrupole moment.
As discussed in Section~\ref{sec:pe}, however, \gwTenEleven exhibits significant radiation in the $(\ell, m) = (3, \pm 3)$ mode sourced by the current quadrupole and mass octupole moments.
General relativity fixes the relative amplitudes of gravitational radiation received in different spherical harmonic modes.
The strong detection of multiple modes in a gravitational-wave signal, as in \gwTenEleven, offers an opportunity to test these predictions~\citep{Capano:2020dix, Puecher:2022sfm, Mahapatra:2023hqq, gwtc4-tgr}.

We repeat inference on the properties of \gwTenEleven, but now introduce a parameter $\delta A_{33}$ that allows for deviations in the signal's $(\ell, m) = (3, \pm 3)$ mode amplitudes, relative to the $(2, \pm 2)$ mode content~\citep{Puecher:2022sfm, gwtc4-tgr}.
Higher-order modes, such as $(4, \pm 4)$ spherical harmonics, are not detected in \gwTenEleven, and so are not tested here.
The resulting posterior distribution on $\delta A_{33}$ is shown in Figure~\ref{fig:tgr_sma}.
Posteriors on $\delta A_{33}$ are characteristically bimodal, due to degeneracies between a binary's inclination, orbital phase, and expected $(3,\pm 3)$ mode amplitude~\citep{Mills:2020thr, Puecher:2022sfm}.
The dominant mode is consistent with GR, with $\delta A_{33} = \gwTenElevenSMAUpperModeMedian^{+\gwTenElevenSMAUpperModeUpperError}_{-\gwTenElevenSMAUpperModeLowerError}$, while the subdominant mode has $\delta A_{33} = \gwTenElevenSMALowerModeMedian^{+\gwTenElevenSMALowerModeUpperError}_{-\gwTenElevenSMALowerModeLowerError}$.
Taking both posterior modes together, we find $\gwTenElevenSMALowerBound \leq \delta A_{33}  \leq \gwTenElevenSMAUpperBound$ at $90\%$ credibility.
This is the best measurement to date of $\delta A_{33}$.
Among binaries in GWTC-4, the next-best constraints are provided by GW190814 and GW190412~\citep{LIGOScientific:2020stg, LIGOScientific:2020zkf, Puecher:2022sfm}, which give $\gwOhEightFourteenSMALowerBound \leq \delta A_{33}  \leq \gwOhEightFourteenSMAUpperBound$ and $\gwOhFourTwelveSMALowerBound \leq \delta A_{33}  \leq \gwOhFourTwelveSMAUpperBound$, respectively~\citep{gwtc4-tgr}.
\gwTenEleven therefore confirms that gravitational waves radiated in $(3,\pm 3)$ spherical harmonic modes have amplitudes consistent with expectations from GR.
Further details are elaborated in Appendix~\ref{app:tgr:sma}.

\subsection{Ultralight bosons}
\label{sec:physics:boson}

\begin{figure}[t!]
    \centering
    \includegraphics[width=0.48\textwidth]{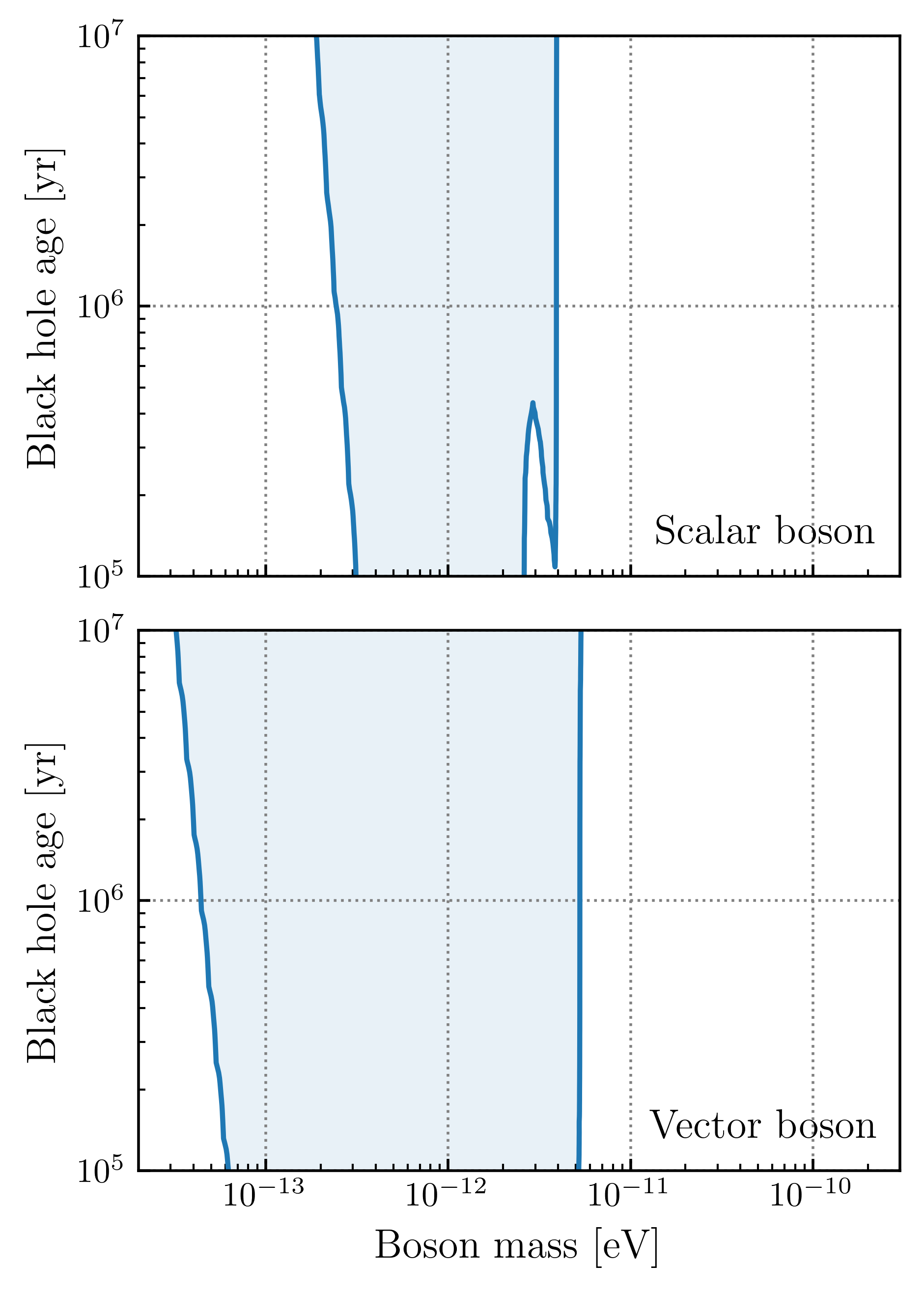}
    \caption{
    Masses of novel ultralight scalar (top) and vector (bottom) bosons that are excluded at $90\%$ credibility by the non-zero spin measurements of \gwTenEleven.
    If bosons with masses lying in the shaded regions existed, then the primary black hole of \gwTenEleven would have undergone the superradiance instability, spontaneously generating a cloud of bound boson particles and driving the black holes' spins below their observed values.
    The superradiance instability can deplete a black hole's spin over an astrophysically-brief timescale, but this timescale becomes significantly longer outside a narrow range of boson masses.
    The exclusion curves therefore depend on the presumed age of \gwTenEleven's primary black hole.
    }
    \label{fig:bosons}
\end{figure}

Spinning black holes may prodigiously and spontaneously source particle production through the superradiant instability mechanism~\citep{Press:1972zz, Damour:1976kh, Brito:2015oca}.
Ingoing waves may acquire energy as they scatter off a rotating black hole.
In the vicinity of a rotating black hole, an oscillating, gravitationally bound fluctuation in a bosonic field may grow exponentially, spontaneously turning an initially small perturbation into a macroscopic boson cloud surrounding the black hole.
Cloud growth is powered at the expense of the black hole's rotational energy and may occur on timescales as short as days or minutes.
This superradiance instability is specifically relevant for bosons with Compton wavelengths comparable to the black hole's size.
If $m_b$ is the boson mass, $G$ is the gravitational constant, and $M$ the black hole's mass, then superradiance requires $m_b \sim \hbar c^2/ (GM)$.
Thus, the observation of a rapidly-rotating black hole of mass $M$ immediately excludes the existence of novel bosons with masses near $m_b$;
if such a particle existed, it should have long since depleted the black hole's spin.
Constraints can, in principle, be performed using both electromagnetic and gravitational-wave observations, and for both supermassive and stellar-mass black holes~\citep[e.g.,][]{Arvanitaki:2009fg, Arvanitaki:2010sy, Pani:2012vp, Baryakhtar:2017ngi, Fernandez:2019qbj, Ng:2019jsx, Ng:2020ruv, Stott:2020gjj}.

The confidently rapid primary spin of \gwTenEleven excludes the existence of ultralight bosons with masses between approximately $10^{-13}$ and $3\times 10^{-12}\,\mathrm{eV}$.
Figure~\ref{fig:bosons} shows the boson masses excluded by the primary spin measurement of \gwTenEleven, as a function of the presumed age of each binary's primary black hole and computed using the \textsc{SuperRad} package~\citep{Siemonsen:2022yyf, May:2024npn}; see Appendix~\ref{app:tgr:bosons} for details.
The top and bottom panels correspond to scalar and vector bosons, respectively.
The filled contours indicate regions in which the primaries of both binaries should have undergone the superradiance instability, yielding final present-day spins that are inconsistent with observation at $90\%$ credibility.
Since all relevant azimuthal modes are included for the black hole ages considered, the narrow exclusion region near $4\times10^{-12}\,\mathrm{eV}$ in the scalar case arises from cloud growth with a higher azimuthal number $m=3$.
Conservatively assuming an age of $10^5$ years for the primary black hole of \gwTenEleven, this signal excludes the existence of scalar bosons with masses in the interval $[\TenElevenScalarLowerConstraint, \TenElevenScalarUpperConstraint]\times10^{\SuperradPower}\,\mathrm{eV}$.
The signal excludes vector bosons, meanwhile, with masses in the $[\TenElevenVectorLowerConstraint, \TenElevenVectorUpperConstraint]\times10^{\SuperradPower}\,\mathrm{eV}$ interval.
Due to its more uncertain spin measurements, \gwElevenTen does not appreciably constrain the existence of ultralight bosons.

\gwTenEleven rules out the existence of bosons at higher masses than those excluded by previous gravitational-wave observations.
An analysis considering the population of black holes comprising the LIGO--Virgo--KAGRA GWTC-2 catalog strongly disfavored scalar boson masses between $[2.2, 2.7]\times 10^{-13}\,\mathrm{eV}$ when assuming $10^5$ year old black holes~\citep{Ng:2020ruv}.
Under the same age assumption, recent analysis using GW231123 and GW190517 excluded scalar and vector boson masses in the intervals $[0.6, 11]\times10^{-13}$ and $[0.1, 18]\times10^{-13}\,\mathrm{eV}$, respectively, at $90\%$ credibility~\citep{Aswathi:2025nxa}.
A complementary analysis of GW231123 with a relativistic model for self-interacting scalars excludes axion masses in the interval $[0.6, 5]\times10^{-13}\,\mathrm{eV}$ with decay constants $\gtrsim 10^{14}\,\mathrm{GeV}$~\citep{Caputo:2025oap}.

\section{Conclusions}
\label{sec:conc}

In this paper, we have presented gravitational-wave signals from two binary black hole coalescences -- \gwTenEleven and \gwElevenTen{} -- discovered during the second part of the fourth observing run of the LIGO, Virgo, and KAGRA observatories.
The measured properties of these events naturally suggest consideration as a pair.
The spins of the more massive black holes in \gwTenEleven and \gwElevenTen are respectively situated at either extreme of the binary black hole population.
The primary black hole of \gwTenEleven possesses one of the largest and most precisely measured black hole spins observed via gravitational waves, and is spinning in a direction primarily (but not exactly) aligned with its orbital angular momentum.
Conversely, the primary black hole of \gwElevenTen is rapidly spinning in a direction confidently antiparallel to its orbit, the first confidently anti-aligned black hole spin measured to date.
At the same time, these two binary black holes exhibit nearly identical masses, with each event favoring a primary mass in the range $15$--$20\,M_\odot$ and an unequal mass ratio between their component black holes.

Taken together, the mass ratios, large primary spins, and significant spin-orbit misalignment angles of \gwTenEleven and \gwElevenTen are strongly suggestive of hierarchical binary black hole mergers in dense stellar environments, such as globular, nuclear, or young stellar clusters.
Under this interpretation, the primary black holes of both binaries are themselves the remnants of past black hole mergers.
However, although the properties of \gwTenEleven and \gwElevenTen are in strong tension with predictions from isolated binary evolution, from gravitational-wave data alone we cannot rule out formation by massive stellar binaries (or systems of higher multiplicity).
\gwTenEleven and \gwElevenTen nevertheless suggest that at least some merging binary black holes merge dynamically in dense environments, and that these environments are sufficiently massive to retain remnant black holes and foster repeated mergers.

The large and precisely measured primary spin of \gwTenEleven furthermore enables myriad tests of fundamental physics.
In particular, we find that this event provides the best measurements to date of a black hole's spin induced quadrupole moment, confirming the Kerr prediction to within a factor of two (or to within $10\%$, depending on the parametrization used, an improvement in precision by two orders of magnitude; see Section~\ref{sec:physics:siqm}).
The nature of this test is independent from and complementary to the ringdown spectroscopy performed on the extremely loud event GW250114; analysis of GW250114's ringdown identified overtones with frequencies constrained to within $30\%$ of their Kerr predictions~\citep{gw250114, gw250114_companion}.
\gwTenEleven's strong radiation in multiple spherical harmonic modes furthermore enables the best constraint to date on the relative amplitudes of $(\ell, m) = (2, \pm 2)$ and $(3, \pm 3)$ modes, confirming the expected structure of gravitational-wave emission beyond the quadrupole approximation.
Finally, the large spins of both \gwTenEleven and \gwElevenTen rule out the existence of novel boson particles with masses in the range $10^{-13}$ to $10^{-12}\,\mathrm{eV}$.

Growing gravitational-wave catalogs, enabled by the concurrent operation of increasingly sensitive gravitational-wave observatories, continue to yield individually-interesting sources that expand our knowledge of the compact binary landscape.
Gravitational waves detected during the fourth observing run of the LIGO, Virgo, and KAGRA observatories have thus far enabled novel tests of relativity and gravitational-waveform models~\citep{gw230814, gw250114, gw250114_companion} and provided sources in new and unexpected regions of parameter space~\citep{LIGOScientific:2024elc, LIGOScientific:2025rsn}.
We expect discoveries to continue through the remainder of the fourth LIGO--Virgo--KAGRA observing run and beyond.

\section*{acknowledgments}
This material is based upon work supported by NSF's LIGO Laboratory, which is a
major facility fully funded by the National Science Foundation.
The authors also gratefully acknowledge the support of
the Science and Technology Facilities Council (STFC) of the
United Kingdom, the Max-Planck-Society (MPS), and the State of
Niedersachsen/Germany for support of the construction of Advanced LIGO 
and construction and operation of the GEO\,600 detector. 
Additional support for Advanced LIGO was provided by the Australian Research Council.
The authors gratefully acknowledge the Italian Istituto Nazionale di Fisica Nucleare (INFN),  
the French Centre National de la Recherche Scientifique (CNRS) and
the Netherlands Organization for Scientific Research (NWO)
for the construction and operation of the Virgo detector
and the creation and support  of the EGO consortium. 
The authors also gratefully acknowledge research support from these agencies as well as by 
the Council of Scientific and Industrial Research of India, 
the Department of Science and Technology, India,
the Science \& Engineering Research Board (SERB), India,
the Ministry of Human Resource Development, India,
the Spanish Agencia Estatal de Investigaci\'on (AEI),
the Spanish Ministerio de Ciencia, Innovaci\'on y Universidades,
the European Union NextGenerationEU/PRTR (PRTR-C17.I1),
the ICSC - CentroNazionale di Ricerca in High Performance Computing, Big Data
and Quantum Computing, funded by the European Union NextGenerationEU,
the Comunitat Auton\`oma de les Illes Balears through the Conselleria d'Educaci\'o i Universitats,
the Conselleria d'Innovaci\'o, Universitats, Ci\`encia i Societat Digital de la Generalitat Valenciana and
the CERCA Programme Generalitat de Catalunya, Spain,
the Polish National Agency for Academic Exchange,
the National Science Centre of Poland and the European Union - European Regional
Development Fund;
the Foundation for Polish Science (FNP),
the Polish Ministry of Science and Higher Education,
the Swiss National Science Foundation (SNSF),
the Russian Science Foundation,
the European Commission,
the European Social Funds (ESF),
the European Regional Development Funds (ERDF),
the Royal Society, 
the Scottish Funding Council, 
the Scottish Universities Physics Alliance, 
the Hungarian Scientific Research Fund (OTKA),
the French Lyon Institute of Origins (LIO),
the Belgian Fonds de la Recherche Scientifique (FRS-FNRS), 
Actions de Recherche Concert\'ees (ARC) and
Fonds Wetenschappelijk Onderzoek - Vlaanderen (FWO), Belgium,
the Paris \^{I}le-de-France Region, 
the National Research, Development and Innovation Office of Hungary (NKFIH), 
the National Research Foundation of Korea,
the Natural Sciences and Engineering Research Council of Canada (NSERC),
the Canadian Foundation for Innovation (CFI),
the Brazilian Ministry of Science, Technology, and Innovations,
the International Center for Theoretical Physics South American Institute for Fundamental Research (ICTP-SAIFR), 
the Research Grants Council of Hong Kong,
the National Natural Science Foundation of China (NSFC),
the Israel Science Foundation (ISF),
the US-Israel Binational Science Fund (BSF),
the Leverhulme Trust, 
the Research Corporation,
the National Science and Technology Council (NSTC), Taiwan,
the United States Department of Energy,
and
the Kavli Foundation.
The authors gratefully acknowledge the support of the NSF, STFC, INFN and CNRS for provision of computational resources.
This work was supported by MEXT,
the JSPS Leading-edge Research Infrastructure Program,
JSPS Grant-in-Aid for Specially Promoted Research 26000005,
JSPS Grant-in-Aid for Scientific Research on Innovative Areas 2402: 24103006,
24103005, and 2905: JP17H06358, JP17H06361 and JP17H06364,
JSPS Core-to-Core Program A.\ Advanced Research Networks,
JSPS Grants-in-Aid for Scientific Research (S) 17H06133 and 20H05639,
JSPS Grant-in-Aid for Transformative Research Areas (A) 20A203: JP20H05854,
the joint research program of the Institute for Cosmic Ray Research,
University of Tokyo,
the National Research Foundation (NRF),
the Computing Infrastructure Project of the Global Science experimental Data hub
Center (GSDC) at KISTI,
the Korea Astronomy and Space Science Institute (KASI),
the Ministry of Science and ICT (MSIT) in Korea,
Academia Sinica (AS),
the AS Grid Center (ASGC) and the National Science and Technology Council (NSTC)
in Taiwan under grants including the Science Vanguard Research Program,
the Advanced Technology Center (ATC) of NAOJ,
and the Mechanical Engineering Center of KEK.

We are grateful for the valuable feedback provided by anonymous reviewers.
Additional acknowledgements for support of individual authors may be found in the following document: \\
\texttt{https://dcc.ligo.org/LIGO-M2300033/public}.
For the purpose of open access, the authors have applied a Creative Commons Attribution (CC BY)
license to any Author Accepted Manuscript version arising.
We request that citations to this article use `A. G. Abac {\it et al.} (LIGO-Virgo-KAGRA Collaboration), ...' or similar phrasing, depending on journal convention.

\software{
Calibration of the LIGO strain data was performed with a \gstlal-based calibration software pipeline~\citep{Viets:2017yvy}.
Data-quality products and event-validation results were computed using the \DMT{}~\citep{DMTdocumentation}, \DQR{}~\citep{DQRdocumentation}, \DQSEGDB{}~\citep{Fisher:2020pnr}, \GWDETCHAR{}~\citep{gwdetchar-software}, \HVETO{}~\citep{Smith:2011an}, \IDQ{}~\citep{Essick:2020qpo}, \OMICRONSCAN{}~\citep{Robinet:2020lbf}, and \PYTHONVIRGOTOOLS{}~\citep{pythonvirgotools} software packages and contributing software tools.
Analyses in this catalog relied on software from the LVK Algorithm Library Suite~\citep{lalsuite, swiglal}.
The detection of the signals and subsequent significance evaluations were performed with the \gstlal-based inspiral software pipeline~\citep{Messick:2016aqy, Sachdev:2019vvd, Hanna:2019ezx, Cannon:2020qnf, Sakon:2022ibh, Ewing:2023qqe, Tsukada:2023edh, Ray:2023nhx, Joshi:2025nty, Joshi:2025zdu}, with the \mbta pipeline~\citep{Adams:2015ulm, Aubin:2020goo, Allene:2025saz}, and with the \pycbc~\citep{Usman:2015kfa, Nitz:2017lco, Nitz:2017svb, DalCanton:2020vpm, Davies:2020tsx} packages.
Estimates of the noise spectra and glitch models were obtained using \BAYESWAVE{}~\citep{Cornish:2014kda, Littenberg:2014oda, Littenberg:2015kpb, Cornish:2020dwh, Gupta:2023jrn}.
Low-latency source localization was performed using \BAYESTAR{}~\citep{Singer:2015ema}.
Source-parameter estimation was performed with the \BILBY{} library~\citep{Ashton:2018jfp, Romero-Shaw:2020owr}, using the \DYNESTY{} nested sampling package~\citep{Speagle:2019ivv}, and the \RIFT~\citep{Pankow:2015cra, Lange:2017wki, Wysocki:2019grj} package.
\textsc{SEOBNRv5PHM} waveforms used in parameter estimation were generated using pySEOBNR~\citep{Mihaylov:2023bkc}.
\PESUMMARY{} was used to post-process and collate parameter estimation results~\citep{Hoy:2020vys}.
The various stages of the parameter estimation analysis were managed with the \ASIMOV{} library~\citep{Williams:2022pgn}.
Population inference was performed with the \textsc{GWPopulation} package~\citep{Talbot:2024yqw}.
The \textsc{SuperRad} package~\citep{Siemonsen:2022yyf,May:2024npn} was used to characterize superradiance phenomena.
Plots were prepared with \PLT{}~\citep{Hunter:2007ouj}.
\NUMPY{}~\citep{Harris:2020xlr} and \SCIPY{}~\citep{Virtanen:2019joe} were used for analyses in the manuscript.
}

\textit{Data availability:}
Strain data from the LIGO and Virgo observatories associated with \gwTenEleven and \gwElevenTen are available from the Gravitational Wave Open Science Center.
Datasets generated as part of this study, including posterior samples on the source properties of both events, are available on Zenodo, together with notebooks reproducing figures in this paper~\citep{data-release}.

\appendix
\section{Source Parameter Estimation: Further Results and Details}
\label{app:pe}

\begin{table*}[]
    \footnotesize
    \setlength{\tabcolsep}{4pt}
    \renewcommand{\arraystretch}{1.2}
    \centering
    \caption{
    Inferred source properties of \gwTenEleven under different waveform models.
    }
    \begin{tabular}{r  c c c c c c c c c c c}
    \hline \hline
    Waveform & $m_1\,[M_\odot]$ & $m_2\,[M_\odot]$ & $q$ & $\mathcal{M}\,[M_\odot]$ & $D_\mathrm{L}\,[\mathrm{Mpc}]$ & $\chi_1$ & $\theta_1\,[\mathrm{deg}]$ & $\chi_{1,z}$ & $\chi_{1,\perp}$ & $\chi_\mathrm{eff}$ & $\chi_\mathrm{p}$ \\
    \hline
    \textsc{SEOBNRv5PHM}
    	& $\gwTenElevenSEOBMassOne^{+\gwTenElevenSEOBMassOneUpperError}_{-\gwTenElevenSEOBMassOneLowerError}$
	& $\gwTenElevenSEOBMassTwo^{+\gwTenElevenSEOBMassTwoUpperError}_{-\gwTenElevenSEOBMassTwoLowerError}$
	& $\gwTenElevenSEOBMassRatio^{+\gwTenElevenSEOBMassRatioUpperError}_{-\gwTenElevenSEOBMassRatioLowerError}$ 
	& $\gwTenElevenSEOBChirpMass^{+\gwTenElevenSEOBChirpMassUpperError}_{-\gwTenElevenSEOBChirpMassLowerError}$ 
	& $\gwTenElevenSEOBDistance^{+\gwTenElevenSEOBDistanceUpperError}_{-\gwTenElevenSEOBDistanceLowerError}$ 
	& $\gwTenElevenSEOBSpinOneMagMedian^{+\gwTenElevenSEOBSpinOneMagUpperError}_{-\gwTenElevenSEOBSpinOneMagLowerError}$
	& $\gwTenElevenSEOBSpinOneTiltMedian^{+\gwTenElevenSEOBSpinOneTiltUpperErrorHPD}_{-\gwTenElevenSEOBSpinOneTiltLowerErrorHPD}$
	& $\gwTenElevenSEOBSpinOneZMedian^{+\gwTenElevenSEOBSpinOneZUpperError}_{-\gwTenElevenSEOBSpinOneZLowerError}$
	& $\gwTenElevenSEOBSpinOnePlaneMedian^{+\gwTenElevenSEOBSpinOnePlaneUpperError}_{-\gwTenElevenSEOBSpinOnePlaneLowerError}$
	& $\gwTenElevenSEOBChiEffectiveMedian^{+\gwTenElevenSEOBChiEffectiveUpperError}_{-\gwTenElevenSEOBChiEffectiveLowerError}$
	& $\gwTenElevenSEOBChiPMedian^{+\gwTenElevenSEOBChiPUpperError}_{-\gwTenElevenSEOBChiPLowerError}$  \\
    \textsc{IMRPhenomXPHM-ST}
    	& $\gwTenElevenXPHMMassOne^{+\gwTenElevenXPHMMassOneUpperError}_{-\gwTenElevenXPHMMassOneLowerError}$
	& $\gwTenElevenXPHMMassTwo^{+\gwTenElevenXPHMMassTwoUpperError}_{-\gwTenElevenXPHMMassTwoLowerError}$
	& $\gwTenElevenXPHMMassRatio^{+\gwTenElevenXPHMMassRatioUpperError}_{-\gwTenElevenXPHMMassRatioLowerError}$ 
	& $\gwTenElevenXPHMChirpMass^{+\gwTenElevenXPHMChirpMassUpperError}_{-\gwTenElevenXPHMChirpMassLowerError}$ 
	& $\gwTenElevenXPHMDistance^{+\gwTenElevenXPHMDistanceUpperError}_{-\gwTenElevenXPHMDistanceLowerError}$ 
	& $\gwTenElevenXPHMSpinOneMagMedian^{+\gwTenElevenXPHMSpinOneMagUpperError}_{-\gwTenElevenXPHMSpinOneMagLowerError}$
	& $\gwTenElevenXPHMSpinOneTiltMedian^{+\gwTenElevenXPHMSpinOneTiltUpperErrorHPD}_{-\gwTenElevenXPHMSpinOneTiltLowerErrorHPD}$
	& $\gwTenElevenXPHMSpinOneZMedian^{+\gwTenElevenXPHMSpinOneZUpperError}_{-\gwTenElevenXPHMSpinOneZLowerError}$
	& $\gwTenElevenXPHMSpinOnePlaneMedian^{+\gwTenElevenXPHMSpinOnePlaneUpperError}_{-\gwTenElevenXPHMSpinOnePlaneLowerError}$
	& $\gwTenElevenXPHMChiEffectiveMedian^{+\gwTenElevenXPHMChiEffectiveUpperError}_{-\gwTenElevenXPHMChiEffectiveLowerError}$
	& $\gwTenElevenXPHMChiPMedian^{+\gwTenElevenXPHMChiPUpperError}_{-\gwTenElevenXPHMChiPLowerError}$  \\
\textsc{IMRPhenomXO4a}
    	& $\gwTenElevenXOFourAMassOne^{+\gwTenElevenXOFourAMassOneUpperError}_{-\gwTenElevenXOFourAMassOneLowerError}$
	& $\gwTenElevenXOFourAMassTwo^{+\gwTenElevenXOFourAMassTwoUpperError}_{-\gwTenElevenXOFourAMassTwoLowerError}$
	& $\gwTenElevenXOFourAMassRatio^{+\gwTenElevenXOFourAMassRatioUpperError}_{-\gwTenElevenXOFourAMassRatioLowerError}$ 
	& $\gwTenElevenXOFourAChirpMass^{+\gwTenElevenXOFourAChirpMassUpperError}_{-\gwTenElevenXOFourAChirpMassLowerError}$ 
	& $\gwTenElevenXOFourADistance^{+\gwTenElevenXOFourADistanceUpperError}_{-\gwTenElevenXOFourADistanceLowerError}$ 
	& $\gwTenElevenXOFourASpinOneMagMedian^{+\gwTenElevenXOFourASpinOneMagUpperError}_{-\gwTenElevenXOFourASpinOneMagLowerError}$
	& $\gwTenElevenXOFourASpinOneTiltMedian^{+\gwTenElevenXOFourASpinOneTiltUpperErrorHPD}_{-\gwTenElevenXOFourASpinOneTiltLowerErrorHPD}$
	& $\gwTenElevenXOFourASpinOneZMedian^{+\gwTenElevenXOFourASpinOneZUpperError}_{-\gwTenElevenXOFourASpinOneZLowerError}$
	& $\gwTenElevenXOFourASpinOnePlaneMedian^{+\gwTenElevenXOFourASpinOnePlaneUpperError}_{-\gwTenElevenXOFourASpinOnePlaneLowerError}$
	& $\gwTenElevenXOFourAChiEffectiveMedian^{+\gwTenElevenXOFourAChiEffectiveUpperError}_{-\gwTenElevenXOFourAChiEffectiveLowerError}$
	& $\gwTenElevenXOFourAChiPMedian^{+\gwTenElevenXOFourAChiPUpperError}_{-\gwTenElevenXOFourAChiPLowerError}$  \\ [1pt]
    \hline
    \hline
    \end{tabular}
    \label{tab:pe-params-241011}
\end{table*}

\begin{table*}[]
    \footnotesize
    \setlength{\tabcolsep}{4pt}
    \renewcommand{\arraystretch}{1.2}
    \centering
    \caption{
    Inferred source properties of \gwElevenTen under different waveform models.
    }
    \begin{tabular}{r  c c c c c c c c c c c}
    \hline \hline
    Waveform & $m_1\,[M_\odot]$ & $m_2\,[M_\odot]$ & $q$ & $\mathcal{M}\,[M_\odot]$ & $D_\mathrm{L}\,[\mathrm{Mpc}]$ & $\chi_1$ & $\theta_1\,[\mathrm{deg}]$ & $\chi_{1,z}$ & $\chi_{1,\perp}$ & $\chi_\mathrm{eff}$ & $\chi_\mathrm{p}$ \\
    \hline
    \textsc{SEOBNRv5PHM}
    	& $\gwElevenTenSEOBMassOne^{+\gwElevenTenSEOBMassOneUpperError}_{-\gwElevenTenSEOBMassOneLowerError}$
	& $\gwElevenTenSEOBMassTwo^{+\gwElevenTenSEOBMassTwoUpperError}_{-\gwElevenTenSEOBMassTwoLowerError}$
	& $\gwElevenTenSEOBMassRatio^{+\gwElevenTenSEOBMassRatioUpperError}_{-\gwElevenTenSEOBMassRatioLowerError}$ 
	& $\gwElevenTenSEOBChirpMass^{+\gwElevenTenSEOBChirpMassUpperError}_{-\gwElevenTenSEOBChirpMassLowerError}$ 
	& $\gwElevenTenSEOBDistance^{+\gwElevenTenSEOBDistanceUpperError}_{-\gwElevenTenSEOBDistanceLowerError}$ 
	& $\gwElevenTenSEOBSpinOneMagMedian^{+\gwElevenTenSEOBSpinOneMagUpperError}_{-\gwElevenTenSEOBSpinOneMagLowerError}$
	& $\gwElevenTenSEOBSpinOneTiltMedian^{+\gwElevenTenSEOBSpinOneTiltUpperErrorHPD}_{-\gwElevenTenSEOBSpinOneTiltLowerErrorHPD}$
	& $\gwElevenTenSEOBSpinOneZMedian^{+\gwElevenTenSEOBSpinOneZUpperError}_{-\gwElevenTenSEOBSpinOneZLowerError}$
	& $\gwElevenTenSEOBSpinOnePlaneMedian^{+\gwElevenTenSEOBSpinOnePlaneUpperError}_{-\gwElevenTenSEOBSpinOnePlaneLowerError}$
	& $\gwElevenTenSEOBChiEffectiveMedian^{+\gwElevenTenSEOBChiEffectiveUpperError}_{-\gwElevenTenSEOBChiEffectiveLowerError}$
	& $\gwElevenTenSEOBChiPMedian^{+\gwElevenTenSEOBChiPUpperError}_{-\gwElevenTenSEOBChiPLowerError}$  \\
    \textsc{IMRPhenomXPHM}
    	& $\gwElevenTenXPHMMassOne^{+\gwElevenTenXPHMMassOneUpperError}_{-\gwElevenTenXPHMMassOneLowerError}$
	& $\gwElevenTenXPHMMassTwo^{+\gwElevenTenXPHMMassTwoUpperError}_{-\gwElevenTenXPHMMassTwoLowerError}$
	& $\gwElevenTenXPHMMassRatio^{+\gwElevenTenXPHMMassRatioUpperError}_{-\gwElevenTenXPHMMassRatioLowerError}$ 
	& $\gwElevenTenXPHMChirpMass^{+\gwElevenTenXPHMChirpMassUpperError}_{-\gwElevenTenXPHMChirpMassLowerError}$ 
	& $\gwElevenTenXPHMDistance^{+\gwElevenTenXPHMDistanceUpperError}_{-\gwElevenTenXPHMDistanceLowerError}$ 
	& $\gwElevenTenXPHMSpinOneMagMedian^{+\gwElevenTenXPHMSpinOneMagUpperError}_{-\gwElevenTenXPHMSpinOneMagLowerError}$
	& $\gwElevenTenXPHMSpinOneTiltMedian^{+\gwElevenTenXPHMSpinOneTiltUpperErrorHPD}_{-\gwElevenTenXPHMSpinOneTiltLowerErrorHPD}$
	& $\gwElevenTenXPHMSpinOneZMedian^{+\gwElevenTenXPHMSpinOneZUpperError}_{-\gwElevenTenXPHMSpinOneZLowerError}$
	& $\gwElevenTenXPHMSpinOnePlaneMedian^{+\gwElevenTenXPHMSpinOnePlaneUpperError}_{-\gwElevenTenXPHMSpinOnePlaneLowerError}$
	& $\gwElevenTenXPHMChiEffectiveMedian^{+\gwElevenTenXPHMChiEffectiveUpperError}_{-\gwElevenTenXPHMChiEffectiveLowerError}$
	& $\gwElevenTenXPHMChiPMedian^{+\gwElevenTenXPHMChiPUpperError}_{-\gwElevenTenXPHMChiPLowerError}$  \\
\textsc{IMRPhenomXO4a}
    	& $\gwElevenTenXOFourAMassOne^{+\gwElevenTenXOFourAMassOneUpperError}_{-\gwElevenTenXOFourAMassOneLowerError}$
	& $\gwElevenTenXOFourAMassTwo^{+\gwElevenTenXOFourAMassTwoUpperError}_{-\gwElevenTenXOFourAMassTwoLowerError}$
	& $\gwElevenTenXOFourAMassRatio^{+\gwElevenTenXOFourAMassRatioUpperError}_{-\gwElevenTenXOFourAMassRatioLowerError}$ 
	& $\gwElevenTenXOFourAChirpMass^{+\gwElevenTenXOFourAChirpMassUpperError}_{-\gwElevenTenXOFourAChirpMassLowerError}$ 
	& $\gwElevenTenXOFourADistance^{+\gwElevenTenXOFourADistanceUpperError}_{-\gwElevenTenXOFourADistanceLowerError}$ 
	& $\gwElevenTenXOFourASpinOneMagMedian^{+\gwElevenTenXOFourASpinOneMagUpperError}_{-\gwElevenTenXOFourASpinOneMagLowerError}$
	& $\gwElevenTenXOFourASpinOneTiltMedian^{+\gwElevenTenXOFourASpinOneTiltUpperErrorHPD}_{-\gwElevenTenXOFourASpinOneTiltLowerErrorHPD}$
	& $\gwElevenTenXOFourASpinOneZMedian^{+\gwElevenTenXOFourASpinOneZUpperError}_{-\gwElevenTenXOFourASpinOneZLowerError}$
	& $\gwElevenTenXOFourASpinOnePlaneMedian^{+\gwElevenTenXOFourASpinOnePlaneUpperError}_{-\gwElevenTenXOFourASpinOnePlaneLowerError}$
	& $\gwElevenTenXOFourAChiEffectiveMedian^{+\gwElevenTenXOFourAChiEffectiveUpperError}_{-\gwElevenTenXOFourAChiEffectiveLowerError}$
	& $\gwElevenTenXOFourAChiPMedian^{+\gwElevenTenXOFourAChiPUpperError}_{-\gwElevenTenXOFourAChiPLowerError}$  \\
    \hline
    \hline
    \end{tabular}
    \label{tab:pe-params-241110}
\end{table*}

\begin{figure*}[t!]
    \centering
    \includegraphics[width=0.9\textwidth]{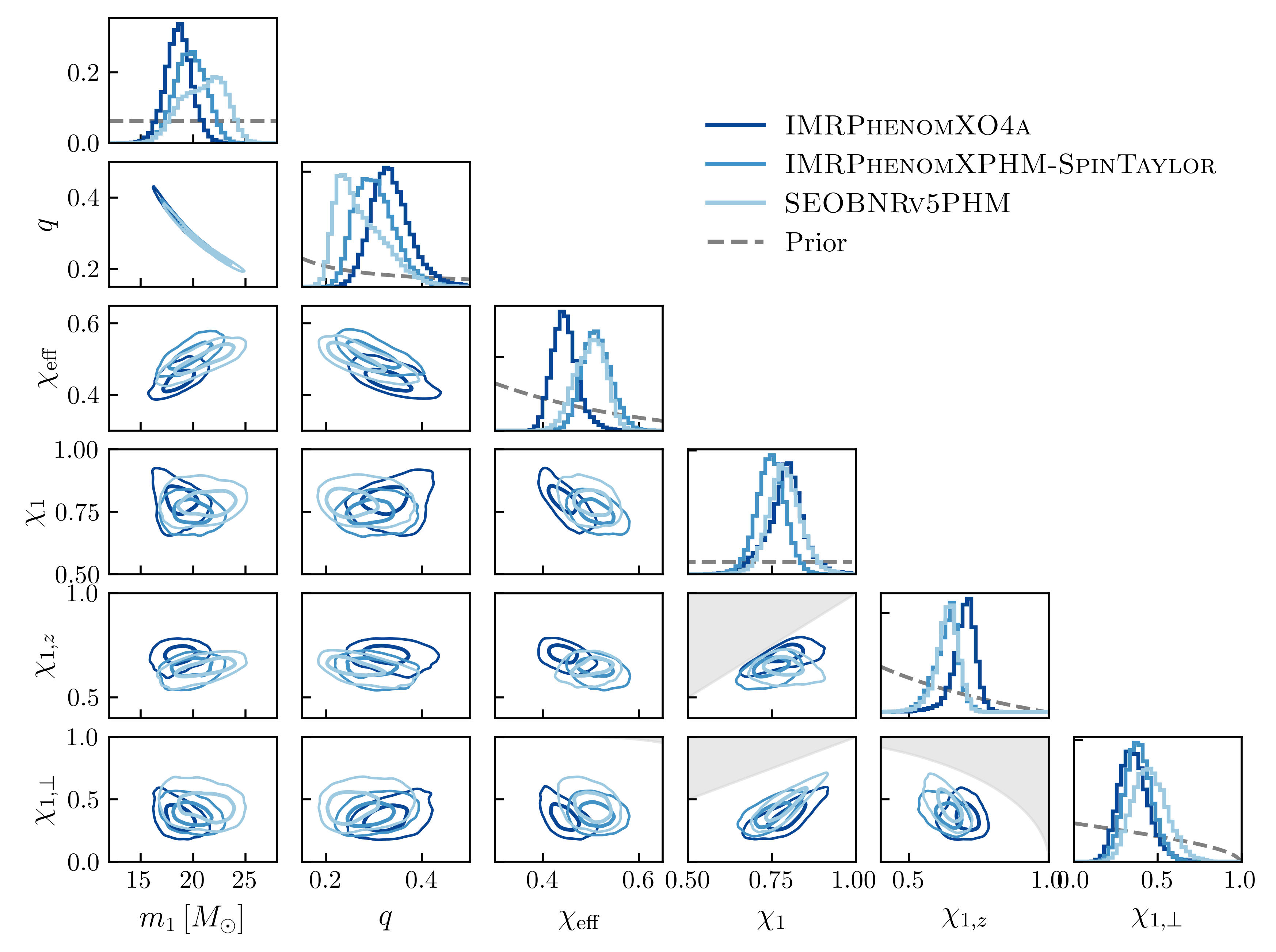}
    \caption{
    Posterior on the source properties of \gwTenEleven under each individual waveform model considered.
    Different waveform models yield qualitatively similar conclusions, although waveforms exhibit non-negligible systematic differences due to the strong precession and unequal masses of the source binary.
    }
    \label{fig:pe-corner-241011}
\end{figure*}

\begin{figure*}[t!]
    \centering
    \includegraphics[width=0.9\textwidth]{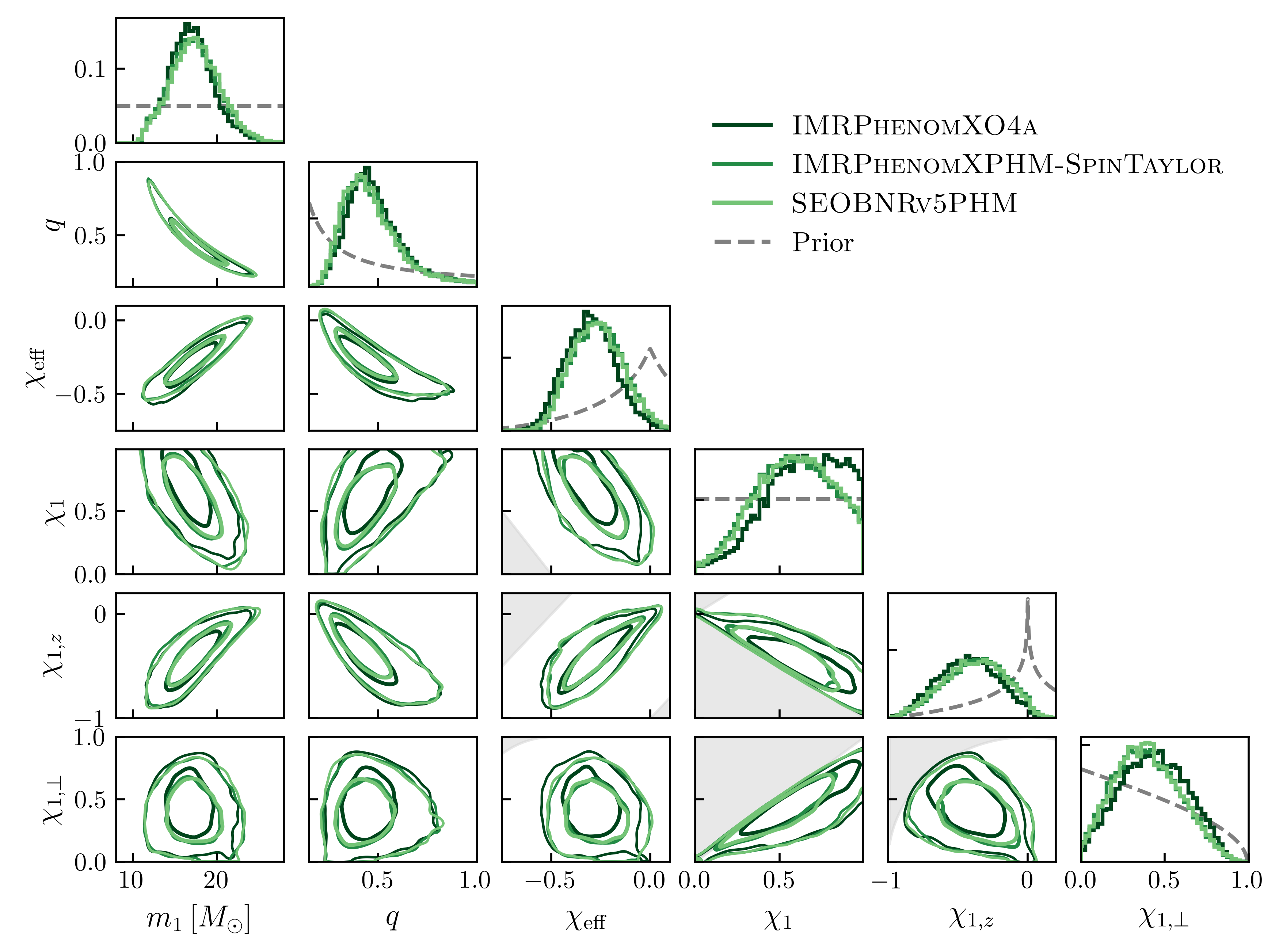}
    \caption{
    As in Figure~\ref{fig:pe-corner-241011}, but for \gwElevenTen.
    }
    \label{fig:pe-corner-241110}
\end{figure*}

Results quoted in the main text are given by the union of posterior samples under three different gravitational waveform models: \textsc{SEOBNRv5PHM}~\citep[\textsc{SEOBNR};][]{Ramos-Buades:2023ehm}, \textsc{IMRPhenomXPHM-SpinTaylor}~\citep[\textsc{XPHM};][]{Pratten:2020ceb,Colleoni:2024knd} and \textsc{IMRPhenomXO4a}~\citep[\textsc{XO4a};][]{Thompson:2023ase}.
Although all of these models describe quasi-circular precessing binaries and include higher-order multipole moments, they differ in the approach used to model the waveforms.
The \textsc{SEOBNR}, \textsc{XPHM}, and \textsc{XO4a} models are each constructed from a combination of analytical and numerical information.
They offer a complete description of binary inspiral, merger, and ringdown, and are therefore applicable to systems of any mass.
The \textsc{SEOBNR} model computes the signal in the time domain.
The \textsc{XPHM} and \textsc{XO4a} models each calculate signals in the frequency domain, but differ in their treatment of spin precession.
Whereas the \textsc{XPHM} model numerically solves the post-Newtonian spin-precession dynamics, the \textsc{XO4a} model adopts a phenomenological ansatz that is fit to numerical-relativity simulations, calibrating evolution of precession angles, the coprecessing frame, and modal asymmetries between positive and negative $m$ modes.

\begin{figure*}[t!]
    \centering
    \includegraphics[width=0.49\textwidth]{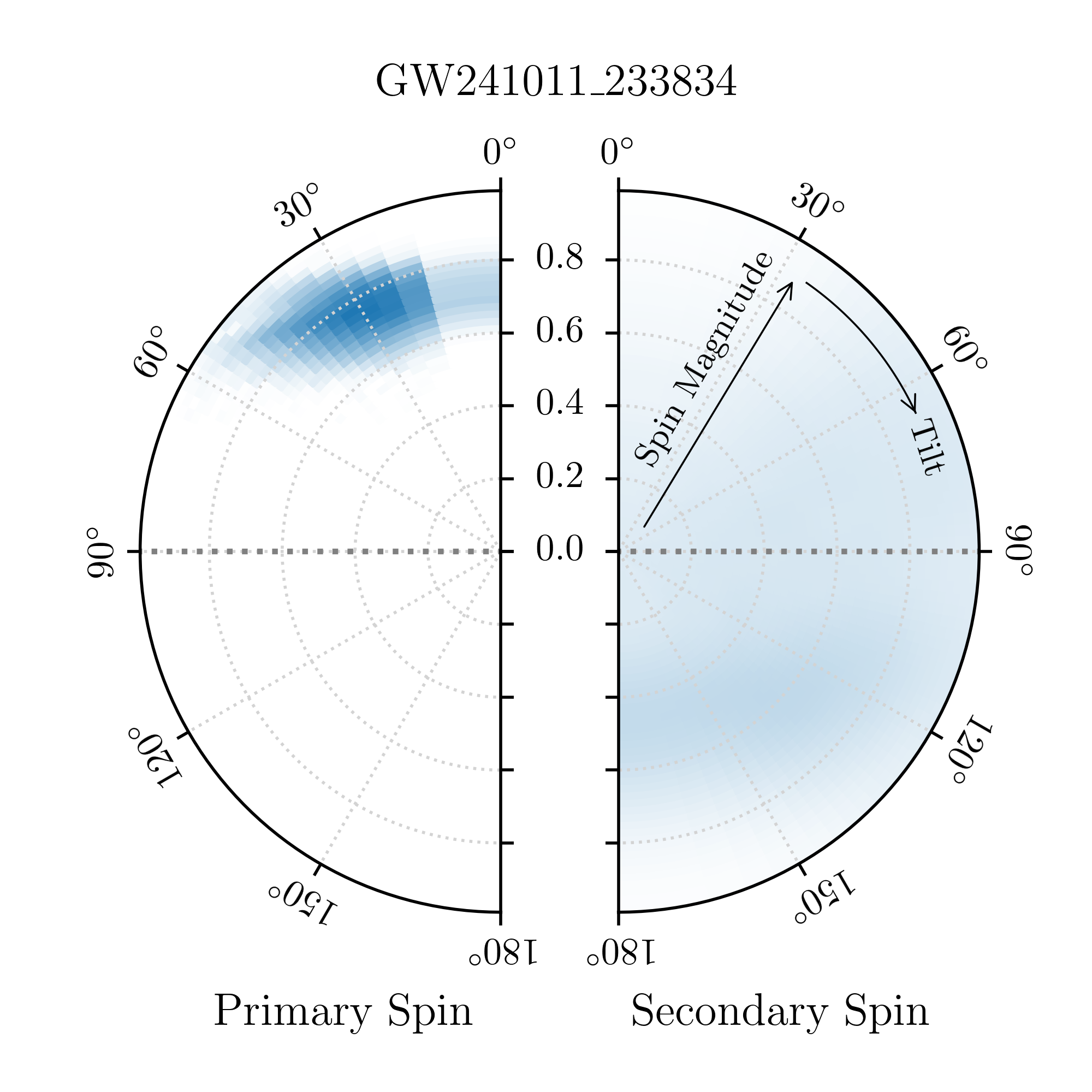}
    \hfill
     \includegraphics[width=0.49\textwidth]{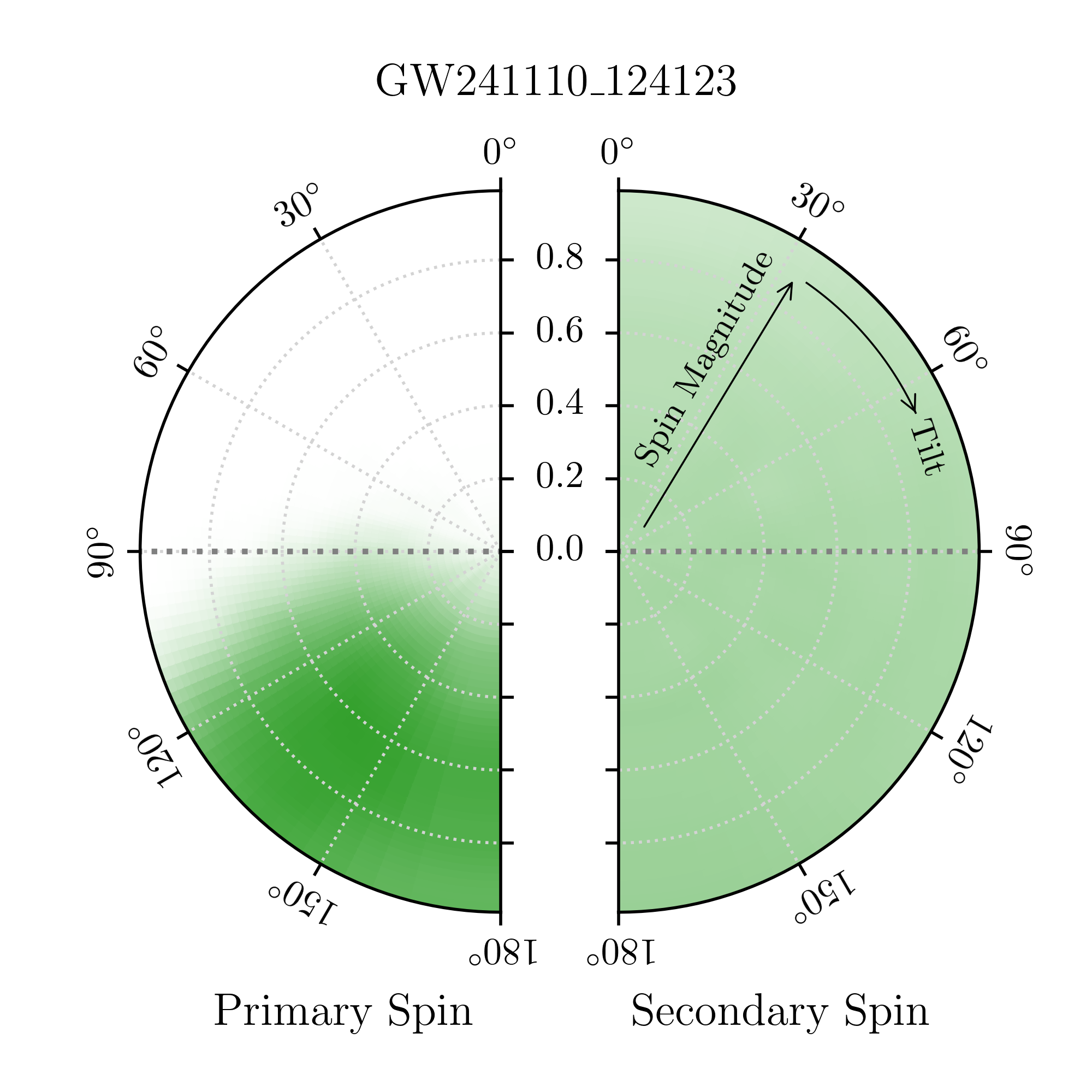}
    \caption{
   Posterior on both the primary and secondary spin vectors of \gwTenEleven (left) and \gwElevenTen (right).
   As in Fig.~\ref{fig:spin-disk}, radial and polar coordinates correspond to dimensionless spin magnitude vectors and spin-orbit misalignment angles, respectively.
   Pixels spaced uniformly in spin magnitude and in cosine-tilt angles, such that each pixel contains equal prior probability.
   For both events, the properties of the secondary spin vectors are unconstrained.
    }
    \label{fig:full-spin-disks}
\end{figure*}

The main text presented only primary spin measurements with \gwTenEleven and \gwElevenTen.
For completeness, in Figure~\ref{fig:full-spin-disks} we present posteriors on both the primary and secondary dimensionless spin magnitude of each binary black hole.
The magnitude and orientation of the events' secondary spin vectors are unconstrained; neither the secondary spin magnitudes, the secondary spin--orbit misalignment angles, nor the azimuthal angles between component spins constrained away from the boundaries of their respective priors.

In Table~\ref{tab:pe-params-241011} and Figure~\ref{fig:pe-corner-241011}, we present posteriors on the source properties of \gwTenEleven obtained independently with each waveform.
All models recover nearly identical binary chirp masses.
At the same time, they each provide slightly different (although statistically consistent) estimates of the binary's primary mass, mass ratio, and primary spin.
Such systematic differences between waveform models are not unexpected; the unequal mass ratio, high spin, significant spin precession, and high SNR of this source are likely to exacerbate differences between waveform models~\citep{Dhani:2024jja, MacUilliam:2024oif, Akcay:2025rve}.
Table~\ref{tab:pe-params-241110} and Figure~\ref{fig:pe-corner-241110}, in turn, shows properties of \gwElevenTen inferred under the previously listed waveform models.
We recover good agreement between the different waveforms, with only small differences in the width and mean of the posterior for parameters like the chirp mass and the spin parameters.

\section{Eccentricity Measurements: Further Details}
\label{app:eccentricity}

\begin{figure*}[t!]
    \centering
    \includegraphics[width=0.96\textwidth]{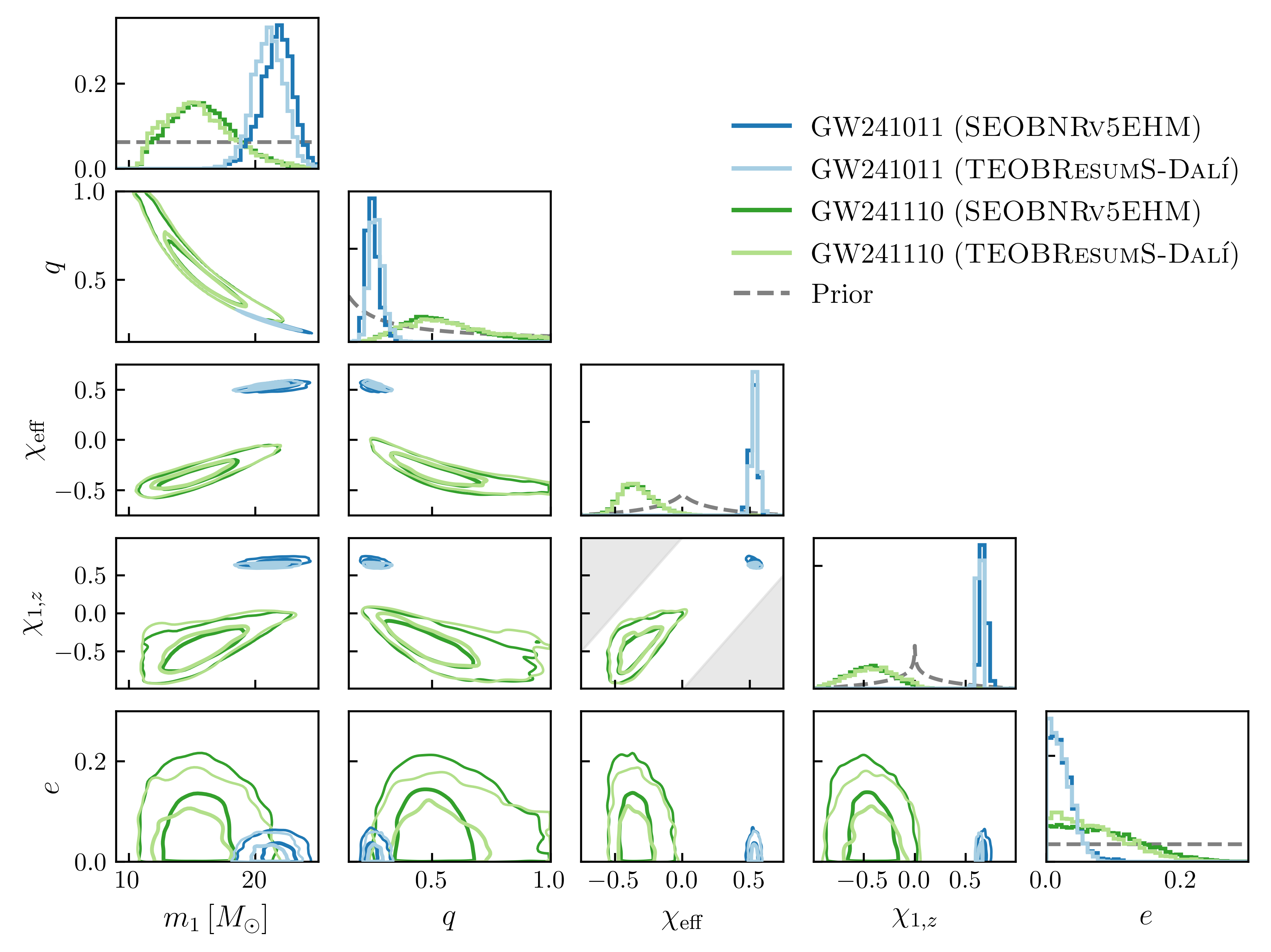}
    \caption{
   Posteriors on the source properties of \gwTenEleven and \gwElevenTen, obtained using the eccentric and spin-aligned \textsc{SEOBNRv5EHM}~\citep{Gamboa:2024hli} and \textsc{TEOBRESUMS-DAL\'I}~\citep{Nagar:2024dzj} waveform models.
   Orbital eccentricities, quoted at a reference frequency of $13.33\,\mathrm{Hz}$, are consistent with $e=0$, while other source parameters are consistent with estimates obtained elsewhere using quasicircular and precessing waveform models.
    }
    \label{fig:eccentricity-corner}
\end{figure*}

Gravitational-wave measurements of orbital eccentricity have long been challenging, as the dynamics of eccentric orbits vary rapidly on short orbital timescales and give rise to complex waveform morphologies.
Recently, however, a number of mature models have been developed that describe gravitational-wave emission from eccentric compact binaries through binary inspiral, merger, and ringdown.
We study the eccentricity of \gwTenEleven and \gwElevenTen using two waveform models: the \textsc{SEOBNRv5EHM}~\citep{Gamboa:2024hli} and \textsc{TEOBRESUMS-DAL\'I}~\citep{Nagar:2024dzj} waveforms.
Both models define eccentricity based a Keplerian parametrization of the orbit~\citep{1959RSPSA.249..180D}, in which the deformation of the orbit is measured by the Keplerian eccentricity $e$ and the relative position of the binary along the orbit at a specific reference frequency is determined by the relativistic anomaly (\textsc{SEOBNRv5EHM}) or mean anomaly (\textsc{TEOBRESUMS-DAL\'I}).
Each model restricts spin to lie parallel (or antiparallel) to a binary's orbital angular momentum, and therefore neglects precessional effects due to in-plane spin components.
We adopt Bayesian priors that are uniform in detector-frame component masses, uniform in comoving volume and detector-frame time, and isotropic in source position and orientation.
The prior on aligned spin components is taken to be the projection of uniform-in-magnitude and isotropic spin priors, consistent with the prior adopted in Sec.~\ref{sec:pe}.
We adopt uniform priors on the relativistic anomaly and on the eccentricity at a reference frequency of $13.33\,\mathrm{Hz}$~\citep{Ramos-Buades:2023yhy}.
Sampling of the \textsc{SEOBNRv5EHM} waveform model is performed using both the \BILBY~\citep{Ashton:2018jfp, Romero-Shaw:2020owr} and \RIFT~\citep{Pankow:2015cra, Lange:2017wki, Wysocki:2019grj} packages, the former invoking the \DYNESTY~\citep{Speagle:2019ivv} nested sampler; posterior samples from each software are combined in equal proportion.
Sampling of the \textsc{SEOBNRv5EHM} waveform model is performed with \RIFT.

Figure~\ref{fig:eccentricity-corner} shows more complete posteriors on the source properties of \gwTenEleven and \gwElevenTen using both waveform models.
The component masses, effective inspiral spin, and primary aligned spin of both sources are consistent with results presented above using quasicircular and precessing waveform models, with little correlation between these parameters and orbital eccentricity.
This suggests, although does not prove, that eccentricity limits for \gwTenEleven and \gwElevenTen are minimally biased by the lack of precessional effects, and conversely that measurements of spin-orbit precession are likely robust despite neglecting eccentricity.

Care must be taken when comparing eccentricities among waveform families and to predictions from the literature.
First, the orbital eccentricity for inspiraling compact binaries is not uniquely defined; different waveform families may adopt distinct, gauge-dependent choices for $e$.
Growing efforts exist to standardize the definition of compact binary eccentricity using waveform-based descriptions to remove gauge-ambiguity~\citep{Shaikh:2023ypz, Islam:2025oiv, Shaikh:2025tae}, but in this paper we have not attempted to unify the \textsc{SEOBNRv5EHM} and \textsc{TEOBRESUMS-DAL\'I} definitions in this fashion.
We nevertheless find strong consistency between eccentricity limits placed by both waveform models.
Second, orbital eccentricity rapidly evolves under gravitational-wave radiation, and thus must be quoted at a specific time.
As in our case above, orbital eccentricity is typically quoted at a fixed gravitational-wave reference frequency, to be taken as a proxy for time.
The instantaneous frequency of an eccentric binary inspiral is not monotonic in time, however, and so there exist different conventions with which to define reference frequencies.
In this work, we define the reference frequency as the orbit-averaged detector-frame frequency of a binary's $(\ell, m) = (2, \pm 2)$ quadrupole radiation.
Astrophysical predictions, in contrast, tend to adopt reference frequencies defined by the frequency of the \textit{instantaneously loudest frequency harmonic}~\citep{Wen:2002km}.
Eccentricities quoted at numerically identical reference frequencies can, under both prescriptions, correspond to eccentricities at very distinct times in a binary's evolution~\citep{Vijaykumar:2024piy}.

\section{Binary black hole population inference with \gwTenEleven and \gwElevenTen}
\label{app:pop}

In Section~\ref{sec:astro:pop}, we noted that \gwTenEleven and \gwElevenTen do not appear to be clear outliers with respect to the population of merging binary black holes.
This appendix elaborates on this statement, presenting and discussing updated measurements of the binary black hole population using \gwTenEleven and \gwElevenTen.

\subsection{The binary black hole spin distribution}
\label{subsec:pop_models}

\begin{figure*}[t!]
    \centering
    \includegraphics[width=0.49\textwidth]{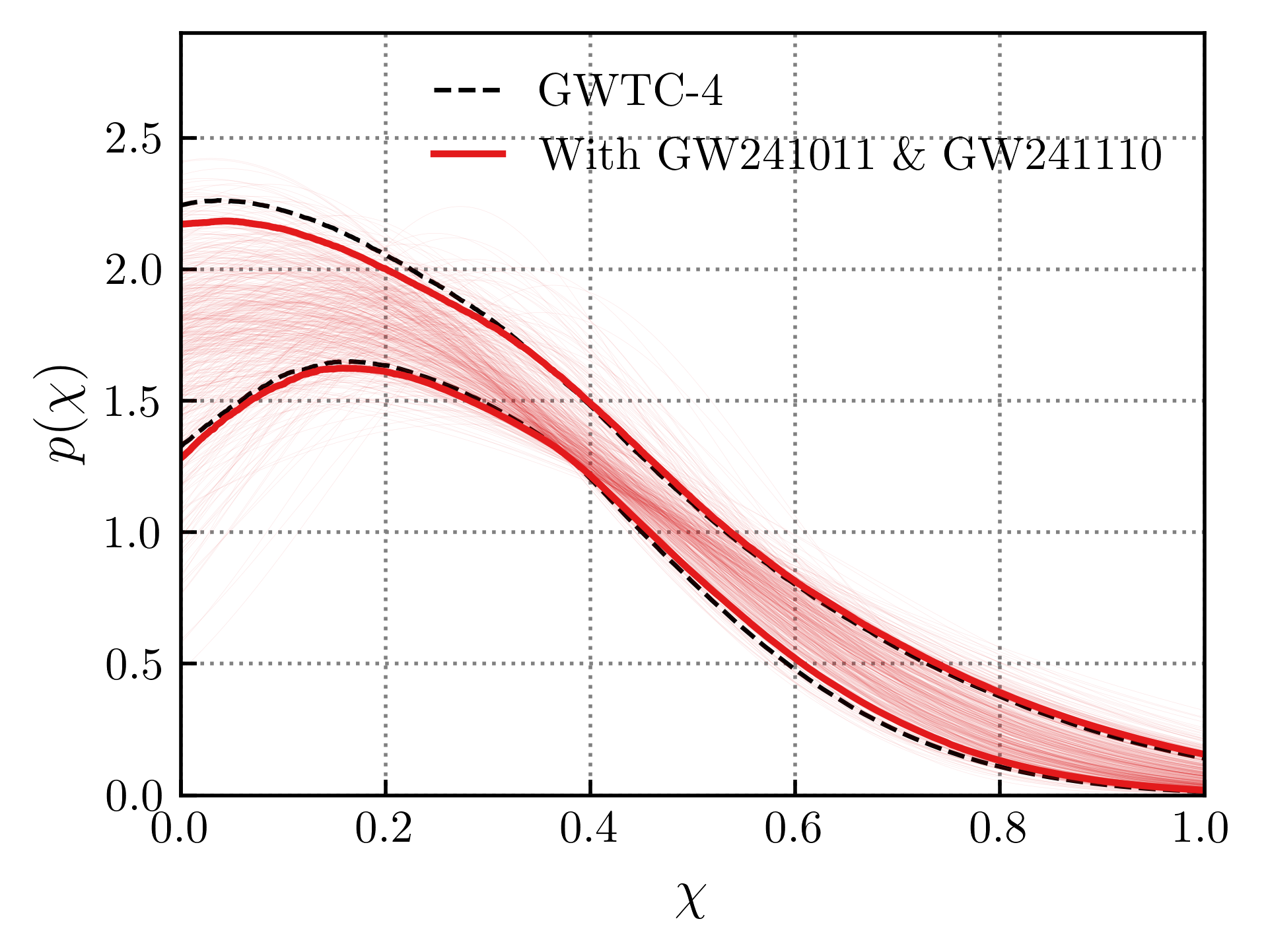}
    \includegraphics[width=0.49\textwidth]{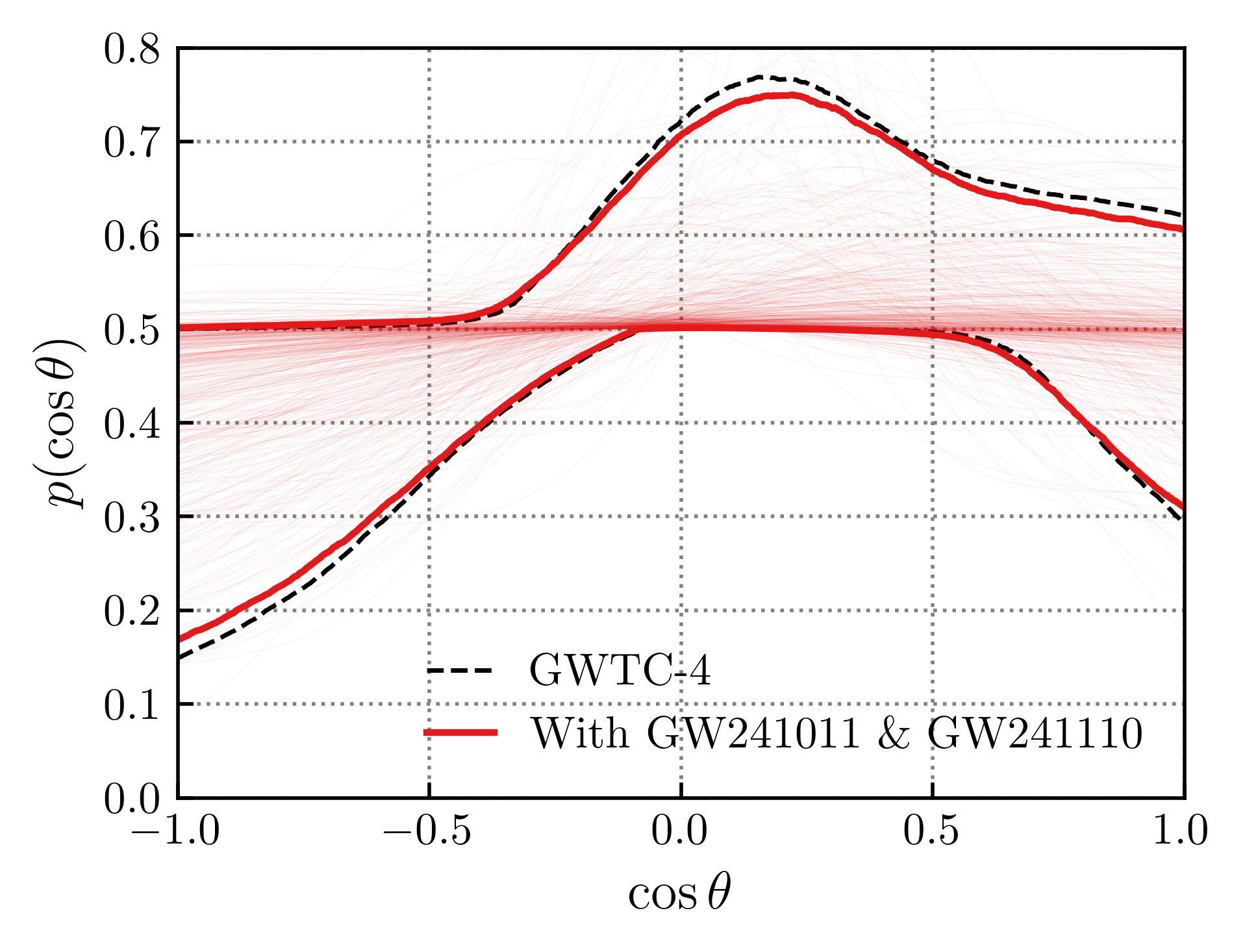}
    \caption{
    Inferred distribution of component spin magnitudes (left) and cosine spin--orbit misalignment angles (right) of merging binary black holes, with and without \gwTenEleven and \gwElevenTen.
    We use the \textsc{Gaussian Component Spin} population model from~\cite{gwtc4-astrodist}, in which spin magnitudes are Gaussian-distributed, and cosine spin--orbit angles follow a mixture between Gaussian and uniform components.
     Dashed lines indicate $90\%$ credible bounds on $p(\chi)$ and $p(\cos\theta)$ when using binary black holes from GWTC-4.0~\citep{gwtc4}, while the thick red lines indicate updated bounds when additionally including \gwTenEleven and \gwElevenTen.
     The ensemble of thin red lines shows the probability distributions corresponding to individual draws on our population posterior, when including \gwTenEleven and \gwElevenTen.
     The inclusion of \gwTenEleven and \gwElevenTen negligibly affects the inferred spin magnitude and spin--orbit tilt distributions.
    }
    \label{fig:population-default}
\end{figure*}

We hierarchically measure the population properties of binary black holes following the methodology described in~\cite{gwtc4-astrodist}.
We select all binary black holes among GWTC-4.0~\citep{gwtc4}, detected with a false-alarm rate below $1\,\mathrm{yr}^{-1}$ by at least one search algorithm, and exclude the events GW190814~\citep{LIGOScientific:2020zkf}, GW190917\_114630~\citep{LIGOScientific:2021usb}, and GW230529~\citep{LIGOScientific:2024elc} that contain low-mass objects of unknown nature.
This yields a total of $153$ binary black hole coalescences, plus \gwTenEleven and \gwElevenTen.
Selection biases are estimated using a suite of simulated signals added to data from the first three observing runs (O1, O2, and O3) and the first part of the fourth LIGO--Virgo--KAGRA observing run~\citep[O4a;][]{Essick:2025zed, gwtc4-astrodist}.
We neglect the additional time--volume surveyed early in the second part of the fourth observing run (O4b), in which \gwTenEleven and \gwElevenTen were detected.
This slightly biases our measurements of the binary black hole population, but the short duration and comparable sensitivity of early O4b render this bias negligible.
We perform hierarchical inference using the \textsc{GWPopulation} package~\citep{Talbot:2019okv, Talbot:2024yqw} and the \textsc{Dynesty} nested sampler~\citep{Speagle:2019ivv}.

Figure~\ref{fig:population-default} illustrates the inferred distributions of spin magnitudes (left) and spin--orbit misalignment angles (right) among binary black holes, with and without \gwTenEleven and \gwElevenTen.
We use the \textsc{Gaussian Component Spin} model described in~\cite{gwtc4-astrodist}, in which component spin magnitudes are identically and independently drawn from a truncated normal distribution,
	\begin{equation}
	  p(\chi_1, \chi_2) = N_{[0,1]}(\chi_1|\mu_\chi, \sigma_\chi) N_{[0,1]}(\chi_2|\mu_\chi, \sigma_\chi),
          \label{eq:gauss-mag}
	\end{equation}
and cosine spin--orbit tilts are jointly distributed as a mixture between Gaussian and uniform components:
	\begin{equation}
	p(\cos\theta_1, \cos\theta_2) = \zeta N_{[-1,1]}(\cos\theta_1|\mu_t, \sigma_t) N_{[-1,1]}(\cos\theta_2 |\mu_t, \sigma_t) + (1-\zeta) U_{[-1,1]}(\cos\theta_1) U_{[-1,1]}(\cos\theta_2).
	\label{eq:gauss-tilt}
	\end{equation}
Here, $N_{[a,b]}$ and $U_{[a,b]}$ represent Gaussian and uniform distributions truncated and normalized on the interval $[a, b]$, and the means $\mu_\chi$ and $\mu_t$, standard deviations $\sigma_\chi$ and $\sigma_t$, and mixing fraction $\zeta$ are free parameters inferred from data.
We assume that black hole masses and redshifts follow the default distributions adopted in~\cite{gwtc4-astrodist} and adopt the same Bayesian priors.
We see that the inclusion of events \gwTenEleven and/or \gwElevenTen yields a spin-tilt distribution marginally more consistent with isotropy, but that these events otherwise have negligible effects on the inferred spin distributions.

It is possible that the \textsc{Gaussian Component Spin} model, with unimodal spin magnitude and tilt distributions, may provide a poor description of  events like \gwTenEleven and \gwElevenTen.
We therefore explore two extensions of this model to further study the implications of \gwTenEleven and \gwElevenTen.
\begin{enumerate}
\item First, whereas the $\cos\theta$ distribution in Eq.~\eqref{eq:gauss-tilt} was truncated on the interval $[-1, 1]$, we instead introduce a variable lower truncation bound $\cos(\theta_\mathrm{max})$~\citep{Galaudage:2021rkt, Tong:2022iws}\ (i.e. a maximum spin tilt angle $\theta_\mathrm{max}$) with a prior uniform on the interval $[-1, 1]$, and ask how extreme spin--orbit misalignment angles must be to accommodate events like \gwElevenTen (the \textsc{Max-Tilt} model),
    \begin{center}
        \begin{align}
            p(\cos\theta_1, \cos\theta_2) =~&\zeta N_{[\cos(\theta_\mathrm{max}),1]}(\cos\theta_1|\mu_t, \sigma_t) N_{[\cos(\theta_\mathrm{max}),1]}(\cos\theta_2 |\mu_t, \sigma_t) 
            \nonumber\\ &+~ 
            (1-\zeta) U_{[\cos(\theta_\mathrm{max}),1]}(\cos\theta_1) U_{[\cos(\theta_\mathrm{max}),1]}(\cos\theta_2).
            \label{eq:maxtilt}
        \end{align}  
    \end{center}
\item Second, we allow for the existence of a distinct subpopulation of rapidly spinning black holes with large $\chi$, designed to explore whether events like \gwTenEleven require a multimodal spin distribution (\textsc{High-Spin} model). 
We extend the \ref{eq:gauss-mag} model by introducing a high spin Gaussian,
    \begin{center}
        \begin{align}
        p(\chi_1, \chi_2) =~& \Big(\xi_\chi N_{[0,1]}(\chi_1|\mu_\chi, \sigma_\chi) + (1-\xi_\chi)N_{[0,1]}(\chi_1|\mu_{\chi,\mathrm{high}}, \sigma_{\chi,\mathrm{high}})\Big)
        \nonumber\\ &\times 
        \Big(\xi_\chi N_{[0,1]}(\chi_2|\mu_\chi, \sigma_\chi) + (1-\xi_\chi)N_{[0,1]}(\chi_2|\mu_{\chi,\mathrm{high}}, \sigma_{\chi,\mathrm{high}})\Big),
        \label{eq:highspin}
        \end{align}  
    \end{center}
where $\mu_{\chi,\mathrm{high}}$ is the mean high spin Gaussian with a prior the uniform on the interval $[0.5,1]$, $\sigma_{\chi,\mathrm{high}}$ is the width of the high spin Gaussian with a prior the uniform on the interval $[0.005,1]$, and finally $\xi_\chi$ is the mixing fraction between the low spin and high spin Gaussians with a uniform prior on the interval $[0,1]$.
\end{enumerate}

\begin{figure*}[t!]
    \centering
    \includegraphics[width=0.49\textwidth]{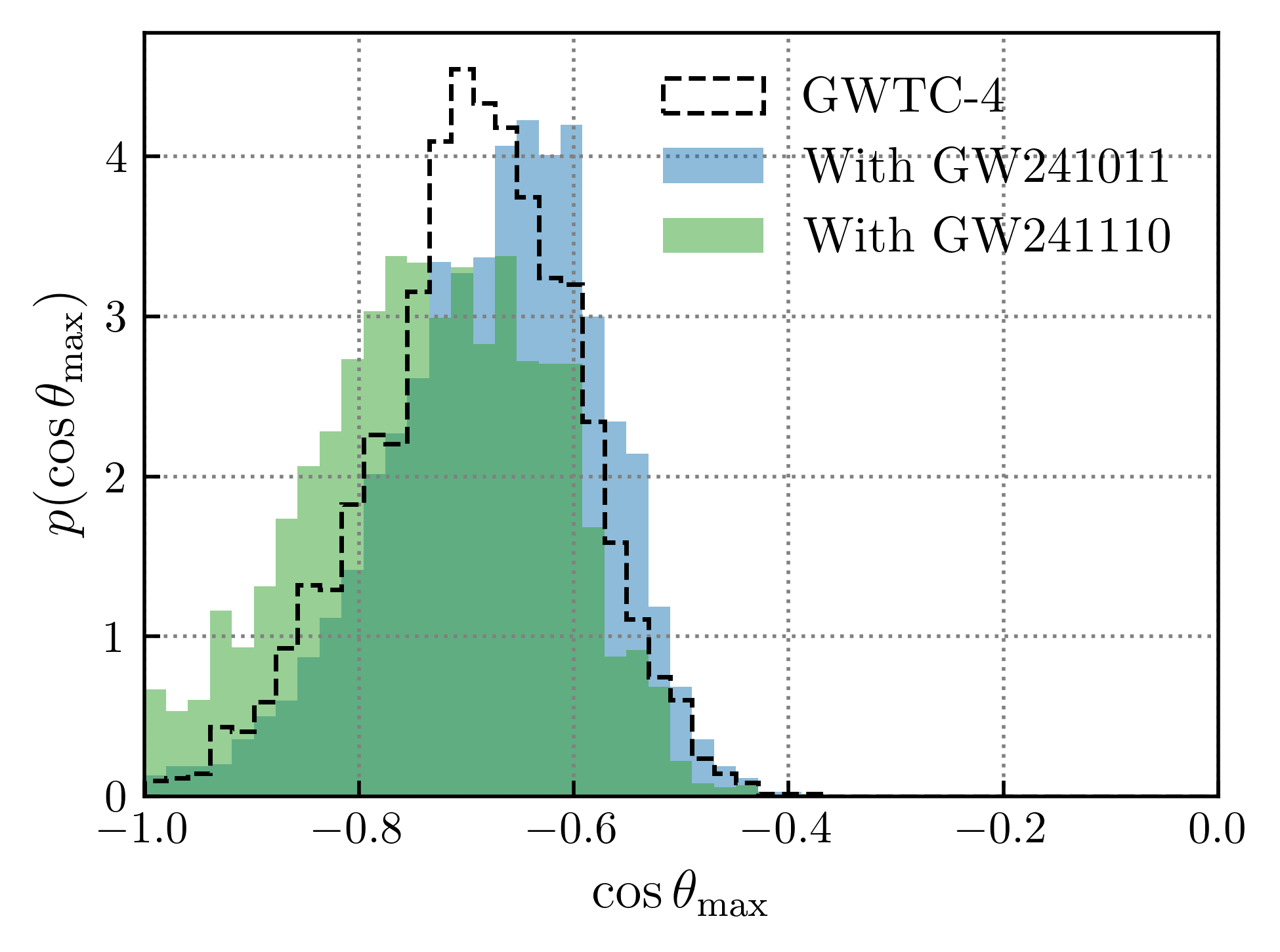}
    \includegraphics[width=0.49\textwidth]{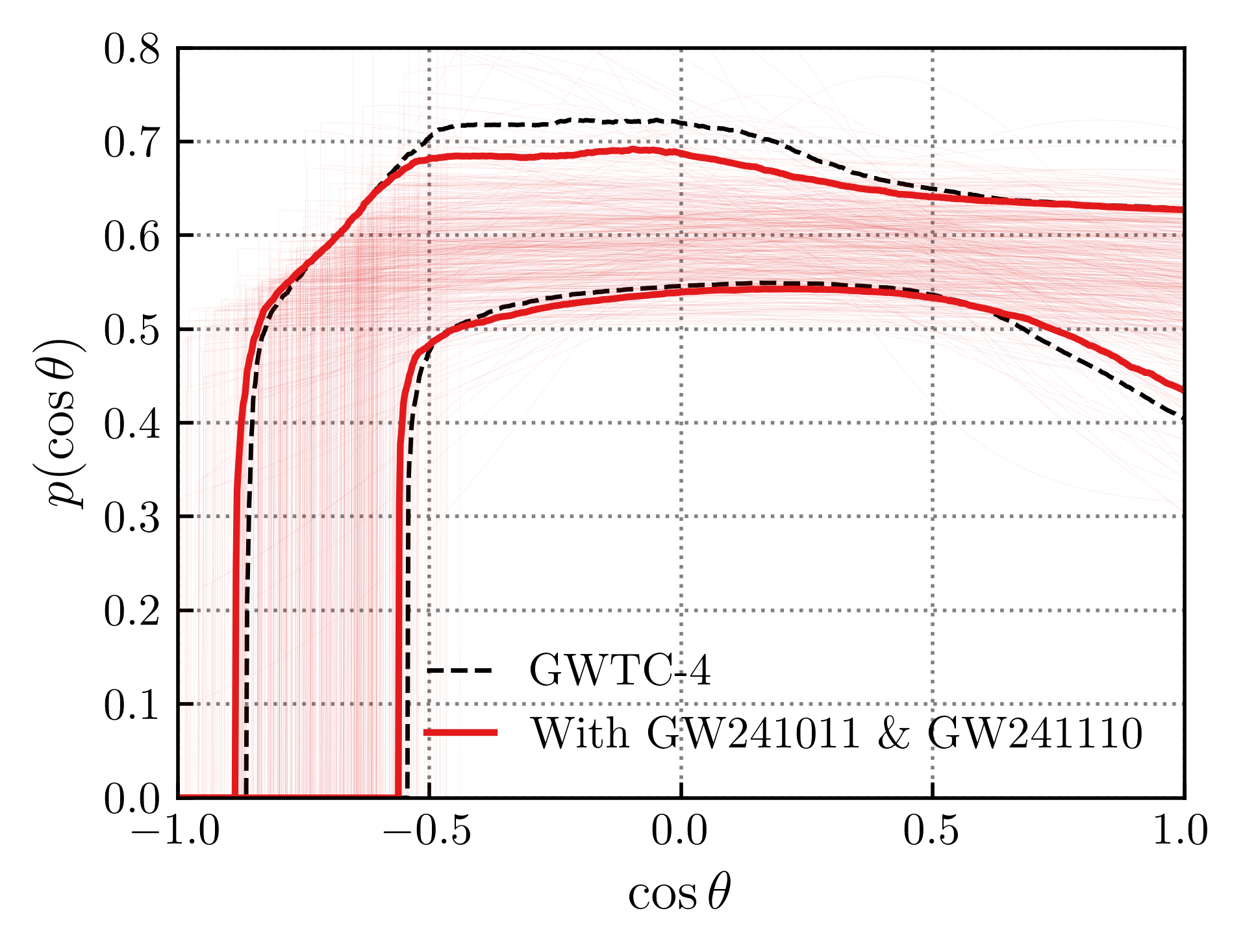}
    \caption{
    \textit{Right}: Inferred distribution of cosine spin--orbit misalignment angles, when additionally inferring the maximum misalignment angle (minimum $\cos\theta$ value) among the binary black hole population (the \textsc{Max-Tilt} model).
    As in Figure~\ref{fig:population-default}, dashed lines indicate $90\%$ credible bounds using black holes from GWTC-4.0~\citep{gwtc4}, thick red lines indicate updated bounds when additionally including \gwTenEleven and \gwElevenTen, and thin red lines illustrate individual draws from our updated population posterior.
    \textit{Left}: Posterior obtained on the minimum value $\cos(\theta_\mathrm{max})$ below which the $\cos\theta$ distribution is truncated.
    The inclusion of \gwTenEleven and \gwElevenTen minimally affects inference of the maximum spin misalignment angle among the binary black hole population.
    Inference using GWTC-4.0 binary black holes requires $\cos\theta_\mathrm{max}<\GWTCFourCosMaxTiltUpperLimit$ ($\theta_\mathrm{max} > \GWTCFourMaxTiltLowerLimit\,\mathrm{degrees}$) at $90\%$ credibility.
    Adding \gwTenEleven and \gwElevenTen gives $\cos(\theta_\mathrm{max}) \leq \GWTCFourPlusTenElevenCosMaxTiltUpperLimit$ and $\GWTCFourPlusElevenTenCosMaxTiltUpperLimit$, respectively.
        }
    \label{fig:population-min-tilt}
\end{figure*}

\begin{figure*}[t!]
    \centering
    \includegraphics[width=0.49\textwidth]{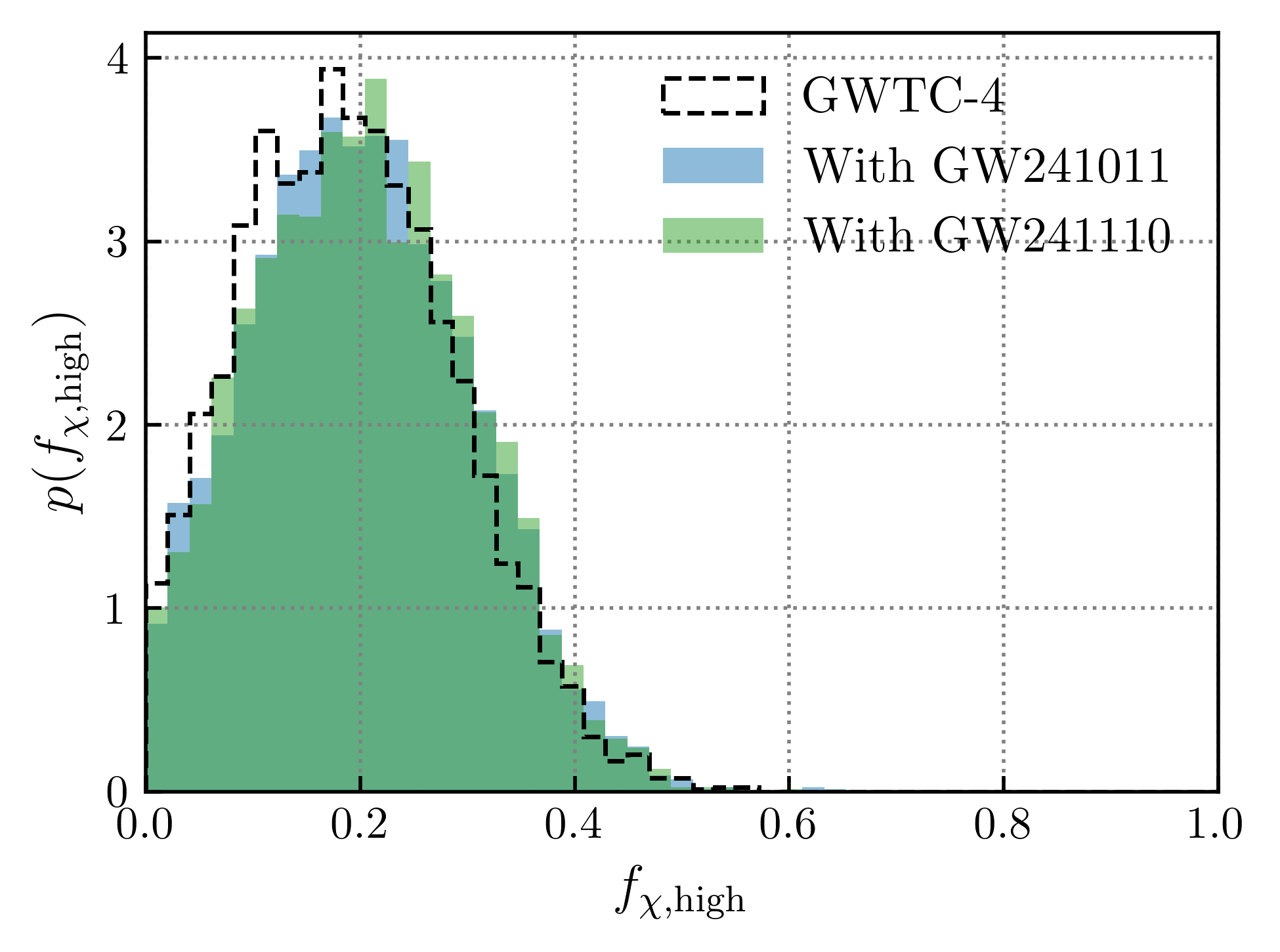}
    \includegraphics[width=0.49\textwidth]{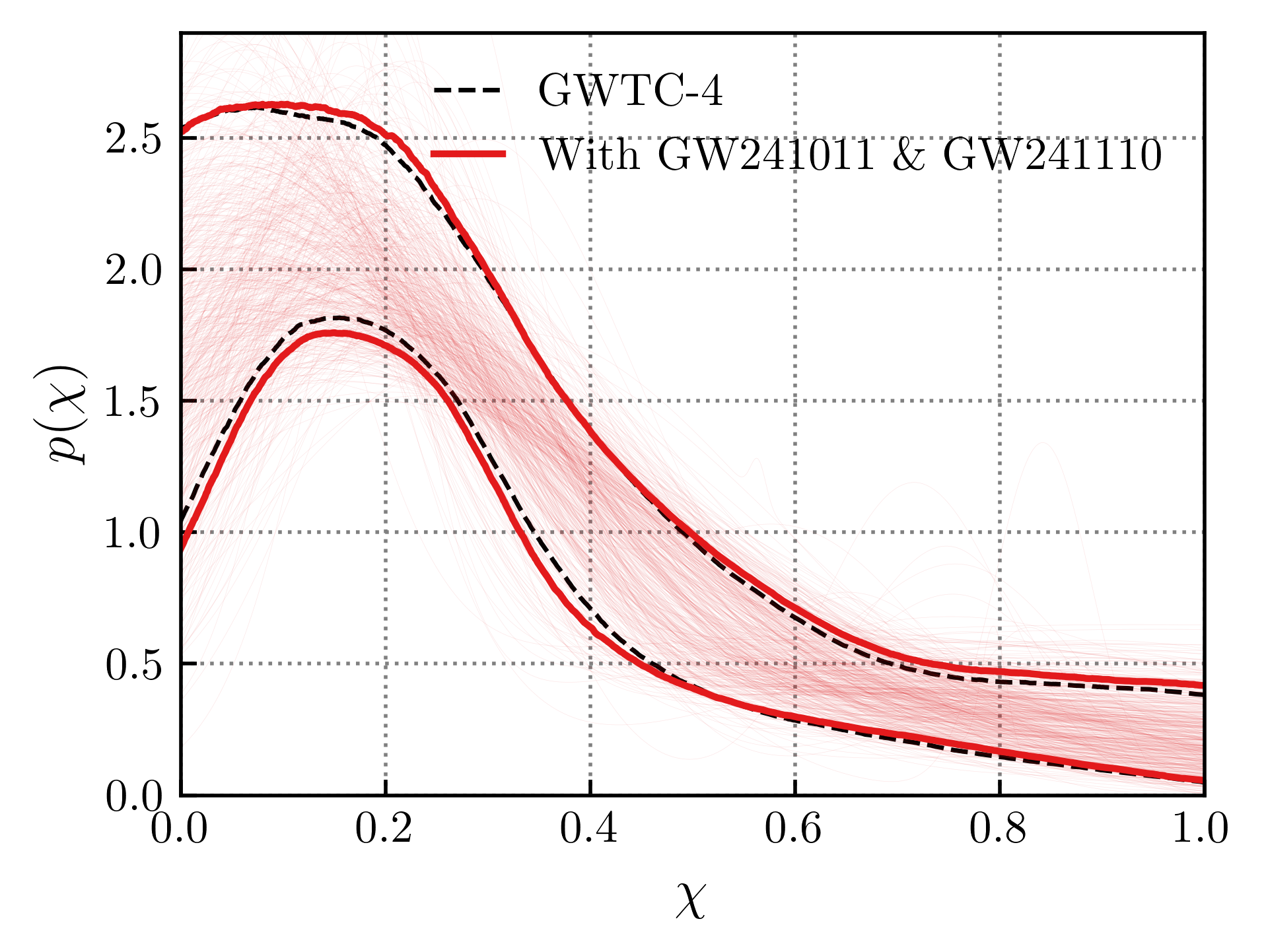}
    \caption{
    \textit{Right}: Inferred distribution of black hole component spin magnitudes when allowing for a distinct subpopulation of rapidly spinning black holes where one or both black hole components occupy this high spin region (the \textsc{High-Spin} model).
    As above, the dashed lines and filled region span $90\%$ credible bounds with and without \gwTenEleven and \gwElevenTen, respectively.
    \textit{Left}: Inferred fraction of black holes comprising a possible highly-spinning subpopulation.
    As above, results are negligibly affected by the inclusion or exclusion of \gwTenEleven and \gwElevenTen.
    In both cases, current data do not require the existence of a distinct population of events with rapidly-spinning components.
    When including \gwTenEleven and \gwElevenTen, we bound $f_{\chi,\mathrm{high}} \leq \GWTCFourPlusSpecialEventsFracHighSpinUpperLimit$ at $90\%$ credibility, consistent with zero.
    }
    \label{fig:population-high-spin}
\end{figure*}

Figure~\ref{fig:population-min-tilt} shows the measured distribution of spin--orbit misalignment angles $\theta$ when inferring the maximum misalignment angle $\theta_\mathrm{max}$.
The left-hand panel shows the posterior on $\cos(\theta_\mathrm{max})$.
Binary black holes among GWTC-4.0 require maximum spin--orbit misalignment angles of $\theta_\mathrm{max} \geq \GWTCFourMaxTiltLowerLimit\,\mathrm{degrees}$ ($\cos\theta_\mathrm{max}\leq \GWTCFourCosMaxTiltUpperLimit$) at $90\%$ credibility.
This result is only marginally affected by the inclusion of \gwTenEleven or \gwElevenTen, which yield $\theta_\mathrm{max} \geq \GWTCFourPlusTenElevenMaxTiltLowerLimit\,\mathrm{degrees}$ and $\GWTCFourPlusElevenTenMaxTiltLowerLimit\,\mathrm{degrees}$, respectively.
Figure~\ref{fig:population-high-spin}, meanwhile, shows the inferred black hole spin-magnitude distribution when allowing for a distinct subpopulation of rapidly spinning black holes, together with the inferred fraction of events $f_{\chi,\mathrm{high}}$ with one or both components occupying a possible high-spin population.
When compared to Figure~\ref{fig:population-default}, it is clear that different models yield slight, systematic differences in the measured spin magnitude distribution;
the \textsc{High-Spin} model gives a spin distribution more concentrated at small $\chi$ with an extended tail to high spin magnitudes.
However, the addition or exclusion of \gwTenEleven and \gwElevenTen again result in negligible differences.
Binary black holes among GWTC-4.0 do not require the existence of a distinct, high-spinning subpopulation, with $f_{\chi, \mathrm{high}} < \GWTCFourFracHighSpinUpperLimit$ at $90\%$ credibility.
When including \gwTenEleven and \gwElevenTen, the high-spin fraction remains consistent with zero, with $f_{\chi, \mathrm{high}} < \GWTCFourPlusSpecialEventsFracHighSpinUpperLimit$.

Taken together, our results indicate that \gwTenEleven and \gwElevenTen are not evident population outliers, requiring neither greater spin--orbit misalignments nor larger spin magnitudes than already afforded by binary black holes among GWTC-4.0.

\subsection{Population Reweighting}
\label{subsec:pop_reweighting}

\begin{figure*}[t!]
    \centering
    \includegraphics[width=0.99\textwidth]{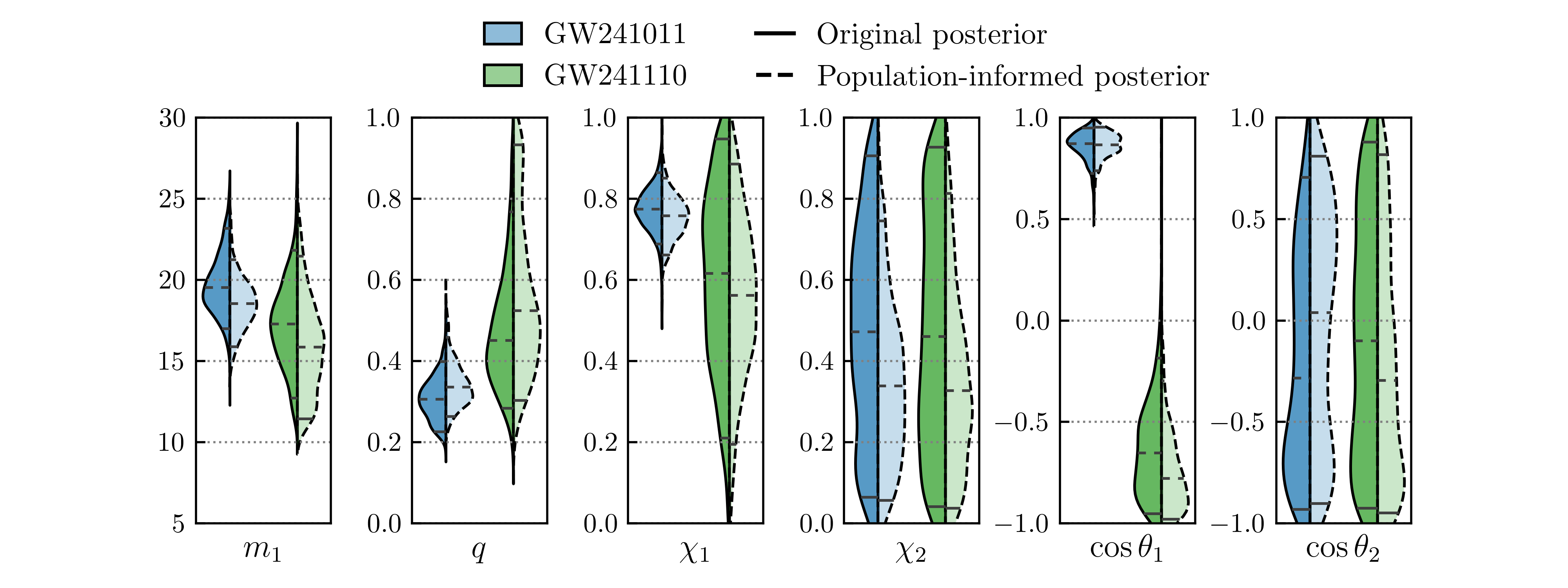}
    \caption{
      Comparison of original and population-informed (using the \textsc{Gaussian Component Spin} population model) posterior distributions for \gwTenEleven and \gwElevenTen. 
    }
    \label{fig:population-reweighted}
\end{figure*}

We reweight the parameter estimation samples using population-informed mass and spin distributions using the three population models.
Reweighted posteriors are obtained with leave-one-out posterior predictive distributions~\citep{Galaudage:2019jdx, Callister:2021note, Essick:2021note}, providing astrophysically motivated priors (such reweighting excludes the contribution of the event in question to the population inference, thus avoiding double-counting effects).
The resulting posteriors for the \textsc{Gaussian Component Spin} population model, for example, are shown in Figure~\ref{fig:population-reweighted}.

For the posterior on $\cos(\theta_1)$ of \gwElevenTen, reweighted by the population using the \textsc{Gaussian Component Spin}, \textsc{Max-Tilt} and \textsc{High-Spin}, the upper limit (or minimum misalignment) is given by $\ElevenTenPrimaryCosTiltUpperLimitDEFAULT$, $\ElevenTenPrimaryCosTiltUpperLimitMAXTILT$ and $\ElevenTenPrimaryCosTiltUpperLimitHIGHSPIN$ respectively at $90\%$ credibility.
The \textsc{Max-Tilt} fit gives the least misaligned result as the model has the flexibility to cut off at values of $\cos(\theta_1)>-1$ whereas the other two models require the fit to end at $\cos(\theta_1)=-1$.
For the posterior on $\chi_1$ of \gwTenEleven, reweighted by the population using the \textsc{Gaussian Component Spin}, \textsc{Max-Tilt} and \textsc{High-Spin}, the lower limit (or minimum spin magnitude) is given by $\TenElevenPrimarySpinMagLowerLimitDEFAULT$, $\TenElevenPrimarySpinMagLowerLimitMAXTILT$ and $\TenElevenPrimarySpinMagLowerLimitHIGHSPIN$ respectively.
The lower limit on the spin magnitude is consistent under all three models.

\section{Cluster models: Further Details}
\label{app:astromodels}

\begin{table*}[]
    \footnotesize
    \setlength{\tabcolsep}{4pt}
    \renewcommand{\arraystretch}{1.2}
    \centering
    \caption{
    Number of binary black hole mergers included in datasets from the \textsc{CMC} and \textsc{cBHBd} catalogs.
    We include the total numbers of first-generation and higher-generation black hole mergers in each catalog, as well as the number of mergers at each stellar metallicity as highlighted in Figure~\ref{fig:cluster-comparison}.
    }
    \begin{tabular}{r r l l l l}
    \hline \hline
    Model & Generation & $Z=0.01\,Z_\odot$ & $Z=0.1\,Z_\odot$ & $Z=Z_\odot$ & Total \\
    \hline
    \textsc{Cluster Monte Carlo}
    	& First-generation
    	& $\CMCNumberOfMergersFirstGenZOOOTwo$
	& $\CMCNumberOfMergersFirstGenZOOTwo$
	& $\CMCNumberOfMergersFirstGenZOTwo$
	& $\CMCNumberOfMergersFirstGenTotal$  \\
    \textsc{Cluster Monte Carlo}
    	& Higher-generation
    	& $\CMCNumberOfMergersNGenZOOOTwo$
	& $\CMCNumberOfMergersNGenZOOTwo$
	& $\CMCNumberOfMergersNGenZOTwo$
	& $\CMCNumberOfMergersNGenTotal$  \\
    \textsc{clusterBHBdynamics}
    	& First-generation
    	& $\CBHBDNumberOfMergersFirstGenZOOOTwo$
	& $\CBHBDNumberOfMergersFirstGenZOOTwo$
	& $\CBHBDNumberOfMergersFirstGenZOTwo$
	& $\CBHBDNumberOfMergersFirstGenTotal$  \\
    \textsc{clusterBHBdynamics}
    	& Higher-generation
    	& $\CBHBDNumberOfMergersNGenZOOOTwo$
	& $\CBHBDNumberOfMergersNGenZOOTwo$
	& $\CBHBDNumberOfMergersNGenZOTwo$
	& $\CBHBDNumberOfMergersNGenTotal$  \\ [1pt]
    \hline
    \hline
    \end{tabular}
    \label{tab:cluster-counts}
\end{table*}

In this appendix, we provide further details regarding the astrophysical population models plotted alongside the properties of \gwTenEleven and \gwElevenTen in Figure~\ref{fig:cluster-comparison}.
For a complete description of the stellar (binary) evolution and cluster dynamics models underlying these results, we direct the reader to~\citet{Kremer:2019iul}, \citet{2022ApJS..258...22R}, and \citet{Antonini:2022vib}.

The \textsc{CMC} catalog~\citep{Kremer:2019iul, 2022ApJS..258...22R} contains a suite of {148} cluster simulations, primarily comprising a grid of 144 models with varying particle number ($N = \{2,4,8,16\} \times 10^5$), virial radius ($r_v = \{0.5,1,2,4\}~\mathrm{pc}$), metallicity ($Z/Z_\odot = \{0.01,0.1,1\}$), and Galactocentric distance ($R_{\rm gc} = \{2,8,20\}~\mathrm{kpc}$).
In addition, the \textsc{CMC} catalog include four more massive clusters with $N=3.2\times10^6$ at $R_{\rm gc}=20~\mathrm{kpc}$, spanning two metallicities ($0.01$ and $1\,Z_\odot$).
Compared to earlier \textsc{CMC} model suites~\citep{2010ApJ...719..915C, 2013MNRAS.429.2881C, 2015ApJ...800....9M}, this set extends the explored ranges of $r_v$ and $Z$, and decouples metallicity from Galactocentric distance.
Cluster models are evolved dynamically over one Hubble time, and binary black hole mergers are identified as those binaries that successfully coalesce over this timescale.
The total numbers of available black hole mergers (both first-generation and hierarchical) are provided in Table~\ref{tab:cluster-counts}.
In Figure~\ref{fig:cluster-comparison} we specifically show the direct union of binary black hole mergers obtained across all simulations at each given metallicity, in order to illustrate the range of binary black hole masses, mass ratios, and spins that can be achieved over a broad range of cluster conditions.
A more detailed prediction for the merger population expected from clusters, in contrast, would likely adopt a weighted combination corresponding to assumptions about the distribution of cluster masses, metallicities, and formation histories.

The \textsc{cBHBd} code is a semi-analytical model that applies cluster dynamics and energy-balance theory to follow the coupled evolution of clusters and their black hole populations, yielding predictions for black hole binary merger rates and properties.
We use the \textsc{cBHBd} catalog published in \citet{Antonini:2022vib}, containing $10^6$ cluster models that are sampled uniformly in the mass range $10^2M_\odot$ to $2\times 10^7M_\odot$.
We select the subset of models with the same metallicities considered above ($Z/Z_\odot = \{0.01,0.1,1\}$) and with initial half-mass densities of $10^5M_\odot\,\mathrm{pc}^{-3}$, yielding a sample of $129$ cluster models with total masses up to $2\times10^6\,M_\odot$.
The cluster models are evolved over one Hubble time, and Figure~\ref{fig:cluster-comparison} contains binary black holes that merge in this time.
The total numbers of available mergers are given in Table~\ref{tab:cluster-counts}.
As with the \textsc{CMC} results above, we take the direct union of binary mergers across all cluster masses, in order to illustrate the range of possible binary properties.

\begin{figure*}[t!]
    \centering
    \includegraphics[width=0.75\textwidth]{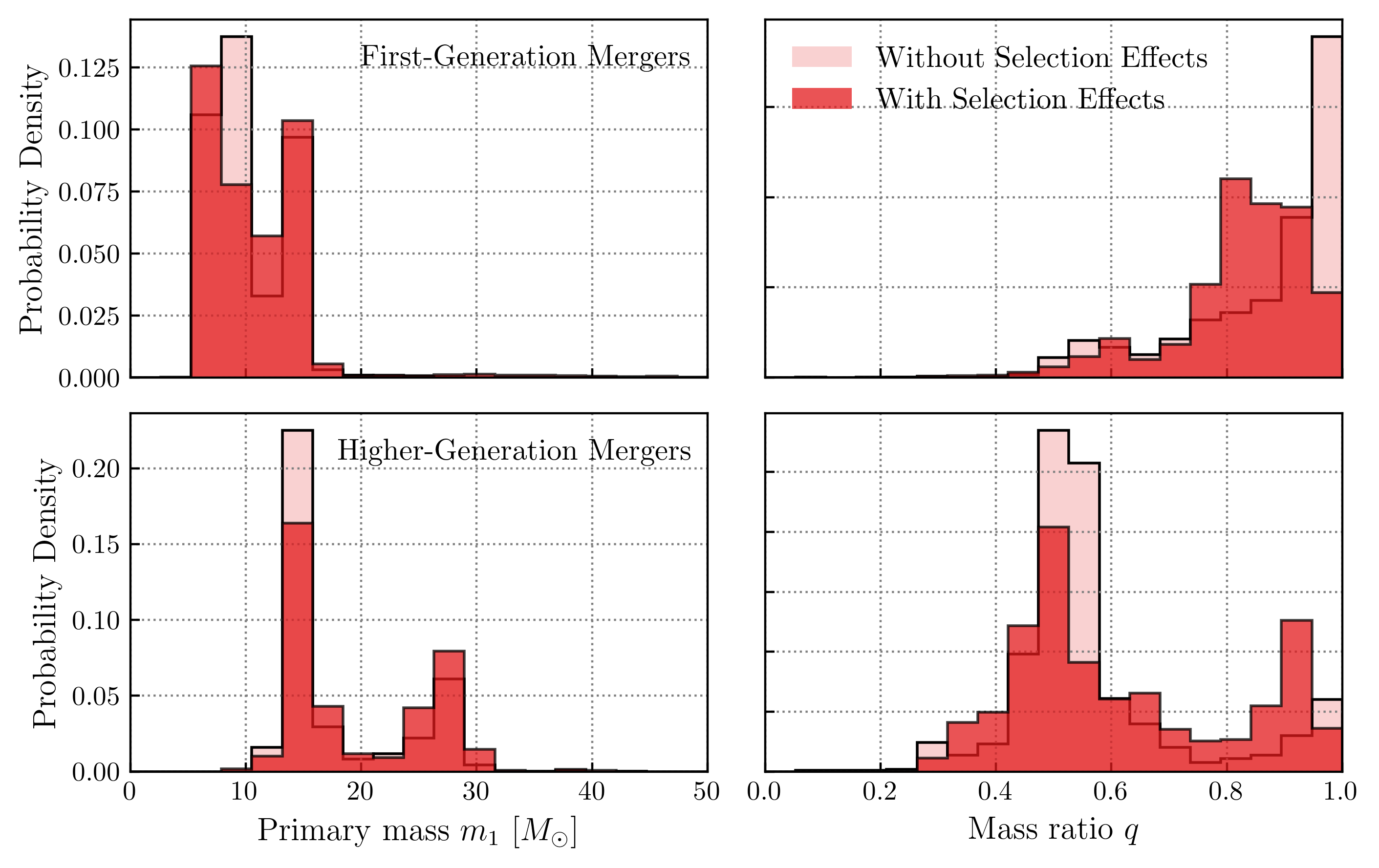}
    \caption{
    The impact of observational selection effects on the predicted properties of binary black hole mergers in dense stellar clusters.
    We specifically show the properties predicted in the \textsc{CMC} catalog for clusters of stellar metallicity $Z=Z_\odot$, corresponding to the right-hand column of Figure~\ref{fig:cluster-comparison}.
    The upper row corresponds to mergers in which both components are first-generation black holes, and the bottom row to mergers in which one or both components are formed from a previous binary black hole coalescence.
    Light red distributions correspond to predictions directly from \textsc{CMC} (and shown previously in Figure~\ref{fig:cluster-comparison}), while dark red distributions have been subjected to an observational selection cut calculated using a campaign of simulated signals injected into LIGO and Virgo data~\citep{Essick:2025zed, LIGOScientific:2025pvj}.
    }
    \label{fig:cluster-comparison-reweighted}
\end{figure*}

Figure~\ref{fig:cluster-comparison} illustrates the range of binary black hole properties predicted under two models of globular cluster evolution.
It does not, however, account for observational selection effects, which may alter the range of binary properties that we expect to successfully detect.
We verify that the inclusion of selection effects does not significantly affect the content of Figure~\ref{fig:cluster-comparison} by using a suite of simulated signals injected into LIGO and Virgo data~\citep{Essick:2025zed, LIGOScientific:2025pvj}.
We reweight these simulated signals from their original proposal distribution to target distributions defined by the \textsc{CMC} and \textsc{cBHBd}.
Selecting only successfully recovered signals then yields the expected distribution of detectable binaries arising from stellar clusters.
The target distributions are themselves obtained by fitting Gaussian mixture models to \textsc{CMC} and \textsc{cBHBd} predictions at each stellar metallicity; the number of mixture components is determined by minimizing the Akaike information criterion~\citep{1974ITAC...19..716A} on reserved testing data.
In order to establish detectability, it is necessary to assume a redshift distribution for binary black hole mergers.
We choose a volumetric merger rate that grows with redshift $z$ as $(1+z)^{2.6}$, following the low-redshift star formation rate of~\citet{Madau:2016jbv}.
Our results do not depend strongly on this particular choice, however.

As an example, Figure~\ref{fig:cluster-comparison-reweighted} shows the original range (light red) of properties among first-generation (top row) and higher-generation (bottom row) black hole mergers, as predicted by \textsc{CMC} for a cluster of stellar metallicity $Z=Z_\odot$.
Imposing an observational selection cut shifts the distributions of binary masses and mass ratios to higher values, at which expected SNRs are largest.
The effect is small, however.
The black hole spin distributions predicted by both \textsc{CMC} and \textsc{cBHBd} are sufficiently narrow (nearly delta functions at spin magnitudes of zero or $\sim 0.7$) that they are unaffected by a selection cut.
Despite the slight shift to larger mass ratios, mergers at the highest mass ratios are conversely suppressed by observational selection effects.
This is due to the fact that binaries with the largest total masses tend to be those with more unequal mass ratios.
Thus, although measured SNR is maximized by increasing mass ratio at fixed total mass, in practice we see that large SNRs most commonly occur for massive sources with somewhat unequal mass ratios.

\section{Further Details on Estimating Progenitor Binary Parameters}
\label{app:progenitor_appendix}

This appendix provides additional details regarding calculation of \gwTenEleven and \gwElevenTen's ancestral binaries, under the hypothesis that these events events contain second-generation  black holes born from a previous binary black hole merger.
As discussed in the main text, we approach this calculation in two ways.
In the \textit{Forward} approach, we adopt a hierarchical Bayesian formalism, placing astrophysically-informed priors on the hypothesized ancestral binaries and marginalizing over the source properties of \gwTenEleven and \gwElevenTen to directly obtain posteriors on ancestral properties.
In the \textit{Backward} approach, we more agnostically adopt the same uninformative priors on the source properties of \gwTenEleven and \gwElevenTen used in standard parameter estimation.
More details about each approach are described in Appendices~\ref{app:progenitor_appendix:method1} and~\ref{app:progenitor_appendix:method2} below.

\begin{table*}[]
    \footnotesize
    \setlength{\tabcolsep}{4pt}
    \renewcommand{\arraystretch}{1.2}
    \centering
    \caption{
    Properties of the ancestral binary black holes of \gwTenEleven and \gwElevenTen, under the hypothesis that the primary mass of each observed merger is itself a remnant from a previous merger.
    Specifically, we include constraints on the primary and secondary masses of the hypothesized ancestors, the ancestors' effective inspiral spins, and the recoil velocity experienced by the remnants following each merger.
    For each event, we show ancestral properties inferred under two different approaches.
    In the \textit{Forward} approach, we place astrophysically-informed priors directly on the ancestral properties \gwTenEleven and \gwElevenTen.
    In the \textit{Backward} approach, we agnostically maintain the source parameters of \gwTenEleven and \gwElevenTen inferred in standard parameter estimation and identify ancestral binaries compatible with these measured parameters.
    }
    \begin{tabular}{r  r  c c c c c}
    \hline \hline
    Event & Method & Ancestral $m_1\,[M_\odot]$ & Ancestral $m_2\,[M_\odot]$ & Ancestral $\chi_\mathrm{eff}$ & $v_\mathrm{recoil}\,[\mathrm{km}\,\mathrm{s}^{-1}]$ \\
    \hline
    \gwTenEleven 
    	& \textit{Forward}
    	& $\gwTenElevenProgMassOneMedianMahapatra^{+\gwTenElevenProgMassOneUpperErrorMahapatra}_{-\gwTenElevenProgMassOneLowerErrorMahapatra}$
	& $\gwTenElevenProgMassTwoMedianMahapatra^{+\gwTenElevenProgMassTwoUpperErrorMahapatra}_{-\gwTenElevenProgMassTwoLowerErrorMahapatra}$
	& $\gwTenElevenProgChiEffMedianMahapatra^{+\gwTenElevenProgChiEffUpperErrorMahapatra}_{-\gwTenElevenProgChiEffLowerErrorMahapatra}$
	& $\gwTenElevenProgRecoilMedianMahapatra^{+\gwTenElevenProgRecoilUpperErrorMahapatra}_{-\gwTenElevenProgRecoilLowerErrorMahapatra}$ \\
    \gwTenEleven 
    	& \textit{Backward}
    	& $\gwTenElevenProgMassOneMedianWong^{+\gwTenElevenProgMassOneUpperErrorWong}_{-\gwTenElevenProgMassOneLowerErrorWong}$
	& $\gwTenElevenProgMassTwoMedianWong^{+\gwTenElevenProgMassTwoUpperErrorWong}_{-\gwTenElevenProgMassTwoLowerErrorWong}$
	& $\gwTenElevenProgChiEffMedianWong^{+\gwTenElevenProgChiEffUpperErrorWong}_{-\gwTenElevenProgChiEffLowerErrorWong}$
	& $\gwTenElevenProgRecoilMedianWong^{+\gwTenElevenProgRecoilUpperErrorWong}_{-\gwTenElevenProgRecoilLowerErrorWong}$ \\
    \hline
    \gwElevenTen
    	& \textit{Forward}
    	& $\gwElevenTenProgMassOneMedianMahapatra^{+\gwElevenTenProgMassOneUpperErrorMahapatra}_{-\gwElevenTenProgMassOneLowerErrorMahapatra}$
	& $\gwElevenTenProgMassTwoMedianMahapatra^{+\gwElevenTenProgMassTwoUpperErrorMahapatra}_{-\gwElevenTenProgMassTwoLowerErrorMahapatra}$
	& $\gwElevenTenProgChiEffMedianMahapatra^{+\gwElevenTenProgChiEffUpperErrorMahapatra}_{-\gwElevenTenProgChiEffLowerErrorMahapatra}$
	& $\gwElevenTenProgRecoilMedianMahapatra^{+\gwElevenTenProgRecoilUpperErrorMahapatra}_{-\gwElevenTenProgRecoilLowerErrorMahapatra}$ \\
    \gwElevenTen
    	& \textit{Backward}
    	& $\gwElevenTenProgMassOneMedianWong^{+\gwElevenTenProgMassOneUpperErrorWong}_{-\gwElevenTenProgMassOneLowerErrorWong}$
	& $\gwElevenTenProgMassTwoMedianWong^{+\gwElevenTenProgMassTwoUpperErrorWong}_{-\gwElevenTenProgMassTwoLowerErrorWong}$
	& $\gwElevenTenProgChiEffMedianWong^{+\gwElevenTenProgChiEffUpperErrorWong}_{-\gwElevenTenProgChiEffLowerErrorWong}$
	& $\gwElevenTenProgRecoilMedianWong^{+\gwElevenTenProgRecoilUpperErrorWong}_{-\gwElevenTenProgRecoilLowerErrorWong}$ \\[1pt]
    \hline
    \hline
    \end{tabular}
    \label{tab:prog-params}
\end{table*}

\begin{figure*}[t!]
    \centering
    \includegraphics[width=\textwidth]{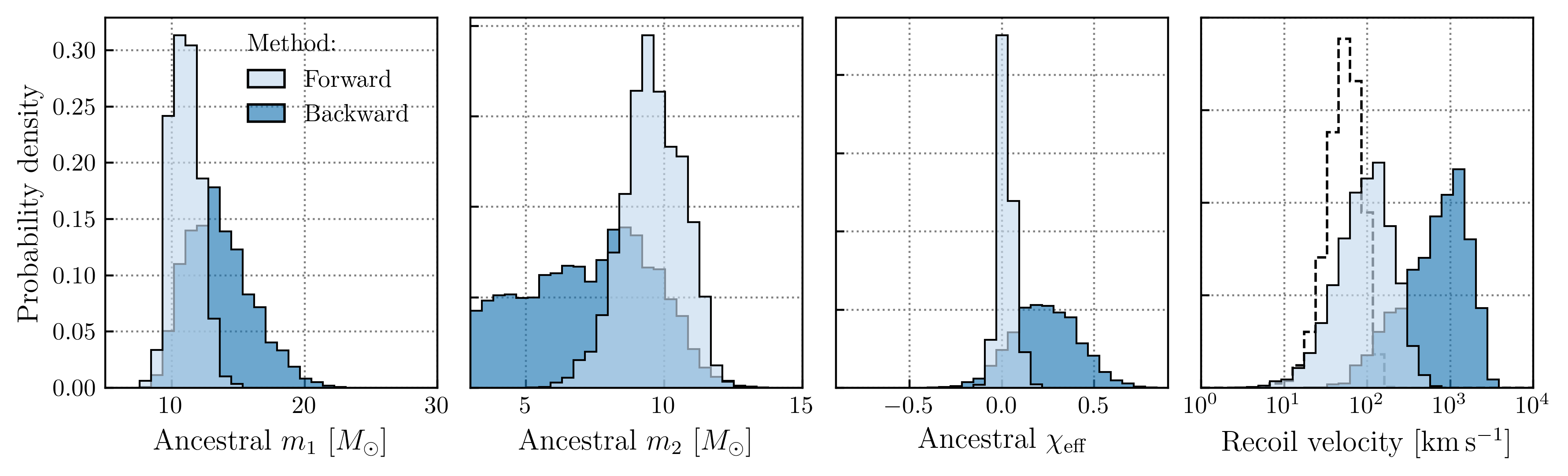} \\
    \includegraphics[width=\textwidth]{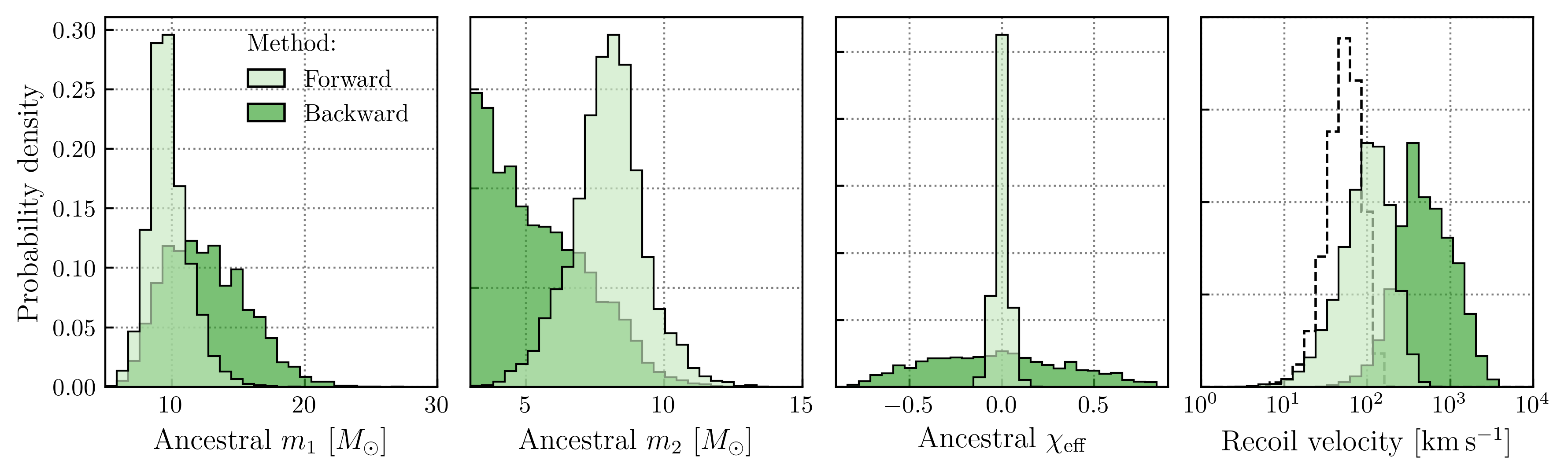}
    \caption{
    Inferred parameters on the ancestral binary black holes that previously merged to form the primary components of \gwTenEleven (top row) and \gwElevenTen (bottom row), under the hypothesis that each system contains a second-generation primary.
    We show results obtained under two sets of priors.
    Darker distributions correspond to an agnostic approach that leaves the priors on \gwTenEleven and \gwElevenTen's source properties unchanged, relative to standard parameter estimation.
    Lighter distributions correspond to an astrophysically-motivated approach, in which priors are placed on the ancestral binaries, inducing priors on \gwTenEleven and \gwElevenTen's primary masses and spins consistent with their hypothesized hierarchical origin.
    For comparison, the dashed histograms illustrate the distribution of cluster escape velocities predicted by \textsc{CMC}.
    In order for the sources of \gwTenEleven and \gwElevenTen to remain consistent with a hierarchical origin in globular clusters, their ancestral recoil velocities should not exceed these escape velocities.
    }
    \label{fig:progenitor-posteriors}
\end{figure*}

\subsection{The \textit{Forward} Approach}
\label{app:progenitor_appendix:method1}

Given the gravitational-wave data $d$ due to a binary black hole merger containing a second-generation remnant, we wish to obtain a probability distribution $p(\vec{\theta}_{\rm 1g} | d)$ on the properties $\vec{\theta}_{\rm 1g}$ of that black hole's first-generation ancestors.
Rather than repeat Bayesian parameter estimation, we proceed using the results of standard parameter estimation performed directly on \gwTenEleven and \gwElevenTen, constructing the posterior~\citep{Mahapatra:2024qsy}
	\begin{equation}
	\label{eq:Bayes-ancestral}
	p( \vec{\theta}_{\rm 1g} | d) = \frac{\pi(\vec{\theta}_{\rm 1g})}{{\mathcal{Z}_{\rm 1g}(d)}} \, \frac{p(\vec{\theta}_{\rm 2g}|d)}{\pi(\vec{\theta}_{\rm 2g})} \bigg\rvert_{\vec{\theta}_{\rm 2g}=\vec{F}(\vec{\theta}_{\rm 1g})}\,.
	\end{equation} 
Here, $\pi(\vec{\theta}_{\rm 1g} )$ is the prior distribution on $\vec{\theta}_{\rm 1g}$.
The quantity $\vec{\theta}_{\rm 2g}$ denotes parameters of the observed binary merger (i.e. \gwTenEleven  and \gwElevenTen themselves); $p(\vec{\theta}_{\rm 2g}|d)$ and $\pi(\vec{\theta}_{\rm 2g})$ are the posterior and prior probability distributions for these events obtained through ordinary parameter estimation.
The normalization constant
	\begin{equation}
	\mathcal{Z}_{\rm 1g}(d) \equiv \int \pi(\vec{\theta}_{\rm 1g})\, \frac{p(\vec{\theta}_{\rm 2g}|d)}{\pi(\vec{\theta}_{\rm 2g})} \bigg\rvert_{\vec{\theta}_{\rm 2g}=F(\vec{\theta}_{\rm 1g})}\, d\vec{\theta}_{\rm 1g}
	\end{equation}
is the evidence for $d$ under the hierarchical merger hypothesis, while $F(\vec{\theta}_{\rm 1g}) = \vec\theta_{\rm 2g}$ is the function mapping the ancestral binary's properties to the final mass and spin of its remnant black hole.
We compute $F(\vec{\theta}_{\rm 1g})$ using the the numerical-relativity remnant surrogate model \textsc{NRSur7dq4Remnant}~\citep{Varma:2019csw}, valid for binary black holes in quasi-circular orbits.
In practice, we reconstruct $p(\vec{\theta}_{\rm 2g}|d)$ and $\pi(\vec{\theta}_{\rm 2g})$ using a Gaussian kernel density estimator fit to samples from each distribution.
To sample possible ancestral properties from $p(\vec{\theta}_{\rm 1g}|d)$, we employ \BILBY~\citep{Ashton:2018jfp} using the \DYNESTY~\citep{Speagle:2019ivv} nested sampler.

We choose a prior distribution $\pi(\vec{\theta}_{\rm 1g}|d)$ that is a close match to the properties of first-generation binary mergers, as predicted in the \textsc{CMC} catalog and shown in Figure~\ref{fig:cluster-comparison}.
Upon combining successful mergers across all simulated clusters, the primary masses of first-generation mergers follow an approximately log-normal distribution, with best-fit mean $\mu_{\ln m_1} = 2.9$ and standard deviation $\sigma_{\ln m_1} = 0.6$, while mass ratios are well-described by a power law $p(q) \propto q^\beta$, with $\beta = 4.3$.
We adopt these distributions as our priors on ancestral primary masses and mass ratios.
In the \textsc{CMC} catalog, first-generation black holes are assumed to have identically zero spins.
We accordingly limit spins to be small but non-zero, adopting a half-normal prior distribution with a standard deviation of $0.1$ on both primary and secondary spin magnitudes.

\subsection{The \textit{Backward} Approach}
\label{app:progenitor_appendix:method2}

The above method sets priors on ancestral black hole parameters which, in turn, lead to posterior distributions on the source properties of the observed gravitational waves that are modified with respect to those obtained in standard parameter estimation.
As an alternative, the method presented in~\cite{Alvarez:2024dpd} preserves the agnostic priors on the source properties of \gwTenEleven and \gwElevenTen themselves, thereby preserving the original posteriors on the observed ``child'' binaries (provided that all child parameters can be realized with non-negligible probability from a chosen prior range of ancestral black holes). 
More precisely, this method yields a joint posterior $p(\vec\theta_\mathrm{1g}, \vec\theta_\mathrm{2g} | d)$ that, when marginalized over $\vec\theta_\mathrm{1g}$, yields a posterior that is proportional to our original posterior $p(\vec{\theta}|d)$, for all $\vec{\theta}_\mathrm{2g}$ satisfying $p(\vec\theta_\mathrm{2g} | \vec\theta_\mathrm{1g}) \neq 0$.
Equality is achieved when every child parameter is compatible with at least one ancestral configuration under the proposed prior.
This method differs from that of Appendix~\ref{app:progenitor_appendix:method1} simply by replacing the term $\pi(\vec\theta_\mathrm{2g})$ in Eq.~\eqref{eq:Bayes-ancestral}, which denotes the prior probability on the child parameters set by our original Bayesian parameter estimation, with $\pi(\vec\theta_\mathrm{1g})$, the prior probability on the child parameters induced by the priors on the ancestral ones.

In practice, we proceed by drawing $2 \times 10^6$ random samples from a broad ancestral prior $p(\vec\theta_\mathrm{1g})$.
Ancestral component masses are drawn uniformly from the range $[3 M_\odot,\,  300 M_\odot]$, with the mass ratio constrained to lie between $1/6$ and unity, and dimensionless spin magnitudes are sampled uniformly from the range $[0,\,0.99]$.
Remnant properties and a recoil kick are computed for each sample, yielding the  induced prior distribution $p(\vec\theta_{2g})$ on possible remnant properties.
For each primary mass and spin posterior sample $\{m_1, \chi_1\}$ obtained from $p(\vec\theta_\mathrm{2g}|d)$ via the original parameter estimation, we draw random samples from $p(\vec\theta_{2g})$ satisfying $\{m_f, \chi_f\} = \{m_1 \pm \delta m_1, \chi_1 \pm \delta \chi_1\}$.
We stress that, depending on the choice of ancestral prior $p(\vec\theta_\mathrm{1g})$, this will only be possible for a fraction of the posterior samples.
The usage of non-zero tolerances $\delta m_1$ and $\delta \chi_1$ is motivated by the fact that the discrete nature of our samples for $p(\vec\theta_\mathrm{1g})$ would naturally prevent us from encountering samples exactly matching the remnant posterior samples.
In our case, we choose $(\delta m_1, \delta \chi_1) = (1, 0.05)$, identified in \cite{Alvarez:2024dpd} to yield stable results.

\subsection{Progenitor Properties: Additional Results}

Table~\ref{tab:prog-params} and Figure~\ref{fig:progenitor-posteriors} present the posterior distributions on ancestral properties, as obtained through both the astrophysically-motivated \textit{Forward} approach and the agnostic \textit{Backward} approach.
Both approaches yield similar estimates for the ancestral primary masses of \gwTenEleven and \gwElevenTen.
The \textit{Forward} approach, with more stringent priors favoring equal mass ratios, yields more precise estimates of the ancestors' secondary masses.
Under an agnostic prior, \gwTenEleven is consistent with an ancestral binary with moderately large, positive effective spin, although both \gwTenEleven and \gwElevenTen are also consistent with non-spinning ancestors.
Recoil velocities are almost entirely dominated by spin priors; the low ancestral spins required in the \textit{Forward} approach limit recoil kicks to lower velocities, while the large spins allowed in the \textit{Backward} approach yield more rapid recoils.

Despite recoil velocity posteriors being prior dominated, for self-consistency it is valuable to check that these posteriors are not inconsistent with expected cluster escape velocities.
The dashed histograms in Fig.~\ref{fig:progenitor-posteriors} show the distribution of cluster escape velocities predicted in the \textsc{CMC} catalog.
If the sources of \gwTenEleven and \gwElevenTen are to be consistent with a hierarchical origin in dense clusters (or at least clusters with masses similar to those explored in the \textsc{CMC} catalog), their inferred ancestral recoil velocities must not be larger than these predicted escape velocities.
The \textit{Forward} approach yields recoil velocity posteriors consistent with the distribution of expected escape velocities.
The \textit{Backward} approach yields a posterior with support in the range of expected escape velocities, although the posterior also extends to much higher recoil velocities; this is expected, due to the deliberately agnostic prior allowing for rapidly spinning ancestors.

\section{Tests of Fundamental Physics: Further Details}
\label{app:tgr}

\subsection{Spin-induced quadrupole moment}
\label{app:tgr:siqm}

In this appendix we provide further detail regarding constraints on \gwTenEleven's spin-induced quadrupole moment, discussed in Sec.~\ref{sec:physics:siqm}.

When testing for the spin-induced quadrupole moment, corrections are added to the inspiral phase of the gravitational waveform, with uniform priors adopted on the correction parameters.
In addition to the results shown in Figure~\ref{fig:siqm}, based on corrections to the \textsc{IMRPhenomXPHM} waveform model~\citep{Pratten:2020ceb, Divyajyoti:2023izl},
we additionally explored results using corrections to the \textsc{SEOBNRv5HM\_ROM}~\citep{Pompili:2023tna} waveform model under the Flexible Theory-Independent framework~\citep{Mehta:2022pcn}.
Besides relying on different base waveform models, these two approaches differ in the treatment of the binary's evolution from inspiral to merger and ringdown.
Additionally, the \textsc{IMRPhenomXPHM} includes spin precession while \textsc{SEOBNRv5HM\_ROM} assumes spin-aligned binaries.
Given the significant spin precession effects in \gwTenEleven, the \textsc{IMRPhenomXPHM}-based results from Figure~\ref{fig:siqm} should therefore be taken as the most physically-meaningful constraints on the spin-induced quadrupole, but the \textsc{SEOBNRv5HM\_ROM} results are helpful in quantifying the degree of systematic uncertainties arising from missing waveform physics.

The \textsc{SEOBNRv5HM\_ROM}-based test gives $\delta \kappa_1 = -0.44^{+1.66}_{-2.01}$ and $\delta \kappa_s = -0.45^{+0.39}_{-1.01}$.
While $\delta \kappa_1$ is consistent with a Kerr black hole in GR, $\delta \kappa_s$ is shifted slightly towards negative values, with the Kerr value lying beyond the $95\%$ credible interval.
This bias, as well as the broadened posteriors relative to \textsc{IMRPhenomXPHM results} is due to the absence of relativistic spin-orbit precession noted above~\citep{Lyu:2023zxv, Divyajyoti:2023izl}.
To verify this, we analyzed simulated signals with source parameters consistent with those of \gwTenEleven.
We considered both signals with misaligned precessing spins and with aligned spin configurations.
When analyzing simulated aligned-spin signals with \textsc{SEOBNRv5HM\_ROM}, posteriors are unbiased and peak near zero.
When instead analyzing simulated precessing signals, posteriors are instead biased towards negative values, as in the case of \gwTenEleven.

\subsection{Subdominant mode amplitude consistency}
\label{app:tgr:sma}

The gravitational-wave strain emitted by compact binary mergers can be decomposed into spin-weighted spherical harmonics of weight $-2$ as:
\begin{equation}
h(t,\theta_{JN},\boldsymbol{\lambda})=\sum_{\ell\geq 2}\sum_{m=-\ell}^{\ell}h_{\ell m}(t,\boldsymbol{\lambda})\,{}_{-2}Y_{\ell m}(\theta_{JN},\phi_0),
\end{equation}
where $(\theta_{JN},\phi_0)$ specify the observer’s orientation in the source frame, and $\boldsymbol{\lambda}$ encodes intrinsic parameters such as masses and spins. 
Following common practice, we fix $\phi_0=0$ so that $\theta_{JN}$ denotes the angle between the binary’s total angular momentum vector and the observer’s line of sight~\citep{Pratten:2020ceb}.

Typically, gravitational-wave signals are dominated by the quadrupole $(\ell,m)=(2,\pm 2)$ multipole. 
However, higher-order multipoles such as $(2,\pm1)$ and $(3,\pm3)$ become increasingly relevant for systems with unequal masses or for viewing angles away from face-on ($\theta_{JN}\neq 0$). 
To quantify potential deviations of these subdominant multipoles from GR, the subdominant multipole amplitude (SMA) test~\citep{Puecher:2022sfm} introduces amplitude deviations $\delta A_{\ell m}$ explicitly into the $(2,\pm1)$ and the $(3,\pm3)$ multipoles in the \textsc{XPHM} waveform model:

\begin{align}
h(t,\theta_{JN},\boldsymbol{\lambda})&=\sum_{m=\pm2}h_{2m}(t,\boldsymbol{\lambda})\,{}_{-2}Y_{2m}(\theta_{JN},0) + \sum_{m=\pm1}(1+\delta A_{21})\,h_{2m}(t,\boldsymbol{\lambda})\,{}_{-2}Y_{2m}(\theta_{JN},0)\notag\\
&+\sum_{m=\pm3}(1+\delta A_{33})\,h_{3m}(t,\boldsymbol{\lambda})\,{}_{-2}Y_{3m}(\theta_{JN},0) + \sum_{\text{other HOM}}h_{\ell m}(t,\boldsymbol{\lambda})\,{}_{-2}Y_{\ell m}(\theta_{JN},0).
\end{align}

The application of the SMA test requires sufficient SNR in the multipole of interest. For each $(\ell, \pm m)$ mode, the mode-specific SNR, $\rho_{\ell m}$, is computed by projecting $h_{\ell \pm m}$ onto the subspace orthogonal to the dominant $(2,\pm2)$ mode and evaluating the optimal SNR of the residual~\citep{Mills:2020thr}. 
In the absence of the $(\ell, \pm m)$ multipole in signal, $\rho_{\ell m}$ follows a $\chi$ distribution with two degrees of freedom in Gaussian noise, portrayed as the null distribution in Figure~\ref{fig:snrs}. 
To ensure sufficient mode content, we adopt a conservative selection threshold: a given mode for an event is included in the SMA analysis only if the lower bound of the 68\% credible interval of its $\rho_{\ell m}$ distribution exceeds 2.145, corresponding to the 90th percentile of the null distribution. 
Among the two considered events, only the $(3,\pm 3)$ mode in \gwTenEleven satisfies this criterion. 
Accordingly, we perform parameter estimation for this mode with a uniform prior on $\delta A_{33}$ in the range $[-10, 10]$.

\subsection{Constraining ultralight bosons through superradiance}
\label{app:tgr:bosons}

In the presence of an ultralight scalar or vector field, a spinning black hole is unstable to the superradiant instability~\citep{Brito:2015oca}. 
Oscillating bosonic modes that satisfy the superradiant condition, $\omega_R<m \Omega_{\rm BH}$, grow exponentially with time at the expense of the black hole's rotational energy.
Here $\omega_R\sim m_b c^2/\hbar$ is the angular frequency, $m$ is the azimuthal number, and $\Omega_{\rm BH}$ is the horizon frequency of the black hole.
The growth of a mode and spindown of the black hole persist until the superradiant condition is saturated~\citep{Arvanitaki:2009fg,Brito:2014wla,East:2017ovw,East:2018glu}. 
Typically, the lowest $m$ mode that is superradiant grows the fastest, but with sufficient time multiple modes can grow and saturate, reducing the black hole's spin to ever smaller values.
The vector boson instability rate is parametrically faster than the scalar rate, such that vector bosons spin down a black hole to lower values in a fixed time.
For example, adopting the median mass and spin values for the primary black hole in \gwTenEleven (Table~\ref{tab:pe-params}), the shortest $e$-folding time for mass growth of a boson cloud is $\sim 8\,\mathrm{s}$ for a vector boson and $9\,\mathrm{hr}$ for a scalar.
This assumes a boson mass optimally matched to the black hole; for smaller boson masses the instability timescales lengthen---roughly as $\propto m_b^{-7}$ and $\propto m_b^{-9}$ for vector and scalar modes, respectively~\citep{Baryakhtar:2017ngi}.

Given a boson mass and a time since black hole formation (or the time since the black hole gained angular momentum), there will be excluded regions of the black hole's mass--spin parameter space where the superradiant instability should have reduced the black hole's spin to lower values.
We calculate this excluded region using the \textsc{SuperRad} package~\citep{Siemonsen:2022yyf,May:2024npn}, based on the linear superradiantly unstable modes and including all relevant azimuthal number modes, for both vectors and scalars. 
For the dominant modes, with $m \leq 2$, we use the relativistic frequencies and instability rates.
For higher azimuthal modes, we use a non-relativistic approximation, which in general underestimates the instability rate and is thus conservative.

Using the posteriors for the source primary masses and spins of \gwTenEleven and \gwElevenTen, we determine the fraction $P$ of each event's posterior samples lying in the region permitted by the given boson mass and black hole age~\citep{Aswathi:2025nxa}.
Sufficiently small $P$ would disfavor this mass--age combination.
It is important, however, to guard against prior effects; it is possible for uninformative data, with spin magnitude samples drawn randomly from a uniform prior, to yield small $P$ for specific ultralight boson masses.
Following \cite{Aswathi:2025nxa}, we therefore also calculate a prior fraction $P'$, obtained by replacing the black hole spin posterior with samples drawn from a uniform distribution $[0,1)$.
The exclusion regions shown in Figure~\ref{fig:bosons} correspond to the requirement that $P<0.1\,P'$; this corresponds to a $90\%$ or better credible bound while also ensuring that constraints are strongly likelihood-driven.

This analysis assumes only a minimally coupled boson with gravitational interactions, but will also apply to scalar or vector bosons with sufficiently weak interactions so as to not disrupt the black hole spindown. 
We also neglect the impact of gravitational effects from the binary companion on the superradiant growth of the boson cloud, implicitly assuming the growth would occur at large enough separation for this to be negligible.

\bibliography{references}{}
\bibliographystyle{aasjournal}

\end{document}